\documentclass[aps,twocolumn,nofootinbib,preprintnumbers,eqsecnum]{revtex4-1}

\usepackage{color}

\usepackage[
      colorlinks=true,
      linkcolor=blue,
      urlcolor=blue,
      filecolor=black,
      citecolor=red,
      pdfstartview=FitV,
      pdftitle={},
        pdfauthor={Thomas Faulkner, Hong Liu, Mukund Rangamani},
        pdfsubject={},
        pdfkeywords={},
        pdfpagemode=None,
        bookmarksopen=true
      ]{hyperref}

\usepackage[normalem]{ulem}
\usepackage{amsmath}
\usepackage{enumerate}
\usepackage{amsfonts}
\usepackage{yfonts}

\usepackage{subfigure}
\usepackage{array}

\usepackage{psfrag}

\usepackage{epsfig}
\usepackage[latin1]{inputenc}
\usepackage{float}
\usepackage{graphicx}
\usepackage{cancel}
\usepackage{mathrsfs}
\usepackage{amssymb}
\usepackage{amsfonts}
\usepackage{amsmath}
\usepackage{slashed}
\usepackage{mathtools}

\newcommand \tr {\mbox{{\bf Tr}}}

\usepackage{graphicx}
\usepackage{bm}

\def\({\left(}
\def\){\right)}
\def\[{\left[}
\def\]{\right]}
\def\<{\langle}
\def\>{\rangle}

\def\tr{\mathop{\rm tr}}

\newcommand\half{{\ensuremath{\frac{1}{2}}}}
\newcommand\p{\ensuremath{\partial}}

\newcommand\vev[1]{{\ensuremath{\left\langle{#1}\right\rangle}}}

\newcommand{\be}{\begin{equation}}
\newcommand{\ee}{\end{equation}}
\newcommand{\bea}{\begin{eqnarray}}
\newcommand{\eea}{\end{eqnarray}}
\newcommand{\bwt}{\begin{widetext}}
\newcommand{\ewt}{\end{widetext}}
\newcommand{\nn}{\nonumber\\}
\newcommand{\bi}{\begin{itemize}}
\newcommand{\ei}{\end{itemize}}
\newcommand{\ben}{\begin{enumerate}}
\newcommand{\een}{\end{enumerate}}
\newcommand{\bca}{\begin{cases}}
\newcommand{\eca}{\end{cases}}
\newcommand{\bln}{\begin{align}}
\newcommand{\eln}{\end{align}}
\newcommand{\bst}{\begin{split}}
\newcommand{\est}{\end{split}}

\newcommand\al{{\alpha}}
\newcommand\ep{\epsilon}
\newcommand\sig{\sigma}
\newcommand\Sig{\Sigma}
\newcommand\lam{\lambda}

\newcommand\om{\omega}
\newcommand\Om{\Omega}

\newcommand\ga{{\ensuremath{{\gamma}}}}
\newcommand\Ga{{\ensuremath{{\Gamma}}}}
\newcommand\de{{\ensuremath{{\delta}}}}
\newcommand\De{{\ensuremath{{\Delta}}}}

\def\th{{\theta}}

\newcommand\ov{\over}
\newcommand\ha{{\half}}

\newcommand\apr{{\ensuremath{{\alpha'}}}}
\def\le{\left}
\def\ri{\right}

\newcommand\sA{{\ensuremath{{\mathcal A}}}}

\newcommand\sE{{\ensuremath{{\mathcal E}}}}

\newcommand\sL{{\ensuremath{{\mathcal L}}}}

\newcommand\sN{{\ensuremath{{\mathcal N}}}}
\newcommand\sO{{\ensuremath{{\mathcal O}}}}
\newcommand\sR{{\ensuremath{{\mathcal R}}}}

\newcommand\sS{{\mathcal S}}

\newcommand{\ka}{{\kappa}}

\newcommand{\ft}{\mathfrak{t}}
\newcommand{\fa}{\mathfrak{a}}
\newcommand{\fR}{\mathfrak{R}}

\newcommand{\eql}{\ell_{\rm eq}}
\newcommand{\Vee}{v_{E}}

\begin{document}

\title {Entanglement growth during thermalization in holographic systems
}

\preprint{MIT-CTP 4510}

\author{ Hong Liu}
\affiliation{Center for Theoretical Physics,
Massachusetts
Institute of Technology,
Cambridge, MA 02139 }
\author{S.~Josephine Suh}
\affiliation{Center for Theoretical Physics, Massachusetts Institute of Technology,
Cambridge, MA 02139 }

\begin{abstract}

We derive in detail several universal features in the time evolution of entanglement entropy and other nonlocal observables in quenched holographic systems. The quenches are such that a spatially uniform density of energy is injected at an instant in time, exciting a strongly coupled CFT which eventually equilibrates. Such quench processes are described on the gravity side by the gravitational collapse of a thin shell that results in a black hole. Various nonlocal observables have a unified description in terms of the area of extremal surfaces of different dimensions.  In the large distance limit, the evolution of an extremal surface, and thus the corresponding boundary observable, is controlled by the geometry around and inside the event horizon of the black hole, allowing us to identify regimes of pre-local- equilibration quadratic growth, post-local-equilibration linear growth, a memory loss regime, and a saturation regime with behavior resembling those in phase transitions.
We also discuss possible bounds on the maximal rate of entanglement growth in relativistic systems.

\end{abstract}

\today

\maketitle

\tableofcontents


\section{Introduction}

Understanding whether and how  quantum matter equilibrate 
is a question of much importance in many different areas of physics. 
Yet such non-equilibrium problems are notoriously difficult to deal with; universal characterizations 
are scarce and far between. 

For a non-integrable system it is expected that a generic (sufficiently excited) non-equilibrium state eventually thermalizes. 
For strongly coupled systems with a gravity dual this expectation is borne out 
as holographic duality maps equilibration from such a state  to black hole formation from a  gravitational collapse,
and  gravitational collapse of a sufficiently massive body is indeed generic in General Relativity.

 Questions related to equilibration then become intimately connected to  those of black hole physics.
 This  on the one hand brings in powerful gravity techniques for studying thermalization processes, and on the other gives new perspectives on the quantum nature of black holes. 

One of the simplest settings for equilibration is the evolution of a system after a global quench, which
can be divided into two types. In the first type  one changes some  parameter(s) of a system at $\ft=0$ within a short interval $\de \ft$. The previous ground state becomes an excited state with respect to the new Hamiltonian 
and evolves to equilibrium under the evolution of the new Hamiltonian. In the second type, 
 one turns on a uniform density of sources for a short interval $\de \ft$ at $\ft =0$ 
and then turns it off. The work done by the source takes the system to an excited state which subsequently equilibrates (under the evolution of the same Hamiltonian before the quench). 
In both situations, the interval $\de \ft$ is taken to be much smaller than any other scale in the system. For convenience we will take $\de \ft$ to be zero in subsequent discussions. 

In $(1+1)$-dimension, by tuning a parameter of a gapped system to criticality Calabrese and Cardy found that~\cite{Calabrese:2005in} 
 the entanglement entropy for a segment of size $2 R$ grows with time linearly as
\be \label{ekko}
\De S (\ft, R) = 2 \, \ft \, s_{\rm eq}, \qquad \ft < R
\ee
 and saturates at the equilibrium value at a sharp saturation time $\ft_s = R$. In the above equation
 $\De S$ denotes difference of the entanglement entropy from that at $\ft=0$ and $s_{\rm eq}$ is the equilibrium thermal entropy density.  Furthermore, they showed that this remarkably simple behavior 
 can be understood from a simple model of 
entanglement propagation using free-streaming quasiparticles traveling at the speed of light.  

Subsequently, the linear behavior~\eqref{ekko} was  found in holographic context for $(1+1)$-dimensional systems dual to a bulk Vaidya geometry~\cite{AbajoArrastia:2010yt} (see also~\cite{Balasubramanian:2010ce,Balasubramanian:2011ur}).  An AdS Vaidya geometry, as we will review in more detail in Sec.~\ref{sec:vaidya}, describes the gravitational collapse of a thin shell of matter to form a black hole. It corresponds to a quench process of the second type in a boundary conformal field theory, where at $\ft =0$, a uniform density of operators are inserted for a very short time. The entanglement entropy is obtained from
the area of an extremal surface in the Vaidya geometry with appropriate boundary conditions~\cite{Ryu:2006bv,Ryu:2006ef,Hubeny:2007xt}.

The agreement of results between the very different setups of ~\cite{Calabrese:2005in} and~\cite{AbajoArrastia:2010yt} is in some sense not surprising.
Both setups involve a homogeneous excited initial state  evolving under a gapless Hamiltonian, and the powerful boundary CFT techniques of~\cite{Calabrese:2005in} 
should apply in both contexts. Behavior similar to that of entanglement entropy has also been found in correlation functions in both contexts~\cite{Calabrese:2006rx,Balasubramanian:2010ce,Balasubramanian:2011ur, Aparicio:2011zy} (see also~\cite{Ryu:2006bv,Takayanagi:2010wp,Balasubramanian:2012tu,Baron:2012fv,Caceres:2012em,Galante:2012pv,Asplund:2011cq,Erdmenger:2012xu,Li:2013sia,Arefeva:2013wma,Caputa:2013eka,kimhuse} for other studies of two-dimensional systems). 

Given the simplicity and elegance of~\eqref{ekko}, it is natural to wonder: (i) whether similar 
linear growth occurs in higher dimensions;  (ii) whether other nonlocal observables such as equal-time 
correlation functions and Wilson loops share similar behavior; (iii) if such linear growth exists, whether it can still 
be understood from free-streaming quasiparticles. 

For entanglement entropy we recently reported the answers to some of these questions for a class of quenched holographic systems~\cite{Liu:2013iza}. Interested in long-distance physics, we focused on entangled regions of large size, and found that the time evolution of entanglement entropy  is characterized by four different scaling regimes: 

\ben 

\item Pre-local-equilibration quadratic growth in time.
 
\item Post-local-equilibration linear growth in time.
 
\item  A saturation regime in which the entanglement entropy saturates its equilibrium value. 
The saturation can be either continuous or discontinuous depending on whether the 
time derivative of the entanglement entropy is continuous at saturation. 
In the continuous case saturation is characterized by a ``critical'' exponent. 

\item When the entangled region is a sphere, there is an additional scaling regime between linear growth and saturation, which we dub ``late time memory-loss", and in which the entanglement entropy only depends on 
time remaining till saturation, and not on the size of the region and time separately.

\een 
These results are generic in the sense that they are insensitive to the specific details of the system as well as those of the quench. 

The above scaling regimes were obtained by  identifying various geometric regimes for the bulk extremal surface. An important observation was the existence of a family of ``critical extremal surfaces'' which lie behind the horizon and separate extremal surfaces that reach the boundary from those which fall into the black hole singularity. 
In the large size limit, one finds that the time evolution of entanglement entropy is controlled by these critical extremal surfaces. In this paper we give a detailed derivation of these results and provide generalizations to other non-local observables such as equal-time correlation functions and Wilson loops. 

Also, with M\'ark Mezei~\cite{speed}, we generalized the free-streaming model of~\cite{Calabrese:2005in} to higher dimensions. It turns out that such a model also exhibits post-local-equilibration linear growth of entanglement entropy, but that intriguingly, the rate of growth of entanglement entropy resulting from free-streaming particles moving at the speed of light is less than what we find here for strongly coupled holographic systems. 

In~\cite{Liu:2013iza}, we argued that the evolution of entanglement entropy can be captured by 
the picture of an entanglement wave propagating inward from the boundary of the entangled region, which we called an ``entanglement tsunami" (see also~\cite{kimhuse}).
There we also suggested a possible upper bound on the rate of entanglement growth in relativistic 
systems. The results of~\cite{Liu:2013iza} and the current paper also have potential applications 
for various issues associated with black hole physics. The fact that the growth of entanglement is controlled by some critical extremal surfaces inside the horizon of a 
collapsing black hole also suggests new avenues for probing physics beyond horizons in holography. 
Similar processes as we consider here were also considered in~\cite{Shenker:2013pqa} to obtain insights into the ``scrambling time'' of a black hole.
We will elaborate more on these issues in the discussion section. 

To conclude this introduction, we note that earlier work on quenches in higher dimensional holographic systems include~\cite{Albash:2010mv,Balasubramanian:2010ce,Balasubramanian:2011ur,Hartman:2013qma} (see 
also~\cite{Baron:2012fv,Keranen:2012tv,Keranen:2011xs,Caceres:2012em,Galante:2012pv,Basu:2011ft,Li:2013sia}). In particular, for $d=3$,  a linear growth toward saturation was mentioned in~\cite{Albash:2010mv}, although it appears that the linear regime mentioned 
in~\cite{Albash:2010mv} is different from that of~\cite{Liu:2013iza} and the current paper.
Ref.~\cite{Albash:2010mv} was also the first to observe discontinuous saturation in various examples. In~\cite{Balasubramanian:2010ce,Balasubramanian:2011ur} 
non-analyticity near saturation was emphasized.  
In a different 
gravity setup, linear growth of entanglement entropy was also observed~\cite{Hartman:2013qma}, whose 
connection to that in~\cite{Liu:2013iza} will be discussed in detail in the main text. 
In~\cite{Caceres:2012em} it was pointed out that the presence of a nonzero chemical potential in the final 
equilibrium state tends to slow the growth of  entanglement.






\section{General setup} \label{sec:setup}

In this paper we consider the evolution of  various non-local observables, including entanglement entropy, equal-time correlation functions, and Wilson loops, after a sharp quench of a strongly coupled {\it gapless} system with a gravity dual. More explicitly, at $\ft =0$ in the boundary system we turn on a spatially uniform density of external sources for an interval $ \de \ft$, creating a spatially homogeneous and isotropic excited state with nonzero energy density, which subsequently equilibrates. 
The precise manner (e.g. what kind of sources are turned on and how) through which the excited state is generated and its microscopic details will not concern us. We are interested in the macroscopic 
behavior of the system at large distances and in extracting ``universal'' behavior in
the evolution of these observables that are insensitive to the  
specific nature of final equilibrium states. 

\begin{figure}[!h]
\begin{center}
\includegraphics[scale=0.3]{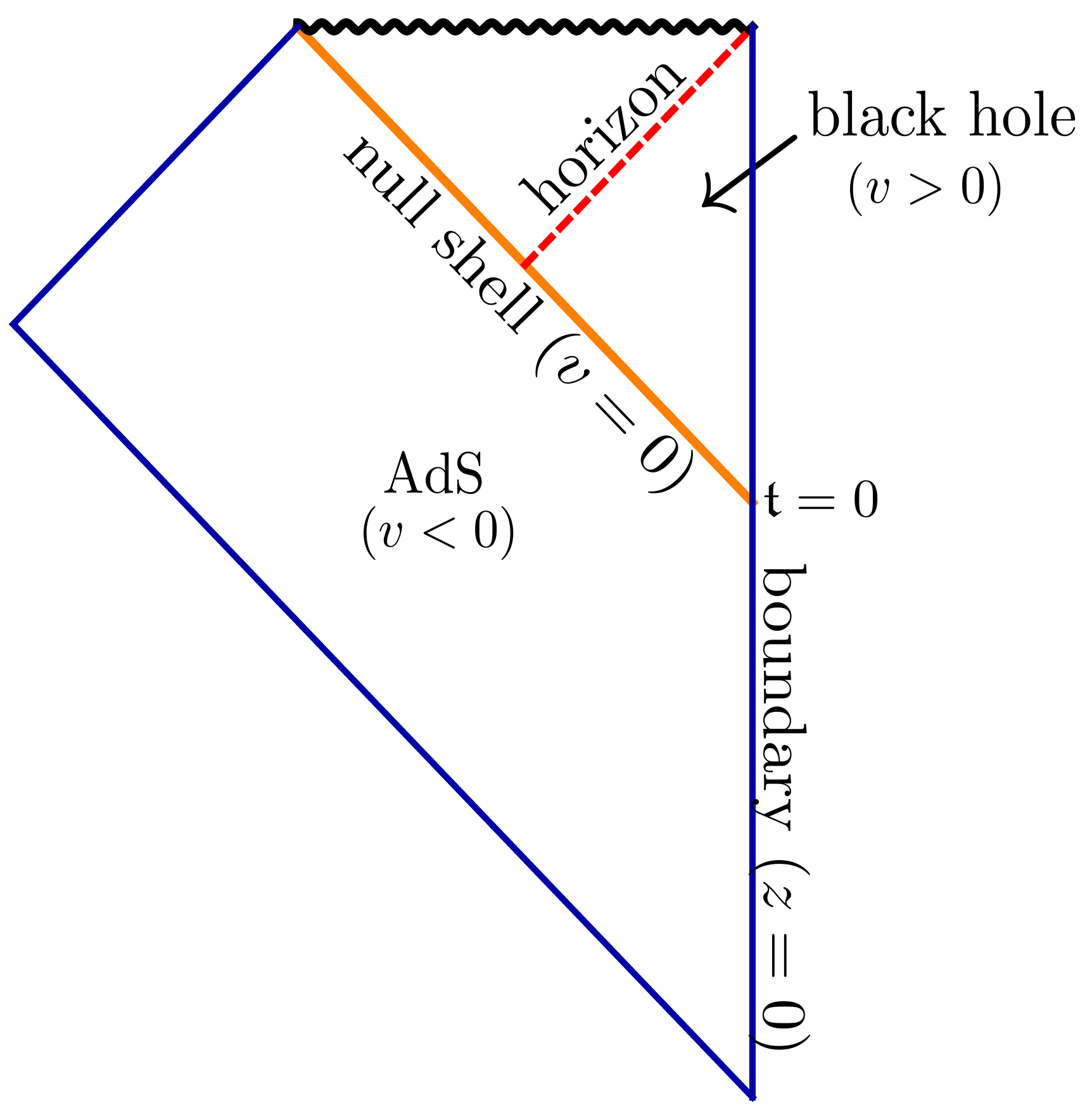}
\end{center}
\caption{Vaidya geometry: One patches pure AdS with a black hole along an in-falling collapsing null 
shell located at $v=0$. We take the width of the shell to be zero which corresponds to the $\de \ft =0$ limit 
of the boundary quench process. 
The spatial directions along the boundary are suppressed in the figure.} \label{fig:vaidya}
\end{figure}

On the gravity side such a quench process is described by a thin shell 
of matter starting from the boundary and collapsing to form a black hole, which can in turn be described by a Vaidya metric, see Fig.~\ref{fig:vaidya}. The matter fields making up the shell and their configuration are determined by the sourcing process in the boundary theory  and are again not important for our purposes.
See e.g.~\cite{Bhattacharyya:2009uu,Lin:2008rw,Garfinkle:2011hm,Baier:2012tc,Wu:2012rib,Lin:2013sga,Wu:2013qi} for more explicit discussions.   
In the classical gravity regime we are working with, which translates to the large $N$ and strongly coupled limit of the boundary theory, all of our observables are only sensitive to the metric of the collapsing geometry.  

In this section we give a detailed description of our setup and review the 
vacuum and equilibrium properties of the class of systems under consideration.

\subsection{Vaidya metric}  \label{sec:vaidya}


We consider a metric of the form
\be \label{vaidya}
ds^2 = {L^2 \ov z^2} \le(- f (v,z) dv^2 - 2 dvdz + d \vec x^2 \ri) \ .
\ee
In the limit the sourcing interval $\de \ft$ goes to zero, the width of the collapsing shell goes to zero and $f(v,z)$ can be expressed in terms of a step function 
\be
f (v,z) = 1-  \th (v) g(z) \ . 
\ee
For $v< 0$, the metric is given by that of pure AdS, 
\be 
ds^2 = {L^2 \ov z^2} \le(-dt^2 + dz^2 + d \vec x^2 \ri)
\ee
where
\be \label{tzco1}
v = t - z \ , \qquad t = v + z \ .
\ee
For $v > 0$,~\eqref{vaidya} is given by that of a black hole in Eddington-Finkelstein 
coordinates,
\be \label{BHre1}
ds^2 = {L^2 \ov z^2} \le(- h (z) dv^2 - 2  dvdz + d \vec x^2 \ri)\ , 
\ee
which in terms of the usual Schwarzschild time $t$ can be written as
\be \label{BHre2}
ds^2 = {L^2 \ov z^2} \le(- h (z) dt^2 + {1 \ov h(z)} dz^2 + d \vec x^2 \ri) 
\ee
with
\be \label{tzco2}
h(z) \equiv 1 - g(z)\ , \quad v = t - \sig (z) \ , \quad \sig (z) = \int_0^z {dz' \ov h(z')} \ .
\ee
The functions $h(z)$ in the black hole metric~\eqref{BHre1}--\eqref{BHre2} may be interpreted as ``parameterizing'' 
different types of equilibration processes with different final equilibrium states.  
We assume that~\eqref{vaidya} with some $g(z)$ can always be achieved by choosing an appropriate configuration of matter fields. In following discussions we will not need the explicit form of $h(z)$, and only that it gives rise to a black hole metric. We will work with a general boundary spacetime dimension $d$.

More explicitly, we assume $h(z)$ has a simple zero at the horizon $z = z_h > 0$, 
and that for $z < z_h$,  it
is positive and monotonically decreasing as a function
of $z$ as required by the IR/UV connection.
As we approach the boundary, i.e. as $z \to 0$, $h(z)$ approaches zero with the leading behavior
\be \label{asymMe}
h (z) =1-  M z^d + \cdots 
\ee
where $M$ is some constant.  From~\eqref{asymMe}, one obtains that the energy density of the equilibrium state is
\be \label{endgd}
\sE = {L^{d-1} \ov 8 \pi G_N} {d-1 \ov 2} M \ , 
\ee
while its temperature and entropy density are given by
\be \label{eqent}
T = {|h'(z_h)| \ov 4 \pi} 
\ , \qquad
s_{\rm eq} = {L^{d-1} \ov z_h^{d-1}} {1 \ov 4 G_N}
\ . 
\ee

Representative examples of~\eqref{BHre1} include the AdS Schwarzschild black hole with
\be \label{schw}
h(z) = 1- {z^d \ov z_h^d}
\ee
which describes a neutral final equilibrium state, 
and the AdS Reissner-Nordstrom (RN) black hole  with
\be \label{eq:RN}
h(z) =  1- M z^d + Q^2 z^{2d-2}\ ,­
\ee
which describes a final equilibrium state with a nonzero chemical potential for some conserved charge.  

A characteristic scale of the black hole geometry~\eqref{BHre1}--\eqref{BHre2} is the horizon size\footnote{Note that while the horizon location is a coordinate dependent quantity, in the particular radial coordinate used in~\eqref{BHre1}--\eqref{BHre2} $z_h$ corresponds to a meaningful boundary scale as for example indicated by~\eqref{mfp}.}
$z_h$ which from~\eqref{eqent} can be expressed in terms of the entropy density $s_{\rm eq}$ as
\be \label{mfp}
z_h  =  \le({L^{d-1} \ov 4 G_N} {1 \ov s_{\rm eq}}\ri)^{1 \ov d-1}  \ . 
\ee
Were we considering a gas of quasiparticles, the prefactor ${L^{d-1} \ov 4 G_N}$ in~\eqref{mfp} could be interpreted as the number of internal degrees of freedom of a quasiparticle, 
and $z_h$ would then be the average distance between quasiparticles, or mean free path. 
Here of course we are considering  strongly coupled systems which do {\it not} have a quasiparticle description. Nevertheless, $z_h$ provides a characteristic scale of of the equilibrium state. For example, as we will see below it controls the correlation length of equal-time correlation functions and Wilson loops in equilibrium. 

For the collapsing process described by~\eqref{vaidya} we can also identify $z_h$ as 
 a ``local equilibrium scale'' $\ell_{\rm eq}$, 
which can be
defined as the time scale when the system has ceased production of thermodynamic entropy, or in other words, has achieved local equilibrium at distance scales of order the ``mean free path'' of the equilibrium state. We will discuss further support for this identification at the end of Sec.~\ref{sec:dcal}.

We note that in the AdS Schwarzschild case~\eqref{schw}, the temperature $T$ is the only scale and controls both the local equilibrium scale $z_h$ and energy density $\sE$ (given by~\eqref{endgd}), 
\be \label{sche}
 T = {d \ov 4 \pi z_h} \ , \qquad M = {1 \ov z_h^d} = \le({4 \pi T \ov d}\ri)^d \ ,
\ee 
but that in a system with more than one scale as in the Reissner-Nordstrom case, $z_h$ and $\sE$ (or $M$) do not depend only on $T$. In the Reissner-Norstrom case, it is convenient to introduce a quantity 
\be \label{udef}
u \equiv {4 \pi z_h T \ov d} 
\ee
which decreases monotonically from its Schwarzschild value of unity to $0$, as
the chemical potential  is increased from zero to infinity at fixed $T$. Thus with a large chemical potential (compared to temperature), the local equilibrium scale $\eql \sim z_h$ can be much smaller than the thermal wave length $1/T$. In this regime, the system is controlled by finite density physics which gives rise to the scale $z_h$. For recent related discussions, see~\cite{Davison:2013bxa}.

Finally, we note that the metric~\eqref{vaidya} is not of the most general form describing a spatially homogenous and isotropic
equilibration process. If the equilibrium state has a nontrivial expectation value for (or sourced by) some scalar operators, the metric has the form 
\be \label{vaidya1}
ds^2 = {L^2 \ov z^2} \le(- f (v,z) dv^2 - 2 q(v, z) dvdz + d \vec x^2 \ri) 
\ee
with $f (v,z) = 1-  \th (v) g(z)$ and $q(v,z) = 1 - \th (v) m(z)$. The black hole part of the spacetime now has a metric of the form 
\be \label{BHre3}
ds^2 = {L^2 \ov z^2} \le(- h (z) dv^2 - 2 k (z) dvdz + d \vec x^2 \ri) 
\ee
with $h(z) \equiv 1 - g(z)$ and $k(z) \equiv 1 - m (z)$, and can also be written as 
\be \label{bhnem}
ds^2 = {L^2 \ov z^2} \le(- h (z) dt^2 + {dz^2 \ov l(z)}  + d \vec x^2 \ri) , \;
k^2 (z) = {h(z) \ov l (z)} \ . 
\ee
We will restrict our discussion
mostly to~\eqref{vaidya}, but it is straightforward to generalize our results 
to~\eqref{vaidya1} as will be done in various places below.

\subsection{Extremal surfaces and physical observables} \label{sec:varob}

We are interested in finding the area $\sA_\Sig$ of an $n$-dimensional  extremal surface $\Ga_\Sig$  in the Vaidya geometry~\eqref{vaidya} which ends at an $(n-1)$-dimensional {\it spatial} surface $\Sig$ lying at some time $\ft$ in the boundary theory. 
We will use $A_{\Sig}$ to denote the area of $\Sig$. Since~\eqref{vaidya} is not invariant under time translation, $\Ga_\Sig$ and therefore $\sA_\Sig$ will depend on $\ft$. 

$\sA_\Sig$ can be used to compute various observables in the boundary theory:

\ben

\item For $n=1$, we take $\Sig$ to be two points separated by some distance $2R$. $\Ga_\Sig$ is then the geodesic connecting the two points, and its length $\sA (R, \ft)$ gives the equal-time two-point correlation function of an operator with large dimension,
\be \label{corre}
G( 2R, \ft) \propto e^{-m \sA (R, \ft)} ,
\ee 
where $m$ is the mass of the bulk field dual to the operator. 

\item For $n=2$, we take $\Sig$ to be a closed line, which defines the contour of a spacelike Wilson loop. The area $\sA_\Sig (\ft)$ then gives the expectation value of the Wilson loop operator~\cite{Rey:1998ik,Maldacena:1998im},
\be \label{wilson}
\vev{W_{\Sig} (\ft)} \propto e^{-  \sA_\Sig (\ft)/2 \pi \apr},
\ee
where $(2 \pi \apr)^{-1}$ is the bulk string tension.

\item For $n=d-1$, we take $\Sig$ to be a closed surface which separates space into two regions. The area $\sA_\Sig (\ft)$ then gives the entanglement entropy associated with the region bounded by $\Sig$~\cite{Ryu:2006bv,Hubeny:2007xt},
\be 
S_{\Sig}(\ft) = {\sA_\Sig(\ft) \ov 4 G_N} \ ,
\ee
where $G_N$ is Newton's constant in the bulk. 

\een

When there are multiple extremal surfaces corresponding  to the same boundary data, we will choose the surface with the smallest area. For entanglement entropy, this allows the holographic prescription to satisfy strong sub-additivity conditions~\cite{Headrick:2007km,Wall:2012uf}, while for correlation functions and Wilson loops, the smallest area gives the most dominant saddle point. 

We will often consider as examples the following two shapes for $\Sig$, which are the most symmetric representatives of two types of topologies for the boundary surface:

\bi 

\item a sphere of radius $R$:  with $d \vec x^2$ in~\eqref{vaidya} written in polar coordinates for the first $n$ directions,
\be \label{polat}
d \vec x^2 = d \rho^2 + \rho^2 d \Om_{n-1}^2 + dx_{n+1}^2 + \cdots + dx^2_{d-1}\ ,
\ee
 $\Sig$ is specified by 
\be  \label{spherd}
\rho = R\ , \quad x_a =0\ , \quad a = n+1, \cdots, d-1 \ .
\ee

\item boundary of a strip of half-width $R$: $\Sig$ consists of two $(n-1)$-dimensional hyperplanes located at  
\be \label{stripd}
x_1 = \pm R\ , \quad x_a =0 \ , \quad a = n+1, \cdots, d-1 
\ee
and extended in spatial directions $x_2, \cdots, x_n$. For $n=1$, $\Sig$ consists of two points separated by $2R$. For $n=2$, it defines a rectangular Wilson loop, and for $n=d-1$, it encloses the strip region $x_1 \in (-R, R)$.  
For brevity, we will refer to a $\Sig$ with this second shape as a ``strip".

\ei


\subsection{Vacuum and thermal equilibrium properties of extremal surfaces}  \label{sec:vac}

\subsubsection{Vacuum properties}

Before the quench, our system is in the vacuum state of a strongly coupled CFT with a gravity dual. 
Consider an extremal surface $\Ga_{\Sig}$ (with boundary $\Sig$) in pure AdS, whose area gives the vacuum value  of the corresponding physical observable. When $\Sigma$ is a sphere,\footnote{The following expressions for $\Sigma$ a sphere or strip have appeared 
in many places in the literature. For the case of entanglement entropy with $n=d-1$, they were first obtained in~\cite{Ryu:2006bv}.}
\bea \label{sphex}
\sA_{\rm sphere}&&= {\rm local \; divergences}\nn
&&+{L^n \om_{n-1}} \bca (-1)^{n \ov 2} b_n & n \; {\rm even} \cr
                                    (-1)^{n -1 \ov 2}  b_n \log R   & n \; {\rm odd} 
                                    \eca \quad
\eea
where $\om_{n-1}$ is the area of unit $(n-1)$-dimensional sphere and 
\be 
b_n = {(n-2) !! \ov (n-1) !!} \ .
\ee
When $\Sigma$ is a strip,
\bea \label{stripexm}
\sA_{\rm strip}& =& {\rm local \; divergences} + \bca 2L \log R  & n=1 \cr
                         - {L^n (a_n)^n \ov n-1} {A_{\rm strip} \ov R^{n-1}} \ & n >1 
                         \eca                 \ , \nn
                       a_n&\equiv& {\sqrt{\pi} \Ga (\ha + {1 \ov 2n}) \ov \Ga ({1 \ov 2n})  } 
\eea
where $A_{\rm strip}$ is the area of the strip $\Sigma$ with both sides included. 
The local divergences in~\eqref{sphex} and~\eqref{stripexm} can be interpreted as coming from short-range correlations near $\Sig$ and its leading contributions are proportional to $A_{\Sig}$.

The number of degrees of freedom in a CFT can be characterized by a central charge $s_d$, defined in all dimensions in terms of the universal part of 
the entanglement entropy of a spherical region in the vacuum~\cite{Myers:2010xs},
\be \label{enD}
S_{\rm sphere}^{(\rm vac)} = {\rm local \; divergences} +  \bca (-1)^{d-1 \ov 2} s_d & d \; {\rm odd} \cr
                                    (-1)^{d -2 \ov 2}  s_d \log R   & d \; {\rm even} 
                                    \eca\ ,
\ee
where from~\eqref{sphex},
\be \label{chare}
s_d = {L^{d-1} \ov 4 G_N} \om_{d-2} b_{d-1} =  {\pi^{{d \ov 2}} \ov  \Ga ({d \ov 2})} {L^{d-1} \ov 4 G_N} \times 
\bca 1   & d \; {\rm odd} \cr
{2 \ov \pi}  & d \; {\rm even} 
\eca   \ .
\ee
Note that for $d=2$ the above central charge is related to the standard central charge $c$ as
\be 
s_2 = {c \ov 3}  \ . 
\ee
From the standard AdS/CFT dictionary, $s_d \propto N^2$ where $N$ is the rank of the gauge group(s) of the boundary theory. If we put such a holographic CFT on a lattice, $s_d$ is heuristically the number of degrees of freedom on a single lattice site. 

From~\eqref{wilson} and~\eqref{sphex}--\eqref{stripexm}, a Wilson loop of circular and rectangular 
shape respectively have the vacuum behavior 
\be \label{w1}
W_\Sig \sim \bca 
           e^{- \# \sqrt{\lam} } & {\rm circle} \cr
            e^{- \# \sqrt{\lam} {\ell \ov R} } & {\rm rectangle}
            \eca , \qquad \sqrt{\lam} = {L^2 \ov  \apr}
\ee
where $\ell$ denote the length of the long side of a rectangular Wilson loop. 
Similarly one finds that the two-point correlation function of an operator with large dimension $\De \approx mL \gg1$
is given by
\be \label{c1}
G (2R) \sim {1 \ov R^{2 \De}}  \ .
\ee
\subsubsection{Equilibrium properties} 

After the quench, our system eventually evolves to a final equilibrium state dual to a black hole in the bulk. Here we briefly review properties of an extremal surface $\Ga_{\Sig}$ (with boundary $\Sig$) in the black hole geometry~\eqref{BHre2}, whose area gives the equilibrium value  of the corresponding physical observable. 

To leading order in large size limit, one can show that for $\Sig$ of any shape~\cite{Liu:2013una} (see also Appendix~\ref{app:equ}) 
\be \label{eqarea}
\sA^{\rm (eq)}_\Sig = {L^n V_{\Sig} \ov z_h^n} \equiv {\mathfrak a}_{\rm eq} V_{\Sig} \ , \qquad 
\fa_{\rm eq} = {L^n \ov z_h^n}\ ,
\ee
where $V_{\Sig}$ denotes the volume of the boundary region bounded by surface $\Sig$, and $\fa_{\rm eq}$ can be interpreted as an equilibrium ``density.'' This result has a simple geometric interpretation in the bulk -- in the large size limit, most of the extremal surface simply runs along the 
horizon. In particular, for entanglement entropy,
\be \label{equent}
S^{\rm (eq)}_\Sig = {L^{d-1} \ov 4 G_N} {V_{\Sig} \ov z_h^{d-1}}  = s_{\rm eq} V_\Sig
\ee
where we have used the entropy density $s_{\rm eq}$ from~\eqref{eqent}. For a Wilson loop we have 
\be \label{w2}
W_{\rm eq} \sim e^{- \# \sqrt{\lam} {V_{\Sig} \ov z_h^2}}
\ee
where $V_\Sig$ is now the area of the region enclosed by the loop. 
The two-point correlation function of an operator with dimension $\De \approx m L \gg 1$
is given by
\be \label{c2}
G_{\rm eq} (2R) \sim e^{-  \De {2 R \ov z_h}} \ .
\ee




\subsection{Further comments on the Vaidya setup}

To conclude this section we make some further comments on the Vaidya setup: 

\ben

\item It should be kept in mind that while the final equilibrium state has a temperature and coarse grained thermal entropy density, the Vaidya geometry describes the evolution of a pure state. As a consistency check, one can show that for such a process the entanglement entropy for a region $A$ is the same as that of its complement~\cite{AbajoArrastia:2010yt,Albash:2010mv,Takayanagi:2010wp}. Thus the equilibrium entanglement entropy~\eqref{equent}, despite having a thermal form, reflects 
genuine long-range quantum entanglement.  The reason~\eqref{equent} has exactly the form of a thermal entropy is as follows. We are considering a finite region in a system of infinite size. Thus the number of degrees of freedom outside the region is always infinitely larger than that inside. As a result in a 
typical excited pure state the reduced density matrix for the finite region appears thermal~\cite{Page:1993df}.

\item Before the quench, our system is in a vacuum state of a CFT and thus already has long range correlations, whereas the initial state of~\cite{Calabrese:2005in} only has short-range correlations. 
However, this difference is likely not important for the questions we are interested in, which concern the build-up of the finite density of entanglement entropy in~\eqref{equent}. The long-range entanglement in the vacuum, quantified by the universal part in~\eqref{enD}, is measure zero compared to~\eqref{equent}. Heuristically, for odd $d$, the long-range entanglement entropy in the vacuum, being a $R$-independent constant, amounts to that of a few sites inside the region that are fully entangled with the outside, while in equilibrium, almost all points inside the region become entangled. For even $d$, there is a logarithmic enhancement of the long-range entanglement in the vacuum, but it is still measure zero compared to the final entanglement in the large region limit.




\item From the perspective of entanglement entropy, the equilibration process triggered by the quench builds up long-range entanglement, as can be seen by comparing~\eqref{equent} and~\eqref{enD},
whereas from the perspective of correlation functions~\eqref{corre} and Wilson loops~\eqref{wilson} in which $\sA$ appears in the exponential with a minus sign, the same process corresponds to the destruction of correlations (compare~\eqref{w2}--\eqref{c2} with~\eqref{w1}--\eqref{c1}). More specifically, long range correlations in the latter observables which were present in the vacuum are replaced by short-range correlations with correlation length controlled by $z_h$. However, there is no contradiction, as the process of building up entanglement  also involves redistribution of those in the vacuum -- pre-existing correlations between local operators and over the Wilson loop get diluted by the redistribution process. 
 
 \een
 
\section{Equations of motion for extremal surfaces} \label{sec:eom}

Here we describe equations of motion for $\Ga_{\Sig}$ and its general characteristics when $\Sig$ is a strip or a sphere. 
In such cases  $\Ga_{\Sig}$ can be described by two functions, $z (\rho), v (\rho)$ for a sphere,
or $z (x_1), v(x_1)$ for a strip. For both shapes the functions satisfy the following boundary conditions at the boundary as well as regularity conditions at the tip of the surface,
\be \label{bd0}
z (R) = 0 \ , \quad v (R) = \ft \ , \quad  z' (0)=v' (0) = 0 \ .
\ee 
For a strip we will write $x_1$ simply as $x$. 
It is convenient to introduce the location $(z_t, v_t)$ of the tip of $\Ga_{\Sig}$,
\be \label{tip}
z(0) = z_t \ , \qquad v(0) = v_t \ . 
\ee
The sphere and strip being highly symmetric, specifying $(z_t, v_t)$ completely fixes $\Ga_{\Sig}$.
The relations between $(R, \ft)$ and $(z_t, v_t)$ are in general rather complicated and require solving the full equations for $z(\rho), v(\rho)$ or $z (x), v(x)$. Also, it is possible that a given $(R, \ft)$ corresponds to multiple $(z_t, v_t)$'s, i.e. multiple extremal surfaces have the same boundary data. Then as mentioned earlier we will choose the extremal surface with smallest area. 

For $\Sig$ a sphere or strip we will simply denote $\sA_{\Sig}(\ft)$ as $\sA( R,\ft)$.

\subsection{Strip} \label{app:eomst}

The area of an $n$-dimensional {surface in \eqref{vaidya} ending on the strip $\Sig$ given by~\eqref{stripd}} can be written as 
\be \label{striparea}
\sA = \ha \tilde K  \int_{-R}^R dx \, {\sqrt{Q} \ov z^{n}}, \qquad
Q \equiv 1 - 2 v'z' - f (z,v) v'^2 
\ee
where 
\be \label{coesr}
\tilde K ={L^{n}} 
A_{\rm strip} \ ,
\ee
with $A_{\rm strip}$ being the area of $\Sig$ (both sides of $\Sig$ are included which gives the $\ha$ factor in \eqref{striparea}).  
 $z(x), v(x)$ then satisfy the equations of motion
\bea \label{strip1}
&& z^{n} \sqrt{Q} \p_x \le({z' + f v' \ov z^{n} \sqrt{Q}}\ri) =  \ha {\p f \ov \p v} v'^2 \ , \\
 && z^{n} \sqrt{Q} \p_x \le( {v' \ov z^{n} \sqrt{Q}} \ri) = 
n {Q \ov z} + \ha {\p f \ov \p z} v'^2   \ .
\label{strip2}
\eea
Since the integrand of $\sA$ does not depend explicitly on $x$, there is a first integral 
\be \label{conj0}
z^{n} \sqrt{Q} = J = {\rm const} \ .
\ee
Furthermore, when $\p_v f =0$, equation~\eqref{strip1} can be integrated to give another first integral,
\be \label{cone0}
 z' + f v' =  E = {\rm const} \ . 
 \ee
 
We are mainly interested in {$\Ga_{\Sig}$} which go through both AdS and black hole regions. With reflection symmetry about $x=0$, we only need to consider the $x>0$ half of such a $\Ga_{\Sig}$. We now discuss equations in each region separately:
\ben 

\item  AdS region: From~\eqref{bd0} and~\eqref{cone0} we have 
\be 
E = z'+v'=0 
\ee
and from~\eqref{conj0}
\be \label{eio1}
z ' = -{1 \ov z^{n}} \sqrt{J^2 - z^{2 n}}\ ,  \quad J = z_t^{n} , 
\ee 
which give
\be \label{exe0}
x (z) = \int_{z}^{z_t} {dy\, y^{n} \ov  \sqrt{z_t^{2 n} - y^{2 n}}}, \quad 
v (z) = v_t  + z_t -z \ .
\ee

\item Matching conditions at the shell: Denoting the values of $z$ and $x$ at the intersection of $\Ga_{\Sig}$ and the null shell $v=0$ as $z_c$ and $x_c$, respectively, we have
\be
z_c = z_t+v_t
\ee
and derivatives on the AdS side of the null shell are
\be \label{stripnd}
z_-'=-v_-'=-{1 \ov z_c^n}\sqrt{z_t^{2n}-z_c^{2n}}\ .
\ee
To find derivatives on the other side, we integrate the equations of motion~\eqref{strip1}--\eqref{strip2} across the null shell to find the matching conditions 
\bea 
&&v_+' = v_-' \ , \qquad Q_+ = Q_-\ , \nn
&& z_+' = z_-' + \ha g(z_c) v' = \le(1 - \ha g (z_c) \ri) z_-'  \ . \quad
\label{ntrn1}
\eea
Note we have used the subscript $-$ ($+$) to refer to quantities  on the AdS (black hole) side of the {null} shell. 

\item Black hole region: From matching conditions~\eqref{ntrn1}, $J$ is the same as in the AdS region, i.e. given by~\eqref{eio1}, while $E$ is given by
\be \label{eeval}
E =   \ha g (z_c) z_-'  <0
\ee
implying $t$ is no longer constant.
From~\eqref{cone0},
\be \label{vekp}  
v' = {E- z' \ov h} 
\ee
which can be substituted into~\eqref{conj0} to obtain 
\be \label{eio2}
 z'^2=h(z) \le({z_t^{2n} \ov z^{2 n}} -1\ri) + E^2   \equiv H (z) \ .
\ee
Substituting~\eqref{eio2} back in~\eqref{vekp} we also have
\be  \label{vekp1}
{dv \ov dz} = - {1 \ov h} \le({E \ov \sqrt{H} }+ 1 \ri)  \ .
\ee
\een

Collecting equations in the two regions we find from \eqref{eio1} and \eqref{eio2} 
\be \label{raidu1} 
R =  \int_{z_c}^{z_t} {dz  \ov  \sqrt{{z_t^{2n} \ov z^{2 n}} -1}}  
+  \int_0^{z_c} {dz \ov \sqrt{H(z)}} \ ,
\ee
where we have assumed that $z(x)$ monotonically decreases as $x$ increases (recall we let $x>0$). 
As we will see later $z(x)$ can be non-monotonic in which case the above equation should be suitably modified. 
Similar caveats should be kept in mind for other equations below. 
From integrating~\eqref{vekp1}, 
\be \label{biute}
\ft = \int_0^{z_c} {dz \ov h(z)} \le({E \ov \sqrt{H(z)}} + 1\ri) \ .
\ee
Note that at $z=z_h$, $h(z)^{-1}$ has a pole but the integrand in \eqref{biute} remains finite as the second factor vanishes at $z = z_h$, due to $H(z_h)=E^2$ and $E<0$. 
Finally, from~\eqref{eio1} and~\eqref{eio2} {we have that the area of $\Ga_{\Sig}$ is} given by
\be \label{aver1}
\sA 
=  \sA_{AdS} + \sA_{BH} 
\ee 
{where}
\be \label{aver2}
{1 \ov  \tilde K}\sA_{AdS} = z_t^{1-n}\int^1_{z_c \ov z_t} dy \, {1 \ov y^{n} \sqrt{1 - y^{2 n}}}
\ee
and 
\be \label{aver3}
{1 \ov  \tilde K}\sA_{BH} = z_t^n \int_0^{z_c} dz \,  {1 \ov z^{2n} \sqrt{H(z)}} \ .
\ee
For a given $R$ and $\ft$, we {can} use~\eqref{raidu1} and~\eqref{biute}
to solve for $z_t (R, \ft), z_c (R, \ft)$ {after which}~\eqref{aver1} {can be expressed} in terms of $R$ and $\ft$.

\subsection{Sphere}  \label{app:eomsp}

The area of an $n$-dimensional {surface in~\eqref{vaidya} ending on a sphere $\Sig$ given by~\eqref{spherd}} can be written as 
\be \label{areadf0}
\sA  = K \int_0^R d \rho \, {\rho^{n-1} \ov z^n}\sqrt{Q}\ , \quad 
Q=1- 2 v' z' - f (z,v) v'^2
\ee
where 
\be \label{coesp} 
K =  {L^{n} }
 {A_{\rm sphere} \ov R^{n-1}}\ .
\ee 
It follows that $z(\rho), v(\rho)$ satisfy the equations of motion 
\bea \label{1}
&&{z^{n} \sqrt{Q} \ov \rho^{n-1}}\p_\rho \le[{\rho^{n-1} \ov z^{n}} {1 \ov \sqrt{Q}} v' \ri] = 
 {n Q \ov z} + \ha {\p f \ov \p z} v'^2  \ ,
\\ 
\label{2}
&& {z^{n} \sqrt{Q} \ov \rho^{n-1}}\p_\rho \le[{\rho^{n-1} \ov z^{n}} {1 \ov \sqrt{Q}} (z' + f v') \ri]
= \ha {\p f \ov \p v} v'^2 \ ,
\eea
and boundary conditions \eqref{bd0}. When $\p_v f = 0$, equation~\eqref{2} can be integrated to give 
\be \label{3}
{\rho^{n-1} \ov z^{n}} {1 \ov \sqrt{Q}} (z' + f v') = E = {\rm const}
\ee
which can also be expressed as 
\be \label{4}
{\rho^{n-1} \ov z^{n}} {f \ov \sqrt{Q}} {dt \ov d \rho} = E
\ee
where $t$ is the Schwarzschild time.

Again, we are interested in $\Ga_{\Sig}$ which go through both AdS and black hole regions:

\ben 

\item AdS region: Given~\eqref{bd0}, we again have $E=0$, which implies that the solution 
in the AdS region is the same as that in pure AdS, i.e. is given by~\cite{Ryu:2006ef}
\be \label{adspl2}
z (\rho)= \sqrt{z_t^2 - \rho^2} \ , \quad  v (\rho) =   z_t + v_t  - z (\rho) \ . 
\ee

\item Matching conditions at the shell: Denoting values of $z$ and $\rho$ at the intersection of $\Ga_{\Sig}$ and the null shell $v=0$ as $z_c$ and $\rho_c$, respectively, we have 
\be \label{varp00}
z_c = z_t + v_t \ , \qquad \rho_c = \sqrt{z_t^2 - z_c^2} 
\ee
and derivatives on the AdS side of the null shell are
\be \label{varp0}
 z'_-  =  - v'_- = - {\rho_c \ov z_c} \ . 
\ee
To find the corresponding derivatives on the other side, we integrate~\eqref{1} and~\eqref{2} across the shell, which again leads to the matching conditions \eqref{ntrn1} but with $z'_{-}$, $v'_{-}$ now as in \eqref{varp0}. 
%
\item Black hole region: The matching implies
\be \label{varp1}
E=-{1 \ov 2}\le({\rho_c \ov z_c}\ri)^{n}{g(z_c) \ov z_t} <0
\ee
and {$t$ is no longer constant}. Solving for $v'$ and $Q$ in terms of $z'$ using~\eqref{3}, we obtain
\be \label{usegp}
v' = {1 \ov h(z)} \le(-z' + {EB \sqrt{1 + {z'^2 \ov h}} \ov \sqrt{1 + {E^2 B^2 \ov h}}} \ri) \ ,
\quad B \equiv {z^{n} \ov \rho^{n-1}}   
\ee
which, when substituted in~\eqref{1}, gives the equation for $z$
\bea 
&&\le( h +E^2 B^2\ri)z''  +  \le(h + z'^2  \ri)\le({n-1 \ov \rho} z' + {n h \ov z} \ri)  \nn
&&\qquad +\le(E^2B^2-z'^2\ri) {\p_z h  \ov 2}=0\ .
\label{modeom}
\eea

\een

From integrating~\eqref{usegp}, the boundary time is
\bea \label{texpl}
\ft &=& \int_{\rho_c}^R {d \rho \ov h} \le(-z' + {EB \sqrt{1 + {z'^2 \ov h}} \ov \sqrt{1 + {E^2 B^2 \ov h}}} \ri) \nn
&=& \int_{\rho_c}^R {d \rho \ov h + E^2 B^2}{E^2 B^2 - z'^2 \ov EB \sqrt{h+z'^2 \ov h + E^2 B^2} + z'}
\eea
where the second expression is manifestly well-defined at the horizon, {and the integral is evaluated on shell, with $z(\rho)$ satisfying equation \eqref{modeom} and boundary conditions \eqref{ntrn1} at $\rho=\rho_c$ and $z(R)=0$. Finally, from \eqref{adspl2} and \eqref{usegp}, the area of $\Ga_{\Sig}$ can be written as }
\be \label{speac}
\sA  
= \sA_{AdS} + \sA_{BH} 
\ee
where 
\be \label{adsc}
{1 \ov K} \sA_{AdS} = \int_0^{\rho_c} d\rho \, {\rho^{n-1} \ov z^{n}} \sqrt{1 + z'^2} = \int_0^{\rho_c \ov z_t} dx \, {x^{n-1} \ov (1-x^2)^{n+1 \ov 2}}
\ee
and 
\be \label{bhc}
{1 \ov K} \sA_{\rm BH} = \int_{\rho_c}^R d \rho \, {\rho^{n-1} \ov z^{n}} {\sqrt{1 + {z'^2 \ov h}} \ov \sqrt{1 +{ E^2 B^2 \ov h}}} \ .
\ee

Note the story here is significantly more complicated than for a strip. One needs to first solve the differential equation~\eqref{modeom} with initial condition given by the last equation of~\eqref{ntrn1}. Imposing the boundary condition $z (R) =0$ gives a relation between $\rho_c$ and $z_c$. One then needs to evaluate~\eqref{texpl} to find  {$z_c (R, \ft), \rho_c (R, \ft)$} and finally use~\eqref{speac} to obtain $\sA (R, \ft)$.



\section{General geometric features and strategy} \label{sec:dcal}

We now describe geometric features of $\Ga_\Sig$ during its time evolution, using as examples the case of $\Sig$ being a sphere or a strip. For the two shapes the equations of motion (given in Sec.~\ref{sec:eom}) can be readily solved numerically.
We are interested in long-distance behavior, i.e. we take
\be 
R \gg z_h \ .
\ee



\begin{figure}[!h]
\centering

\subfigure[]{\includegraphics[scale=0.4]{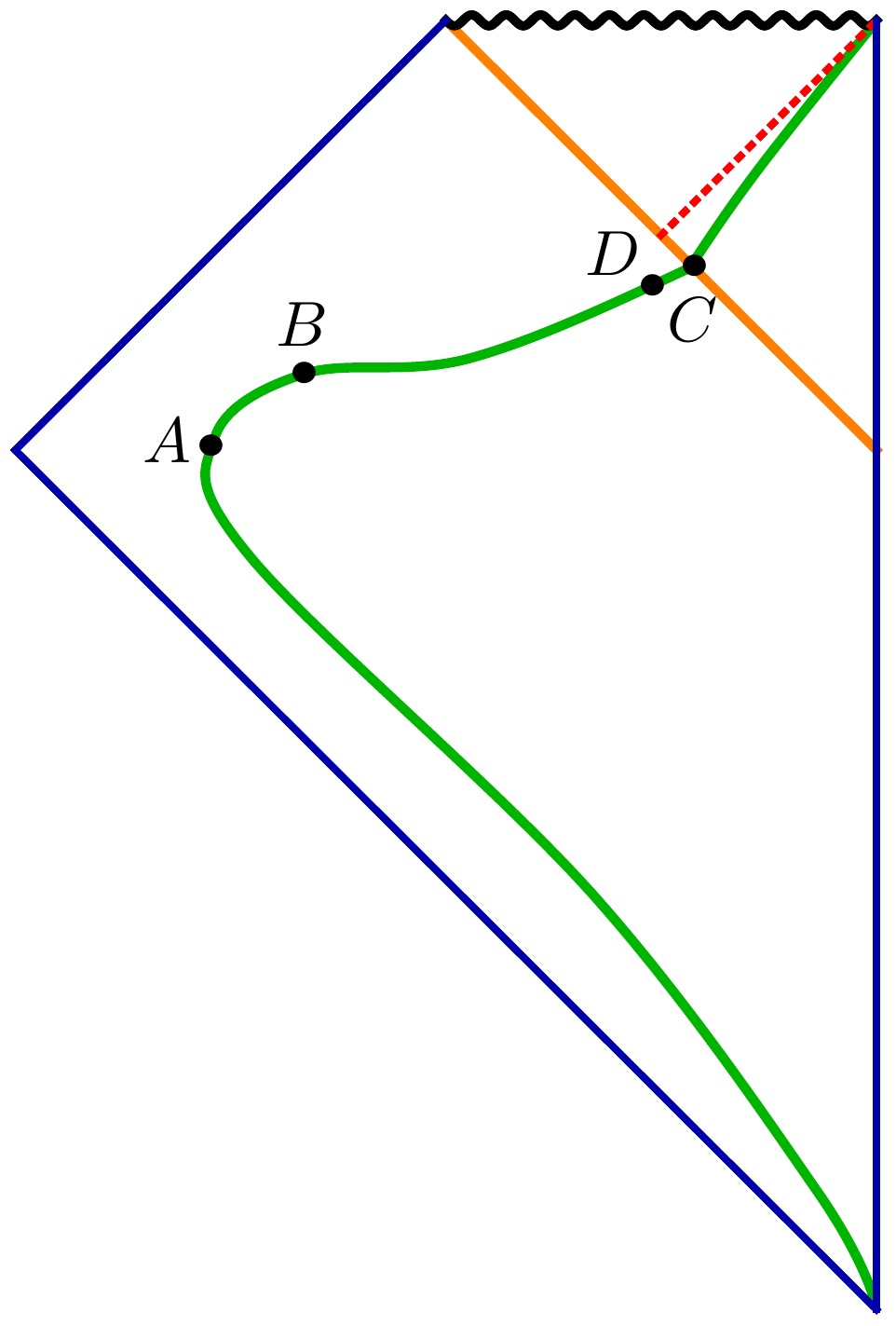} }
\subfigure[]{\includegraphics[scale=0.4]{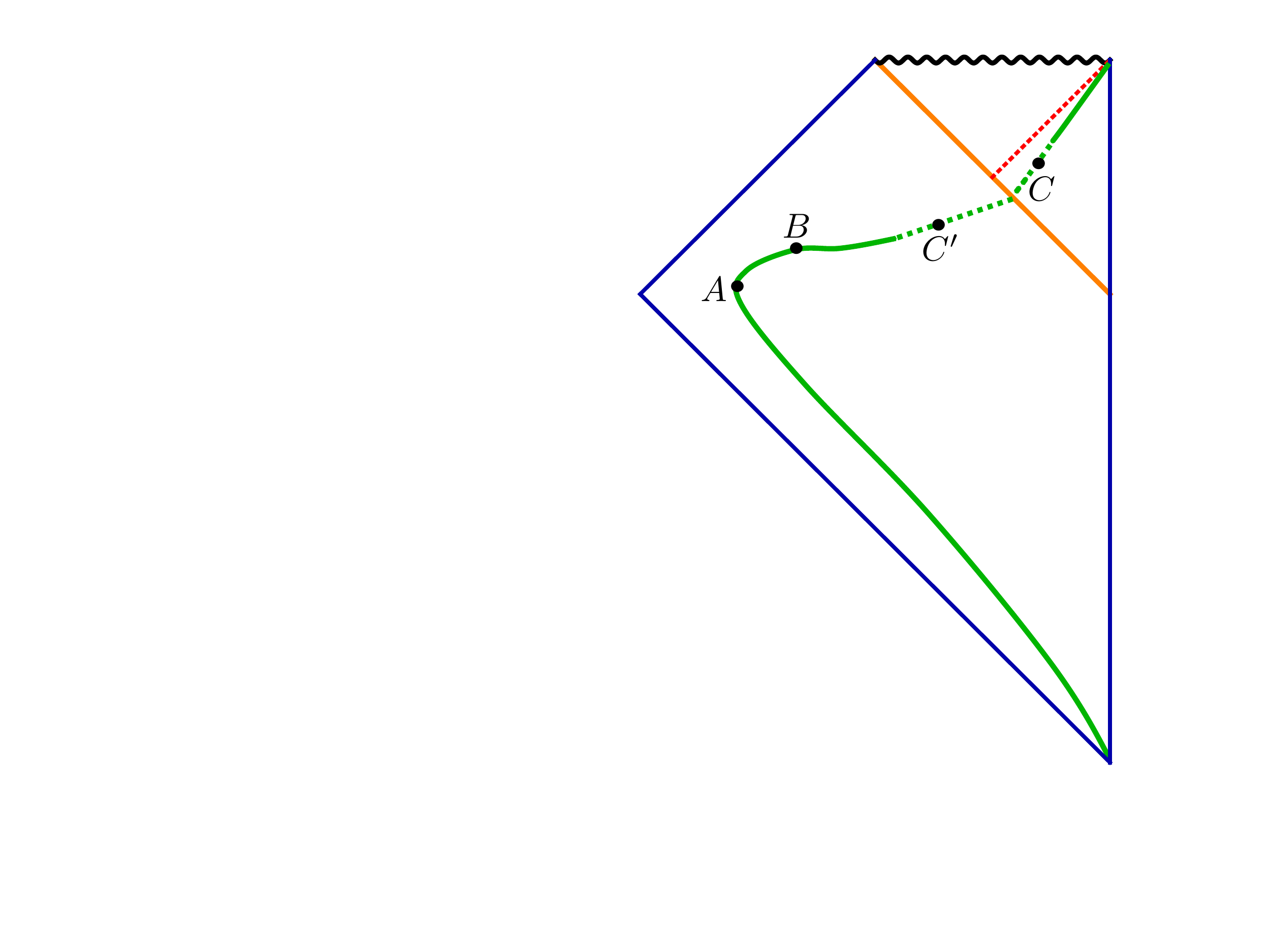}}
\caption{Cartoon of the curve $(z_t (R, \ft), v_t (R, \ft))$ for (a) continuous and (b) discontinuous saturation.
Cartoons of various extremal surfaces whose tip are labelled above are shown in Fig.~\ref{fig:surexm}.
(a): For continuous saturation the whole curve has a one-to-one correspondence to $(R,\ft)$, and saturation happens at point $C$ continuously. (b): Discontinuous saturation happens via a jump of the extremal surface from one with tip at $C'$ to one with tip at $C$. Along the dashed portion of the curve, different points can correspond to the same $(R, \ft)$.}
\label{fig:curves}
\end{figure}

\begin{figure}[!h]
\centering
\subfigure[]{\includegraphics[scale=0.4]{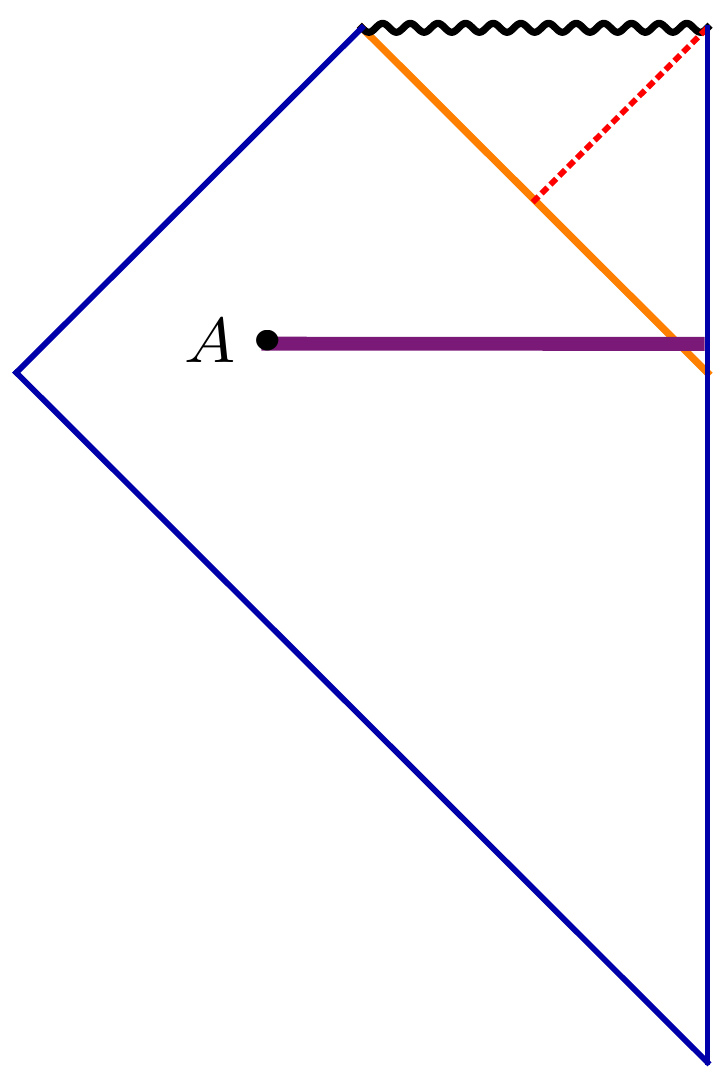} }\quad \quad
\subfigure[]{\includegraphics[scale=0.4]{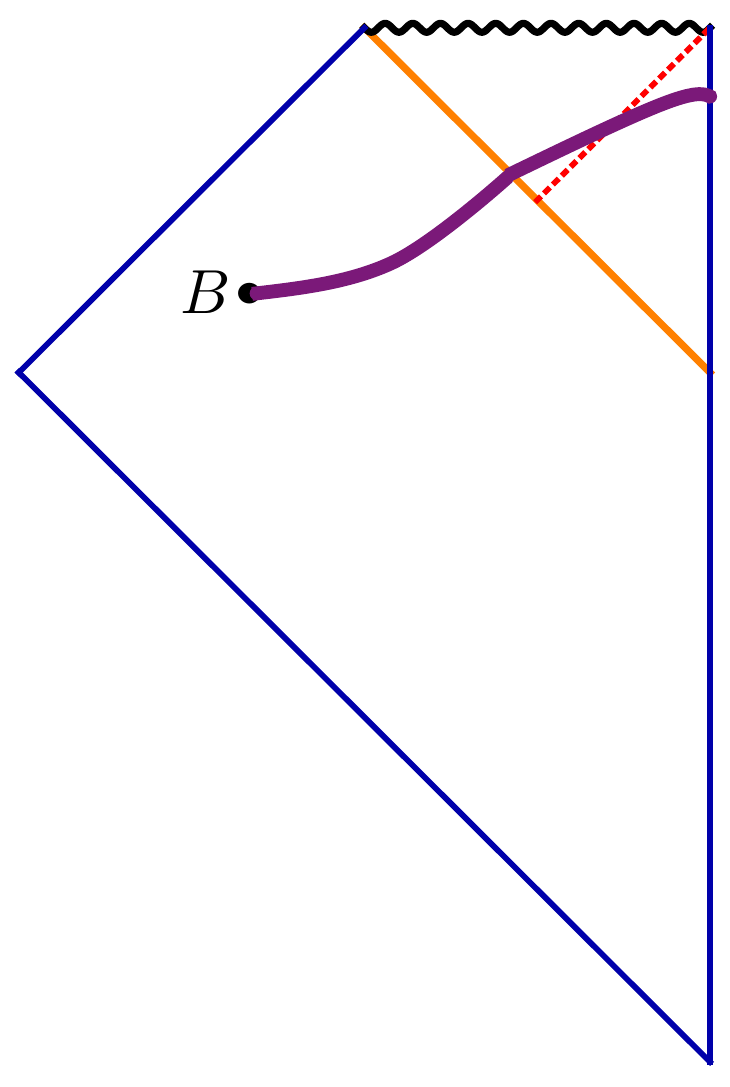} }
\subfigure[]{\includegraphics[scale=0.4]{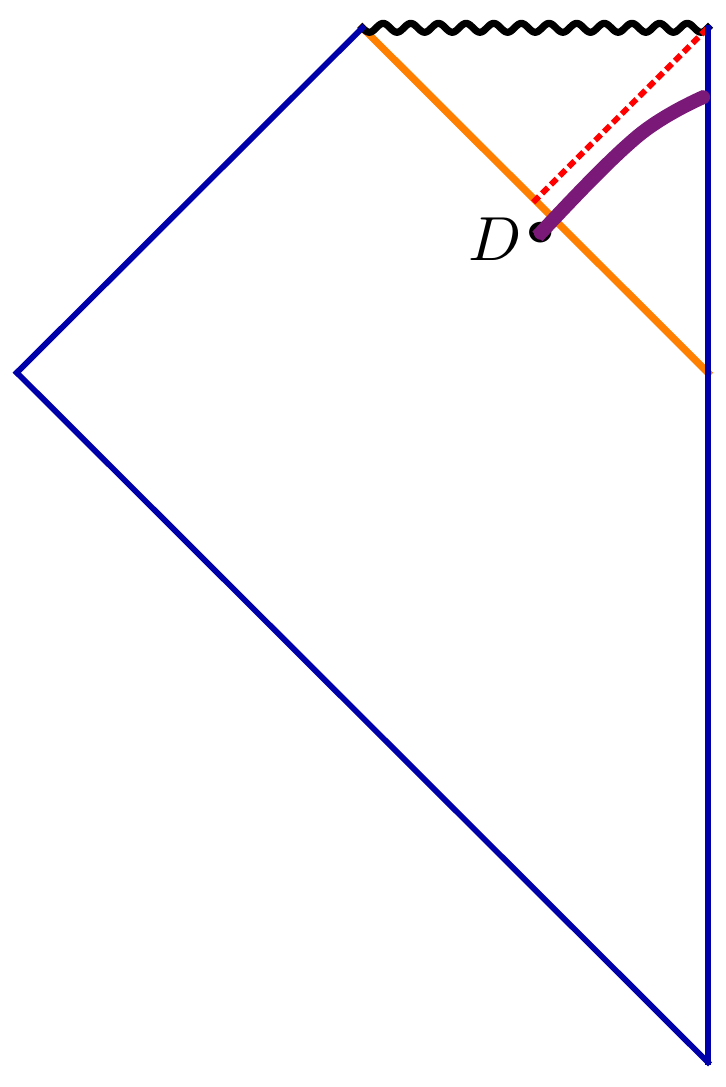} }
\caption{Cartoons of extremal surfaces with tip at various points labelled in Fig.~\ref{fig:curves}.
Spatial directions are suppressed. 
(a): At $\ft=0_+$, the extremal surface starts intersecting the null shell, with $z_c$ very small. (b) When $\ft \gtrsim z_h$, the extremal surface starts intersecting the null shell behind the horizon. (c) The extremal surface close to continuous saturation for which $z_t - z_c$ is small. 
}
\label{fig:surexm}
\end{figure}

At fixed $R$, as $\ft$ is varied, the tip~\eqref{tip} of $\Ga_{\Sig}$ traces out a curve $(z_t (R, \ft), v_t (R, \ft))$ in the Penrose diagram. This provides a nice way to visualize the evolution of $\Ga_{\Sig}$ with $\ft$. See Fig.~\ref{fig:curves}. 

Instead of $(z_t, v_t)$ it is sometimes convenient to    
use $(z_t, z_c)$ or $(z_t, \rho_c)$  to specify $\Ga_{\Sig}$, where $z_c$ and $\rho_c$ 
are the values of $z$ and $\rho$ at which the $\Ga_{\Sig}$ intersects the null shell. 
For both sphere and strip $z_c = z_t + v_t$. For a sphere $\rho_c$ is given by~\eqref{varp00}, while for a strip $x_c$ can be obtained by setting $z=z_c$ in~\eqref{exe0}.

We now elaborate on various stages of the time evolution of $\Ga_\Sig$, and strategies for 
obtaining $\sA (R, \ft)$ in each of them.

For $\ft < 0$, $\Ga_{\Sig}$ lies entirely in AdS, and
\be \label{shep}
z_t  (R, \ft< 0)= \bca R & {\rm sphere} \cr
       {R  \ov a_n} & {\rm strip}
          \eca, \qquad v_t = \ft - z_t
          \ee
where $a_n$ was introduced in~\eqref{stripexm}.  $\sA (R, \ft)$ is independent of $\ft$ and 
is given by its vacuum value. In Fig.~\ref{fig:curves} this corresponds to the part of curve below point $A$. 
Note that as $R \to \infty$, $z_t \to \infty$. 

At  $\ft = 0_+$, or point $A$, $\Ga_{\Sig}$ starts intersecting the null shell  (see Fig.~\ref{fig:surexm}(a)). 
For $\ft \ll z_h$, the point of intersection is  close to the boundary, i.e. $z_c \ll z_h$.  This defines the pre-local-equilibrium stage mentioned in the Introduction. 
In this regime, one can extract $\sA_{\Sigma} (\ft)$ by expanding both $\ft$ and $\sA$ in small $z_c$, which we will do for arbitrary $\Sigma$ in Sec.~\ref{sec:early}. 

When $\ft$ becomes of order $z_h$,  at some point $\Ga_{\Sig}$ starts intersecting the shell behind the horizon, i.e. $z_c > z_h$.  An example is point $B$ in Fig.~\ref{fig:curves}, whose corresponding $\Ga_{\Sig}$ is shown in Fig.~\ref{fig:surexm}(b). 

There exists a sharp  time $\ft_s$ after which $\Ga_{\Sig}$ lies entirely in the black hole region. $\Ga_{\Sig}$ then reduces to that in a static black hole geometry.  It lies on a constant Schwarzschild time $t = \ft$ outside the horizon and is time independent. That is, for $\ft > \ft_s$
\be \label{shep2}
z_t (R, \ft) = z_b (R) < z_h\ , \qquad v_t = \ft - \sig (z_t)  
\ee
where  $z_b$ denotes the location of the tip of $\Ga_{\Sig}$ in the static black hole geometry, and in the second equation we have used~\eqref{tzco2}. This corresponds to the part of the curve above point $C$ 
in Fig.~\ref{fig:curves}. 
 For $\ft > \ft_s$, $\sA (R, \ft)$ is time independent and given by its equilibrium value.

The saturation at the equilibrium value at $\ft_s$ can proceed as a continuous or discontinuous transition, as illustrated in Fig.~\ref{fig:curves}. For a continuous transition, depicted on the left, the entire curve $(z_t, v_t)$ as a function of $\ft$ has one-to-one correspondence  with $(R, \ft)$ and saturation happens at point $C$, with $\ft_s$ given by
\be \label{satcp}
v_t (\ft_s) = 0 \ , \qquad \ft_s (R) = \sig (z_b (R))  = \int_0^{z_b} {dz \ov h(z)} \ .
\ee
In contrast, for a discontinuous saturation, depicted on the right plot of Fig.~\ref{fig:curves}, in the dashed portion of the curve, there are multiple $(z_t, v_t)$ associated with a given $(R, \ft)$. As a result, the minimal area condition requires that the extremal surface jump from point $C'$ to $C$ at some $\ft_s$. In this case there does not exist a general formula for $\ft_s$. For a discontinuous saturation, $\sA_\Sig (\ft) $ is continuous at $\ft_s$, but its first time derivative becomes discontinuous. 

In the case of a continuous saturation, for which the first time derivative of $\sA_\Sig (\ft)$ is continuous, 
one can then define a critical exponent $\ga$ (by definition $\ga > 1$)
\be 
\label{svew}
 \sA_\Sig (\ft)- \sA_\Sig^{\rm (eq)} \propto - (\ft_s - \ft)^\ga \ . 
 \ee
The ``critical'' behavior around saturation can be obtained as follows.  As $\ft \to \ft_s$, the tip of $\Ga_{\Sig}$ approaches the null shell, i.e. $z_t - z_c \to 0$ with $z_t, z_c \to z_b$ (this is depicted by point $D$ in Fig.~\ref{fig:curves} and Fig.~\ref{fig:surexm}(c)). 
Thus one can expand both $\ft - \ft_s$ and $\sA - \sA_{\rm eq}$ in small $z_t-z_c$, as we discuss in detail in Sec.~\ref{sec:satur}.

 



\begin{figure}[!h]
\centering
\subfigure[]{\includegraphics[scale=0.4]{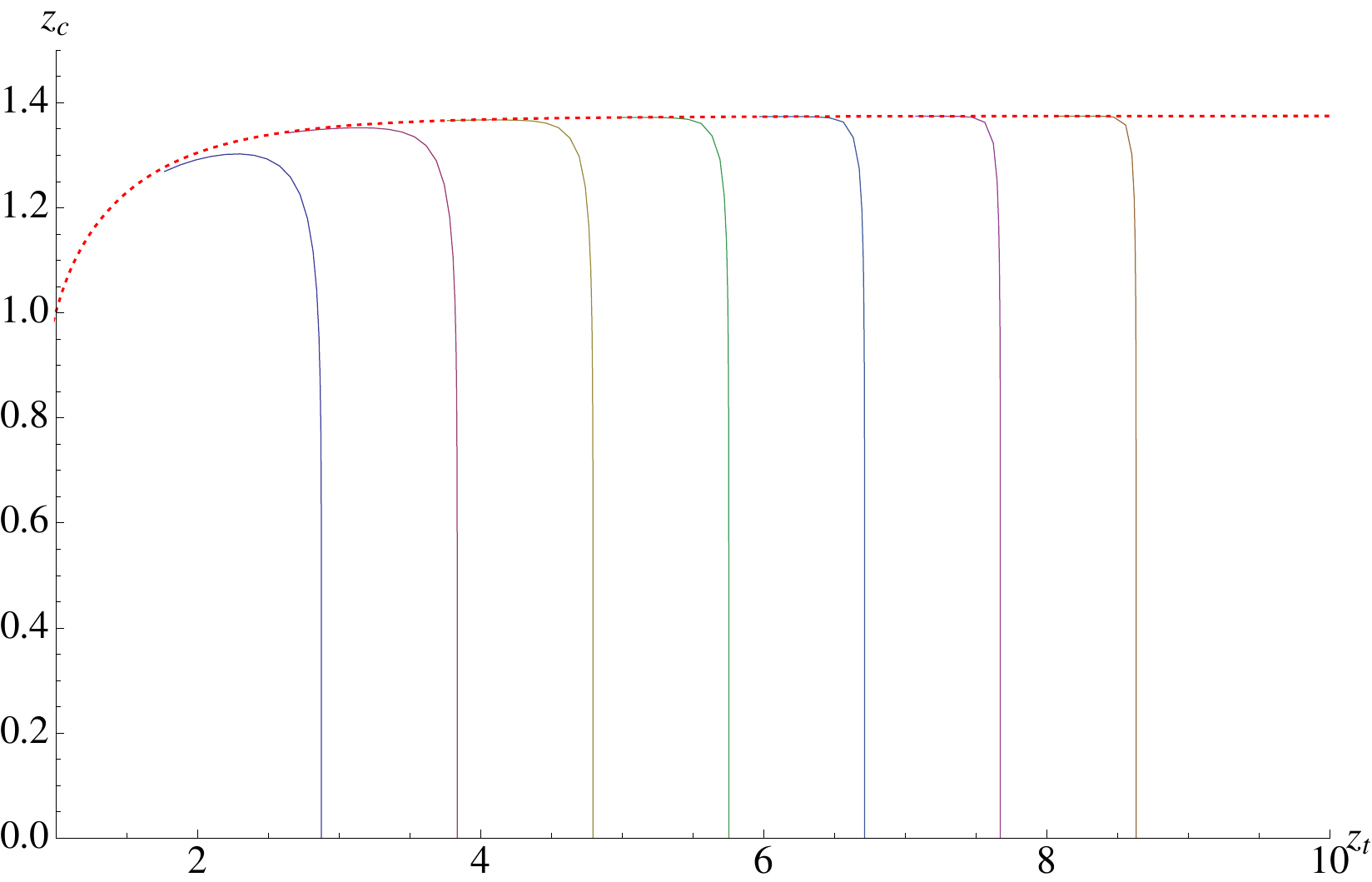}}
\subfigure[]{\includegraphics[scale=0.4]{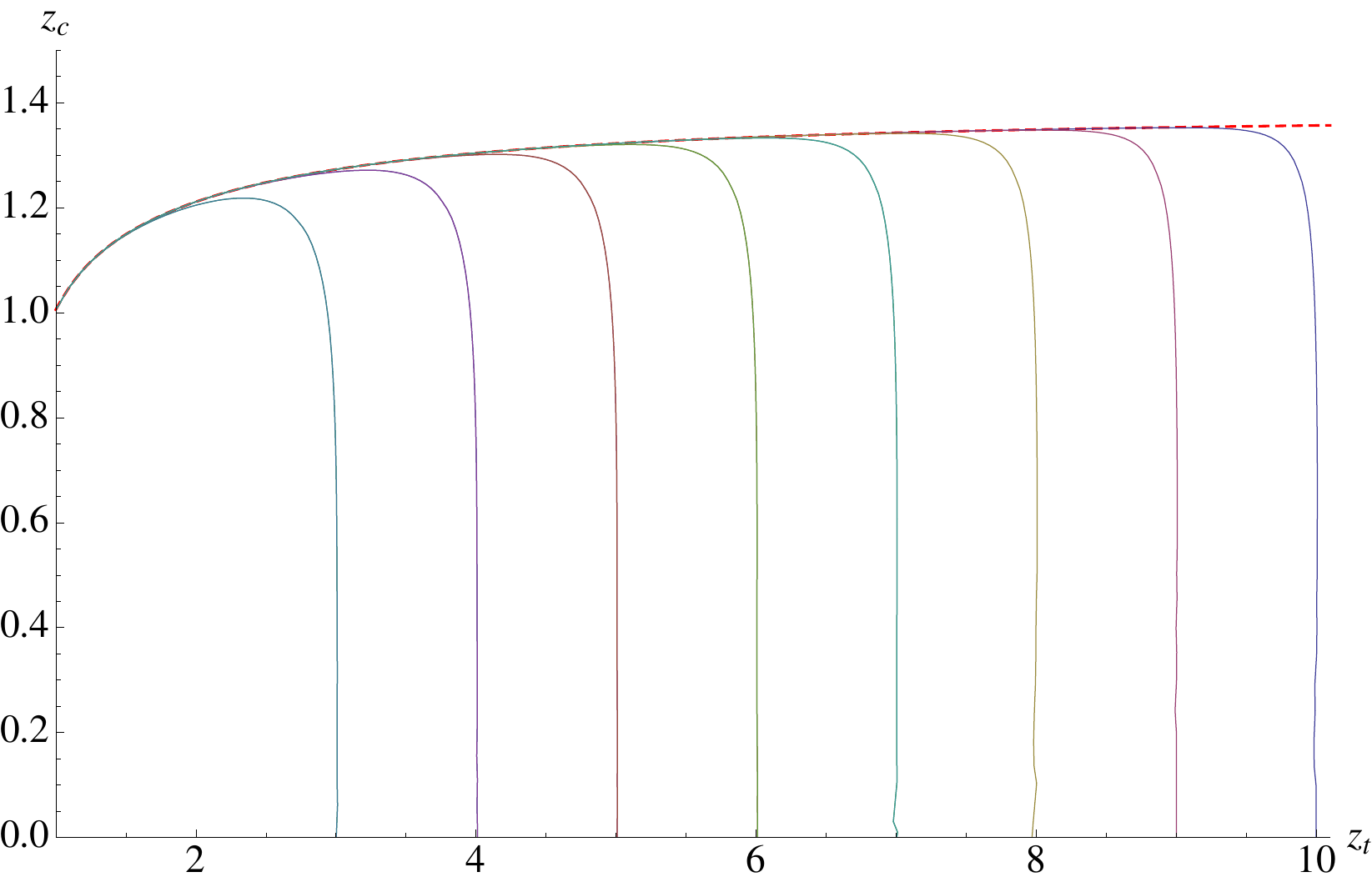} }
\caption{Parametric curves $(z_t (R,\ft), z_c(R, \ft))$ at fixed $R$ and varying $\ft$ for Schwarzschild $h(z)$
in $d=3$.  
Different curves correspond to $R = 2,3, \cdots ,10$. In both plots, we choose units so that the horizon is at $z_h=1$. 
(a): For a strip. Note the saturation is discontinuous with $z_c$ lying behind the horizon at the saturation point where each curve stops.  (b): For a sphere.
The saturation is continuous and $z_c$ lies outside the horizon at the saturation point (in the plot it is too close to
the horizon to be discerned).}
\label{fig:numevo}
\end{figure}

So far we have based our discussion on generic features of bulk extremal surfaces without referring to explicit solutions. To understand what happens during intermediate stages of time evolution, 
i.e. between $B$ and $C$ in the figures of Fig.~\ref{fig:curves}, it is useful to work out specific examples of the evolution of  $(z_t (R, \ft), v_t (R, \ft))$.      
In Fig.~\ref{fig:numevo}, we give the parametric plots of $(z_t (R, \ft), z_c (R ,\ft))$ for various values of $R$, for $\Sig$ a strip and a sphere, for Schwarzschild $h(z)$ with $d=3$. From these plots we see a remarkable phenomenon:  curves of varying $R$, after a brief period of order $O(z_h)$,  all collapse into a single curve $z_c^* (z_t)$ highlighted by the dashed line in each plot.

In Sec.~\ref{sec:critical}, we will show that the universal curve $z_c^* (z_t)$ corresponds to a critical line in $(z_t, z_c)$ space: 
for a given $z_t$, $\Ga_{\Sig}$ reaches the boundary only for $z_c < z_c^*$. In particular, for a $\Ga_{\Sig}$ with  $z_c= z_c^* (z_t)$, to which we will refer as a ``critical extremal surface,''
the surface stretches to $\rho, v = \infty$. As a consequence,  for sufficiently large $R$ and $\ft$, $(z_t, z_c)$ lies very close to the critical line, and the evolution of $\sA (R, \ft)$ is largely governed by properties of the critical extremal surfaces. We will show in Sec.~\ref{sec:polin} and \ref{gscale} that this is responsible for the 
linear growth and memory loss regimes discussed in~\cite{Liu:2013iza}.

To conclude this section we comment on the role of $z_h$ in the evolution. As can be seen from the above discussion, $z_h$ plays the characteristic scale for the evolution of $\Ga_{\Sig}$.   
There is an important geometric distinction between the time evolution of surfaces with $R \lesssim z_h$ and of those with $R \gg z_h$. In the former case, $\Ga_{\Sig}(\ft)$ stays outside the horizon during ts entire evolution, while in the latter case important parts of its evolution are controlled by the geometry near and behind the horizon. This supports the identification of $z_h$ as a ``local equilibrium scale'' as only after such time scale does an extremal surface start probing the geometry around the black hole horizon.

\section{Evolution in $(1+1)$ dimensions} \label{sec:2dsto}

Before going to general dimensions, let us first consider the case where $d=2$ and the final equilibrium state is given by the BTZ black hole, i.e. $g(z)=z^2/z_h^2$. Then $n=1$, and $\Ga_{\Sig}$ is a 
geodesic whose length can be expressed analytically in closed
form~\cite{Balasubramanian:2010ce,Balasubramanian:2011ur}, which enables 
us to directly extract its scaling behavior in various regimes. Related boundary observables are the entanglement entropy of a segment of length $2R$, and equal-time two-point correlation functions of operators with large dimension, at separation $2R$. For definiteness, we consider the entanglement entropy, and show that its evolution exhibits the four regimes discussed in the introduction.

It is convenient to introduce the dimensionless variables 
\be
\tau \equiv 2 \pi T \ft  \ , \qquad \ell \equiv 2 \pi T R\  ,
\ee
where $T$ is the equilibrium temperature. First{,} recall the result for entanglement entropy in a CFT at thermal equilibrium~\cite{Holzhey:1994we,Calabrese:2004eu},
\be \label{them}
S_{\rm eq} (\ell) = {c \ov 3}  \log \le({\sinh {\ell } \ov \ell}  \ri) + {c \ov 3} \log {R \ov \de_0}
= \De S_{\rm eq} + S_{\rm vac} \ .
\ee
Here{,} the second term $S_{\rm vac}$ is the vacuum value (with $\de_0$ a UV cutoff), $c$ is the central charge, and $ \De S_{\rm eq} $ denotes the difference between thermal and vacuum values. Note $\De S_{\rm eq} $ is free of any UV ambiguities, and that for  $\ell \gg 1$, we have 
\be \label{latgel}
 S_{\rm eq} = {c \ov 3} \ell - {c \ov 3} \log (4 \pi T \de_0) + O(e^{-2 \ell})\ .
 \ee
Here we see that the $\log R$ piece in $S_{\rm vac}$ has been replaced by a $\log T$ term, signaling a redistribution of long-range entanglement. Also note that the equilibrium entropy and energy densities are given by
\be \label{ed2d} 
s_{\rm eq} = {\pi c T \ov 3}  \ , \qquad  \sE = {\pi c T^2 \ov 6} \ .
\ee
%

Now, the evolution of entanglement entropy in the Vaidya geometry~\eqref{vaidya} with $g(z) = z^2/z_h^2$ is given by
\be 
S ( R, \ft) = \De S ( R, \ft) + S_{\rm vac}\ , 
\ee
where  (following expressions are obtained from Eqs. (3)-(5) of~\cite{Balasubramanian:2010ce} with a slight rewriting)
\be \label{tievo}
 \De S = {c \ov 3} \log \le({\sinh \tau \ov \ell s (\ell, \tau)}\ri) \ ,
\ee 
and the function $s (\ell, \tau)$ is given implicitly  by 
\be \label{conseq}
\ell = {1 \ov \rho}{c \ov s} + \ha \log \le({2 (1+c) \rho^2 + 2 s \rho - c \ov 2 (1+c) \rho^2 - 2 s \rho - c} \ri) 
\ee
with 
\be \label{vard}
\rho \equiv  \ha \coth \tau + \ha \sqrt{{1 \ov \sinh^2 \tau} +{1-c \ov 1+c}}\ , \quad c = \sqrt{1-s^2} \ .
\ee 
At a given $\ell$, the above expressions only apply for 
\be 
\tau < \tau_s (\ell) \equiv {\ell } \ .
\ee 
At $\tau = \tau_s$, one finds that $c=0$ (i.e. $s=1$), $\rho = \coth \tau_s$, and 
\be
 \De S =  \De S_{\rm eq} \ .
\ee
For $\tau > \tau_s$, $\Delta S$ remains $\Delta S_{\rm eq}$.

To make connections to the discussion in Sec.~\ref{sec:dcal}, note that $\rho$ and $s$ can be related to $z_t$ and $z_c$, locations of the tip of $\Ga_{\Sig}$ and its intersection with the null shell, respectively, as
\be \label{relw}
\rho = {z_h \ov z_c} \ , \qquad s = {z_c \ov z_t} \ . 
\ee
Thus equations~\eqref{conseq}--\eqref{vard} provide {an} explicit mapping between boundary data $(\tau, \ell)$ {and} bulk data $(z_t, z_c)$.
{In the discussions that follow,} it is convenient to introduce an angle $\phi \in [0, \pi/2]$ with 
\be \label{relw1}
c = \cos \phi \ , \qquad s = \sin \phi\ .
\ee
Then {saturation happens at $\phi = \pi/2$, when $z_c = z_t$, while} $\phi \to 0$ corresponds to $z_t/z_c \to \infty$.  At fixed $\tau$, as we vary $\phi$ from $\pi/2$ to $0$, $\ell$ increases monotonically from $\tau$ to $+\infty$. At fixed $\ell$, as we increase $\phi$ from $0$ to $\pi/2$,
 $\tau$ increases monotonically from $0$ to $\tau_s$. {Note we will mostly consider the limit $\ell \gg 1$, as we are interested} in long-distance physics.

\subsection{Early growth} 

{For any $\ell$, in the limit $\tau \ll 1$ $\rho$ is large,} and in order for~\eqref{conseq} to be satisfied we need $s$ to be small (i.e. $\phi$ {small}). We find that 
\be \label{rieb}
\rho = {1 \ov \tau} +{\tau \ov 12} + \cdots, \qquad s = {1 \ov \ell} \le( \tau - {\tau^3 \ov 12} + \cdots \ri)
\ee
and 
\be \label{smalltau}
{3 \ov c} \De S =  {\tau^2 \ov 4} - \le({1 \ov 96} + {1 \ov 16 \ell^2} \ri) \tau^4 + O(\tau^6)  \ .
\ee
Note that {for $z_t$ and $z_c$,~\eqref{rieb} translates to} 
\be
z_c = \ft \le(1 + O(\ft^2) \ri) , \qquad z_t = R \le(1 + O(\ft^2) \ri) \ 
\ee
which is consistent with the regime of early growth outlined in Sec.~\ref{sec:dcal}.
 
Thus {at early times, the entanglement entropy grows quadratically as}
\be  
\De S = {c \ov 3} {\tau^2 \ov 4} + O(\tau^4) =  2 \pi \sE \ft^2 + O(\ft^4) \ ,
\ee
where we have used~\eqref{ed2d}. 
This result was also obtained recently in~\cite{Hubeny:2013hz}.

\subsection{Linear growth} \label{sec:lineh}

We now consider the regime $\ell \gg \tau \gg 1$, which corresponds in~\eqref{conseq}--\eqref{vard} 
to {
\be \label{ri1}
e^{-\tau} \ll \phi \ll e^{-2\tau/5}\ , \qquad {1 \ov \tau} \ll 1
\ee 
with 
\be \label{ri2}
\rho = \ha + {\phi \ov 4} + O\le({e^{-2 \tau} \ov \phi}\ri), \quad \ell = {2 \ov \phi}  + \tau  + \log \phi  + O(1) \ .
\ee
}
Then from~\eqref{tievo} we find that
\bea \label{lien2d}
\De S& =& {c \ov 3}  \tau - {c \ov 3}  \log 4 + O\le({\tau \ov \ell}, {\log \ell  \ov \ell} , e^{-2 \tau} \ri) \nn
&=& 2 s_{\rm eq} \ft - {c \ov 3} \log 4 + \cdots \ .
\eea
The leading term agrees with~\eqref{ekko}. Also note that the subleading term is {\it negative} which is important for the maximal rate conjecture of~\cite{Liu:2013iza}, which we will further elaborate in the conclusion section. 

Note that for $z_t$ and $z_c$, equations~\eqref{ri1}--\eqref{ri2} translate to
\be \label{lzerm}
z_c = 2 z_h + \cdots, \qquad {z_t \ov z_c} = {1 \ov \phi} \gg 1 \ .
\ee
In Sec.~\ref{sec:linstr} and Sec.~\ref{sec:polin} we will see that the linear growth of entanglement entropy in~\eqref{lien2d} is generic for all dimensions and collapsing geometries, being a consequence of the critical surface referred to at the end of Sec.~\ref{sec:dcal}.

\subsection{Saturation} \label{sec:d2sat}

Let us now examine the behavior of entanglement entropy as $\tau \to \tau_s$. For this purpose, consider $\phi = {\pi \ov 2} - \ep$ with $\ep \ll 1$. Then from~\eqref{tievo}--\eqref{vard},
\bea\label{350}
\rho &=& \coth \tau - \ha  \tanh \tau  \ep - {1 \ov 4} \le(\tanh \tau (\tanh^2 \tau - 2 ) \ri) \ep^2 \nn
&&+ O(\ep^3)\ , \\
\label{351}
 \ell &=&   \tau + \ha \tanh \tau \ep^2 + O(\ep^3) \ ,
 \eea
 and
 \be\label{352}
{3\ov c} \De S = \log {\sinh \tau \ov \tau} + \ha \le(1 - {\tanh \tau \ov \tau} \ri) \ep^2 + O(\ep^3)\ .
\ee
Now fix $\ell$ and expand $\tau$ near $\tau_s$, i.e. let $\tau = \tau_s - \de$, $\de \ll1$. We find
\be\label{353}
\de = \ha \tanh \tau_s \ep^2 + {1 \ov 6} \tanh^3 \tau_s \ep^3  
+ (\ep^4)
\ee
and 
\be\label{354} 
{3 \ov c} \de S 
=  
{3 \ov c} \De S_{\rm eq}  -{\sqrt{2} \ov 3} \sqrt{\tanh \tau_s} \de^{3 \ov2} - {1 \ov 6} \tanh^2 
\tau_s \de^2 + O(\de^{5/2})  \ .
\ee
In particular, in the limit $\ell \gg 1$,
\be \label{lhe}
{3 \ov c} \De S = 
{3 \ov c} \De S_{\rm eq}  -{\sqrt{2} \ov 3}  \de^{3 \ov2} - {1 \ov 6}  \de^2 + O(\de^{5/2}, e^{-2 \tau_s} \de^{3 \ov 2}) \ .
\ee
We see that the approach to saturation has a nontrivial exponent ${3  \ov 2}$,
\be 
\De S - \De S_{\rm eq} \propto (\ft_s - \ft)^{3 \ov 2} + \cdots , \qquad \ft \to \ft_s \ .
\ee
This result was also recently obtained in~\cite{Hubeny:2013hz}.

To make {connections to the discussion in Sec.~\ref{sec:dcal}, note that for $z_t$ and $z_c$, equations \eqref{351} and \eqref{352} translate to}
\be 
z_c = z_t \le(1 - {\ep^2 \ov 2} + \cdots \ri) , \qquad z_c = z_h \tanh \tau_s + \cdots  
\ee
which is consistent with the picture of {continuous saturation presented} there.

\subsection{Memory loss regime} \label{sec:2dscal}

We {now} show that for $\tau, \ell \gg 1$ with $\tau < \tau_s$, {$S-S_{\rm eq}$} depends on a single combination of $\tau$ and $\ell$ and interpolates between the linear growth of Sec.~\ref{sec:lineh} and the saturation regime of Sec.~\ref{sec:d2sat}. Thus in this regime the ``memory'' of the size $\ell$ of the region is lost. 

First notice from~\eqref{conseq} and~\eqref{vard}  that for any $\phi$,   
\be \label{2dcrl}
\rho > \rho_* \equiv \ha \le(1 + \tan {\phi \ov 2} \ri), 
\ee
and that
\be
\tau,\ell  \to  \infty \qquad \text{as} \qquad \rho \to \rho_*\ .
\ee
Thus to explore the regime $\tau, \ell \gg 1$, take  $\rho = \rho_* + \ep$ with $\ep \ll 1$. Then 
\bea \label{eie1}
\tau &=&  - \ha \log \ep + \ha \log \le( 1+ \cot {\phi \ov 2} \ri) + O(\ep)\ , \\
\label{eie11}
\ell & = & - \ha \log \ep +  \le(\cot {\phi \ov 2} -1 \ri) + \ha \log \le(1 - \cos \phi + \sin \phi \ov 1 + \cos \phi \ri) \nn
&&+ O(\ep)\label{eie2}\ ,
\eea
and the entropy~\eqref{tievo} can be written as 
\be \label{sace}
{3 \ov c} \De S - {3 \ov c}  \De S_{\rm eq}  = \tau - \ell - \log \le( \sin \phi \ri)+ O\le(e^{-2 \tau}, e^{- 2 \ell}\ri) \ .
\ee
Equations~\eqref{eie1} and~\eqref{eie2} imply that 
\be \label{eorp}
\ell -  \tau =  \chi (\phi) + O(\ep) \ , \quad \chi (\phi) \equiv  \le(\cot {\phi \ov 2} -1 \ri)  + \log \tan {\phi \ov 2} \ ,
\ee
i.e. as $\ep \to 0$, $\tau , \ell \to \infty$ but  $\ell -  \tau$ remains finite. {Inverting \eqref{eorp}} to express $\phi$ in terms of $\ell - \tau$, we can write~\eqref{sace} in the scaling form 
\be  \label{scalf}
 \De S -   \De S_{\rm eq}   = {c \ov 3} \lam(\ell -  \tau)  + O(e^{-2 \tau}) 
\ee
where the scaling function {$\lambda$} is given by
\be \label{scaol}
\lam (y) = -  y - \log \le( \sin h^{-1} (y)\ri) \ .
\ee

Note that $\chi (\phi)$ monotonically decreases  from $+\infty$ to $0$ as $\phi$ {increases} from $0$ to ${\pi \ov 2}$. More explicitly, as $\de \to 0$,
\bea 
\phi &=& \de:  \quad \chi (\phi) = {2 \ov \de} +  \log {\de \ov 2} - 1 + O(\de) \ , \nn
\phi &=& {\pi \ov 2} - \de:  \quad \chi (\phi) = {\de^2 \ov 2} + O(\de^3) \ ,
\eea
from which {$\lam$} has the asymptotic behavior 
\be 
\lam(y) = \bca 
  - y  - \log \le({2 \ov y}\ri) + O\le({\log y \ov y}, y^{-1} \ri) & y \gg 1 \cr
       - {\sqrt{2} y^{3 \ov 2} \ov 3} - {y^2 \ov 6} + O(y^{5/2}) & y \ll 1
   \eca \ .
\ee
Then using the expression for large $y$, we find from~\eqref{scalf} and~\eqref{latgel} that 
for $\ell \gg \tau \gg 1$,
\be 
{3 \ov c} \De S =\tau  - \log 4 + O\le(e^{-2 \tau}, {\tau \ov \ell}, {\log \ell \ov \ell} \ri)\ ,
\ee
which recovers~\eqref{lien2d}, and that for $\de \equiv {\ell } - \tau \ll 1$,
\be\label{374}
{3 \ov c} \De S - {3 \ov c} \De S_{\rm eq} = -  {\sqrt{2} \ov 3} \de^{3 \ov 2} - {\de^2 \ov 6} + O(\de^{5 \ov 2})\ ,
\ee
which {recovers~\eqref{lhe}}.

In Sec.~\ref{sec:critA}, we will show that~\eqref{2dcrl} is precisely the critical line $z_c^* (z_t)$ alluded to near the end of Sec.~\ref{sec:dcal}, and {that} the scaling behavior discussed above is controlled by properties of critical extremal surfaces associated with the critical line. 

Finally, we remark that in higher dimensions, there does not exist a closed expression like~\eqref{tievo}, and we have to rely on  geometric features of bulk extremal surfaces to access the above regimes of evolution, as was outlined in Sec.~\ref{sec:dcal}.

\section{Pre-local-equlibrium quadratic  growth}\label{sec:early}


In this section, we consider the  growth of $\sA_{\Sig}(\ft)$ relative to the area of a minimal surface in AdS with the same boundary $\Sig$ for 
\be \label{opm}
\ft \ll z_h \ .
\ee
Recall our earlier discussion in which we identified $z_h$ as a local equilibrium scale -- at the stage of~\eqref{opm} the system has not yet achieved local equilibrium. Except for the energy density which is conserved in time, equilibrium quantities such as temperature, entropy, or chemical potential are not yet relevant at this stage.

We work in general dimensions, and only assume that $g(z)$ has the asymptotic expansion~\eqref{asymMe}. We will derive a universal result that applies to $\Sig$ of arbitrary shape. 

At early times, the null shell lies in the UV part of the geometry, i.e. near the boundary, and the bulk extremal surface crosses the shell near the boundary, i.e. $z_c \to 0$ as $\ft \to 0_+$~(see Fig.~\ref{fig:surexm}(a)). This implies that: (i) the part of the surface lying in the black hole region is very small, and  (ii) the black hole region can be approximated by perturbing pure AdS. Thus our strategy in finding the small $\ft$ behavior of $\sA$ is to expand {$\ft$ and $\sA$} in small $z_c$. 



A general $(n-1)$-dimensional boundary surface $\Sig$ can be parameterized by
\be 
x_a = x_a (\xi^\al)\ , \quad a = 1, 2, \cdots , d-1\ , \quad \al = 1, 2, \cdots , n-1
\ee
where $x_a$ are spatial coordinates along the boundary and $\xi^\al$ are  coordinates parameterizing the surface. The area $A_\Sig$ of $\Sig$ is given by
\be 
A _\Sig= \int d^{n-1} \xi \, \sqrt{\det h_{\al \beta}} \ , \qquad h_{\al \beta} = {\p x_a \ov \p \xi^\al}{\p x_a \ov \p \xi^\beta} \ .
\ee
The $n$-dimensional bulk extremal surface $\Ga_\Sig$ ending on $\Sig$ 
can be parametrized by
\be 
v (\xi^\al, z)\ , \qquad x_a = X_a (\xi^\al, z) 
\ee
which satisfy the $z=0$ boundary conditions
\be
v (\xi^\al,z=0) = \ft \ , \qquad X_a (\xi^\al, z=0) = x_a (\xi^\al) \ .
\ee
We also require $\Ga_\Sig$ to be smooth at the tip $z_t$. The area $\sA_{\Sig}$ of $\Ga_\Sig$ can be written as 
\bea \label{gbec}
\sA _\Sig (\ft)& =& L^n \int_0^{z_t} dz \int d^{n-1} \xi \, z^{-n} \, \sqrt{\det \ga} \nn
&=& \int_0^{z_t} dz \int d^{n-1} \xi \, \sL  (X_a, v)
\eea
where ${1 \ov z^2} \ga$ is the induced metric on $\Ga_\Sig$,
\bea
\ga_{\al \beta} &=&  {\p X_a \ov \p \xi^\al}{\p X_a \ov \p \xi^\beta} - f (v, z) {\p v \ov \p \xi^\al} {\p v \ov \p \xi^\beta}\  , \\
\ga_{\al z} &=& {\p X_a \ov \p \xi^\al}{\p X_a \ov \p z} - f (v, z) {\p v \ov \p \xi^\al} {\p v \ov \p z} - {\p v \ov \p \xi^\al}\ ,
\\
\ga_{zz} &=& {\p X_a \ov \p z}{\p X_a \ov \p z} - f (v, z) \le({\p v \ov \p z}\ri)^2 -2 {\p v \ov \p z} \ .
\eea
Near the boundary of an asymptotic AdS spacetime, i.e. as $z \to 0$ (or $z/z_h \ll 1$), one can show that 
\be \label{genexl}
X_a (z, \xi^\al) = x_a (\xi^\al) + O(z^2) , \quad v (z, \xi^\al) = \ft - z + O(z^2)  \ .
\ee

Now, we denote the solution in pure AdS ($f=1$) with the same boundary conditions as $\Ga_{\Sig}$ by 
$X^{(0)}_a, v^{(0)}$, and as having tip $z_t^{(0)}$ and area $\sA_\Sig^{(0)}$. Recall that our goal is to work out the difference 
\be \label{ardif}
\De \sA_\Sig (\ft) =  \sA_\Sig (\ft) - \sA_\Sig^{(0)}
\ee 
to leading order in small $\ft$. First, note that the pure AdS solution lies at constant $t$, i.e. 
from~\eqref{tzco1} 
\be \label{oeoe}
v^{(0)} (\xi^\al, z) = \ft - z  \ ,
\ee
and that as discussed earlier, $X_a (\xi, z), v(\xi, z)$ deviate by a small amount from corresponding quantities in pure AdS, i.e.
\be \label{AdSp}
X_a (\xi, z) = X^{(0)}_a + \de X_a\  , \quad v (\xi, z) = v^{(0)} + \de v 
\ee
where from~\eqref{genexl}, lowest order terms in $\de X_a$ and $\de v$ in $z$ should start at $O(z^2)$. 
Solving $v (\xi z_c) = 0$, we then find 
\be \label{ftexp}
\ft=z_c+ O(z_c^2)  
\ee
which in turn implies that expanding $\de X_a$ and $\de v$ in small $\ft$, the lowest order terms 
should start at $O(\ft^2)$. 


Next, to leading order in small $\ft$,~\eqref{ardif} can be found  by varying the action~\eqref{gbec},
\bwt
\be \label{arvar}
\De \sA_\Sig (\ft)= \int_0^{z_t^{(0)}} dz d^{n-1} \xi \, {\de \sL \ov \de f}\biggr|_0 \, \de f  + \int d^{n-1} \xi \, \sL (X^{(0)}, v^{(0)}; z_t^{(0)}) \, \de z_t + \int d^{n-1} \xi \, \le(\Pi^z_A\bigr|_0 \de X_A \ri) \biggr|^{z_t^{(0)}}_0  ,
\ee
\ewt
where $|_0$ denotes that a quantity should be evaluated on the pure AdS solution, $X_A = (X_a, v)$, and 
\be 
\Pi_A^z = {\p \sL \ov  \p \p_z X_A} \ , \qquad \de X_A = X_A - X_A^{(0)} \ .
\ee
In deriving~\eqref{arvar} we have assumed that the boundary terms associated with  integration by part over $\xi_\al$ vanish. This is true when $\Sig$ is compact and there is no boundary in the $\xi_\al$ {directions}, and also when $\Sig$ has no dependence on $\xi_{\al}$, as in the case when $\Sig$ is a strip. We proceed to observe that 
\be 
\sL (X^{(0)}, v^{(0)}; z_t^{(0)})  = 0 
\ee
as the area element vanishes at the tip of the bulk surface, {and that similarly}, regularity conditions at the tip for $\Ga_\Sig^{(0)}$ and boundary conditions at infinity imply that the last term in~\eqref{arvar} vanishes.\footnote{This term has to vanish to ensure $X_A^{(0)}$ is a proper solution to equations of motion.}
Thus {only the first term in~\eqref{arvar} is non-zero}. {Now note} 
\be 
{\de \sL \ov \de f} = {L^n \ov z^n} \ha \sqrt{\det \ga} \tr \le(\ga^{-1} {\de \ga \ov \de f} \ri),
\ee
and from~\eqref{oeoe}
\be 
{\de \ga_{\al \beta} \ov \de f} \biggr|_0 = 0 \ , \qquad {\de \ga_{\al z} \ov \de f} \biggr|_0 = 0 \ , \qquad
{\de \ga_{zz} \ov \de f} \biggr|_0 = -1\ .
\ee
Given that for small $z$,
\be 
X_a^{(0)} (\xi^\al, z) = x_a (\xi^\al) + O(z^2)\ ,
\ee
we find 
\be 
\ga_{\al \beta} = h_{\al \beta} + O(z)\ , \quad 
\ga_{\al z} = O(z)\ , \quad \ga_{zz} = 1 + O(z) \ .
\ee
Thus to leading order
\be
 {\de \sL \ov \de f } \biggr|_0= - {L^n \ov z^n} \ha \sqrt{\det h}
 \ee
 and since 
\be
\de f = -M z^d + \cdots
\ee
is nonvanishing only  for $z \in (0, z_c)$, we find (recall \eqref{ftexp})
\be 
\De \sA_{\Sig}  = L^n A_\Sig {M \ov 2}  \int_0^{z_c} z^{d-n} dz   = {L^n A_\Sig M \ov 2 (d-n+1)}  \ft^{d-n+1} + \cdots 
\ .
\ee

For entanglement entropy, we have $n =d-1$ and thus 
\be \label{quda}
\De S = {\De \sA_\Sig \ov 4 G_N} = {L^{d-1} M \ov 16 G_N} A_\Sig \ft^2 + \cdots ={\pi \ov d-1} \sE A_\Sig \ft^2 + \cdots
\ee
where $\sE$ given in~\eqref{endgd} is the energy density of the system. This expression is free of any UV ambiguities and is universal for any $\Sigma$ and bulk geometry $g(z)$, depending only on the energy density of the state. 

More general metrics~\eqref{vaidya1}--\eqref{bhnem} typically involve scalar fields and the asymptotic 
behavior of the metric components $h(z)$ and $l(z)$ in the black hole region in general depend on the 
falloff of the scalar fields. Furthermore the energy density can also receive contributions from scalar fields. 
Thus it appears likely that~\eqref{quda} may not generalize to such a case. It would be interesting to understand 
this further. 

\section{Critical extremal surfaces} \label{sec:critical}

In this section, using as examples cases of $\Sig$ being a strip or sphere, 
we show that the universal curve $z_c^* (z_t)$ for different $R$'s observed in 
Fig.~\ref{fig:numevo} corresponds to a critical line in $(z_t, z_c)$ space: for a given $z_t$, $\Ga_{\Sig}$ reaches the boundary only if $z_c < z_c^*$. In particular, when $\Ga_{\Sig}$ lies precisely on the critical line  $z_c= z_c^* (z_t)$, in which case we refer to it as a {\it critical extremal surface}, it asymptotes to $\rho, v = \infty$ along some constant $z = z_m \geq z_h$. 


\subsection{Strip} \label{sec:critA}

With $\Sigma$ a strip, the black hole portion of $\Ga_{\Sig}$ is given by $z(\rho)$ satisfying the equation of motion~\eqref{eio2},
\bea \label{yuem}
z'^2 &=&  h(z) \le({z_t^{2n} \ov z^{2 n}} -1\ri) + E^2  (z_t, z_c) \equiv H(z)\ , \nn
E^2& =& {g^2_c \ov 4} \le({z_t^{2n} \ov z_c^{2 n}} -1\ri)  
\eea
and the boundary condition at $z_c$~\eqref{ntrn1},
\be \label{oepp}
z_+'  = - \le(1 - {g_c \ov 2} \ri) \sqrt{{z_t^{2n} \ov z^{2n}_c} -1} \ ,
\ee
where we have denoted
\be 
g_c \equiv g (z_c)
\ee
and $E$ has been obtained from~\eqref{eeval}. As discussed in Sec.~\ref{sec:dcal}, for $\ft \gg z_h$, the extremal surface intersects the shell behind the horizon, i.e. $z_c > z_h$, except possibly near saturation.  

Equation~\eqref{yuem} specifies a one-dimensional classical mechanics problem, with the qualitative behavior of $z (\rho)$ readily deduced from properties of $H(z)$. To {acquire} some intuition on such behavior, {we proceed to work concretely with the Schwarzschild (or Reissner-Norstrom) $g(z)$}. 
{Since our discussion clearly applies to more than the examples of $g(z)$ being examined}, we {maintain the general notation} $g(z)$ and $h(z) = 1- g(z)$ in all expressions. However, we do not attempt to characterize the most general class of $g(z)$ for which $H(z)$ exhibits properties discussed below, nor do we attempt to classify alternative possibilities. 

To begin, note that from~\eqref{oepp}, when $g_c > 2$, $z_+' > 0$ i.e. after entering the black hole region, $\Ga_{\Sig}$ initially moves away from the boundary to larger values of $z$. {We}  introduce $z_s$ as 
\be \label{defzzs}
g(z_s) = 2\ , \qquad z_s > z_h \ .
\ee
$z_+'$ changes sign when $z_c$ crosses  $z_s$. {Next, note that for Schwarzschild $g(z)$}, the first term in~\eqref{yuem} is zero at $z= z_h$ and $z=z_t$, and {negative in between}. Thus  $H(z)$ {has a minimum between $z_h$ and $z_t$ which we denote} $z_m$. Setting $H' (z_m) =0$, {we find $z_m$ satisfies the equation}
\be \label{ejek}
z_t^{2n} = {h' (z_m) z_m^{2n+1} \ov z_m h'(z_m) - 2n h(z_m)} \ .
\ee 
It is easy to see that such a minimum also exists for Reissner-Norstrom {$g(z)$}. {The following discussion only depends} on the existence of such a minimum. 
We now introduce $z_c^*$ {given by}
\be \label{findzc}
H(z_m) \bigr|_{z_c= z_c^*} =0  \ .
\ee
Note $z_c^*$ and $z_m$ are functions of $z_t$ only. {Also note} that there is a special value of $z_t$, which we call $z_t^{(s)}$, where 
 $z_m (z_t^{(s)}) = z_s$. Evaluating~\eqref{ejek} at $z_m = z_s$, we find that
 \be \label{spezt}
z_t^{(s)} = \le( { z_s h' (z_s)  \ov z_s h'(z_s) + 2n} \ri)^{1 \ov 2n} z_s \ .
\ee 
In fact, there are two {additional occurences} at $z_t = z_t^{(s)}$. First, one can check  
\be
z_c^*  =  z_s = z_m \ . 
\ee
Second, by taking the derivative of $E^2$ in~\eqref{yuem} with respect to $z_c$ and {plugging in the values $z_t^{(s)}$ and $z_c^* = z_s$}, we find 
\be \label{eceq}
{dE^2 \ov d z_c} \biggr|_{z_c^*} = 0 \ .
\ee

In the limit $z_t \to \infty$, assuming that $z_m$ remains finite (which is not always true, see e.g.~\eqref{mner1}--\eqref{mner2} below), equations~\eqref{ejek} can be simplified to 
\be \label{ohh1}
{z_m h' (z_m) \ov h (z_m)} = 2n  \ .
\ee
Similarly in the $z_t \to \infty$ limit, assuming that ${z_t \ov z_m}, {z_t \ov z_c^*} \to \infty$, equation~\eqref{findzc}  can be simplified to 
\be \label{ohh2}
{g^2 (z_c^*) \ov 4 z_c^{*2n}} = - {h(z_m) \ov z_m^{2n}} \ .
\ee

In general, for a given $z_t$ there are multiple positive roots to the equation~\eqref{findzc}. In fact, equation~\eqref{eceq} suggests that {two branches of roots of~\eqref{findzc} are converging at $z_t^{(s)}$}. However, for any $g(z)$ which satisfies $g(z_h) = 1$ and $g'(z_h)> 0$, it can be {checked} that 
as $z_t \to z_h$ so that {$z_t = z_h (1 + \ep)$, $\ep \ll 1$}, we have 
\be
z_m =z_t \le(1 - \ha \ep + \cdots\ri) =  z_h \le(1+ \ha \ep + \cdots\ri) 
\ee
and there is a unique $z_c^*$ satisfying 
\be \label{rigro}
z_c^* = z_t \le(1 - O(\ep^2) \ri) \ .
\ee
{Now, increasing $z_t$ and following this root, we note that}: 

\ben

\item {In region I given by} $z_h < z_t < z_t^{(s)}$, 
\be \label{reg1}
z_s > z_c^* > z_m > z_h \ , \qquad  {dE^2 \ov d z_c} \biggr|_{z_c^*} < 0\ ,
\ee
and thus for $z_c < z_c^*$, 
\be 
z_+' < 0 \ . 
\ee

\item {In region II given by} $ z_t > z_t^{(s)}$, 
\be \label{oen2}
z_s < z_c^* < z_m  \ , \qquad  {dE^2 \ov d z_c} \biggr|_{z_c^*} > 0\ ,
\ee
and for $z_s < z_c < z_c^*$, 
\be 
z_+' > 0 \ . 
\ee

\begin{figure}[!h]
\centering
\subfigure[]{}{\includegraphics[scale=0.55]{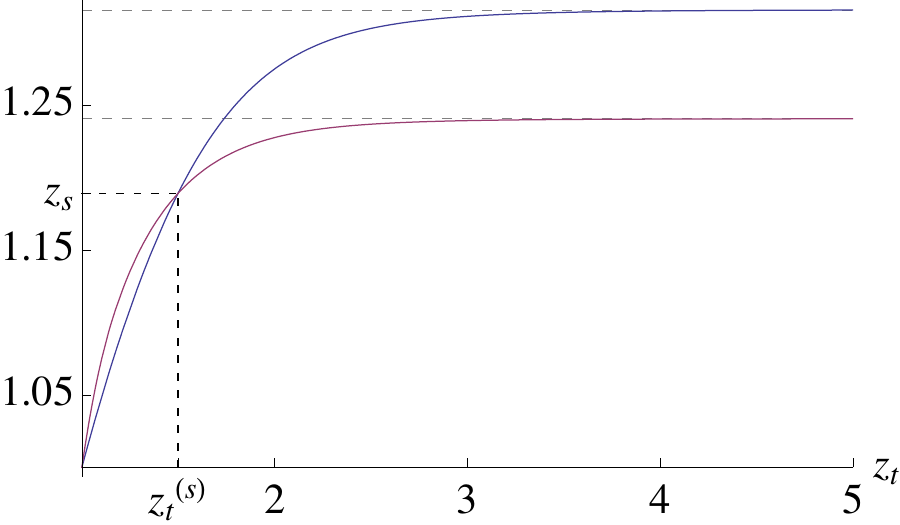}}
\subfigure[]{}{\includegraphics[scale=0.55]{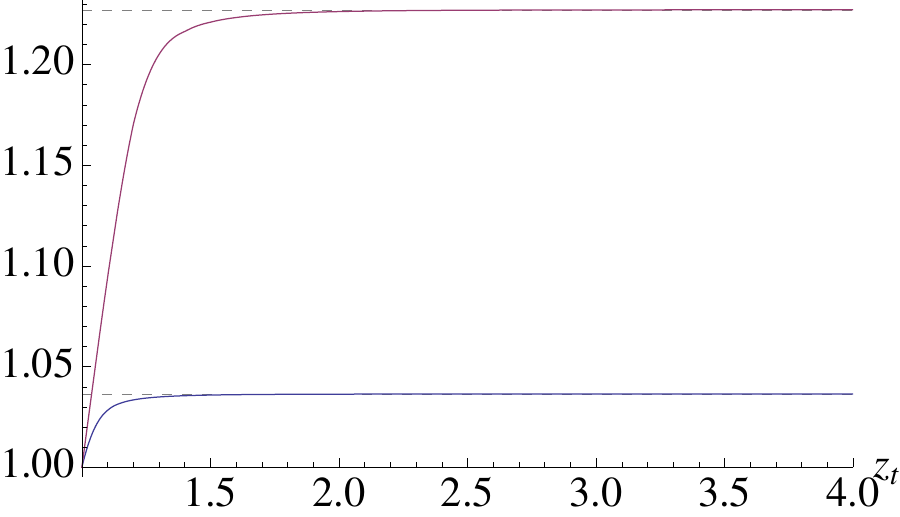}}
\caption{Examples of $z_m$ (blue) and $z_c^*$ (red) as functions of $z_t$ for (a): Schwarzschild $g(z)$ with $d=4$ and $n=3$, (b): RN $g(z)$ with $d=4$, $u=0.2$, and $n=3$. We have fixed $z_h=1$. Note in (b), $z_s$ does not exist and there is only region I~\eqref{reg1}. }
\label{fig:zmzcs} 
\end{figure}

\een
See Fig.~\ref{fig:zmzcs} for plots of $z_c^*$ and $z_m$ as functions of $z_t$ for Schwarzschild $g(z)$ and 
one instance of RN $g(z)$. 

\begin{figure}[h!]
\centering
\subfigure[]{\includegraphics[scale=0.4]{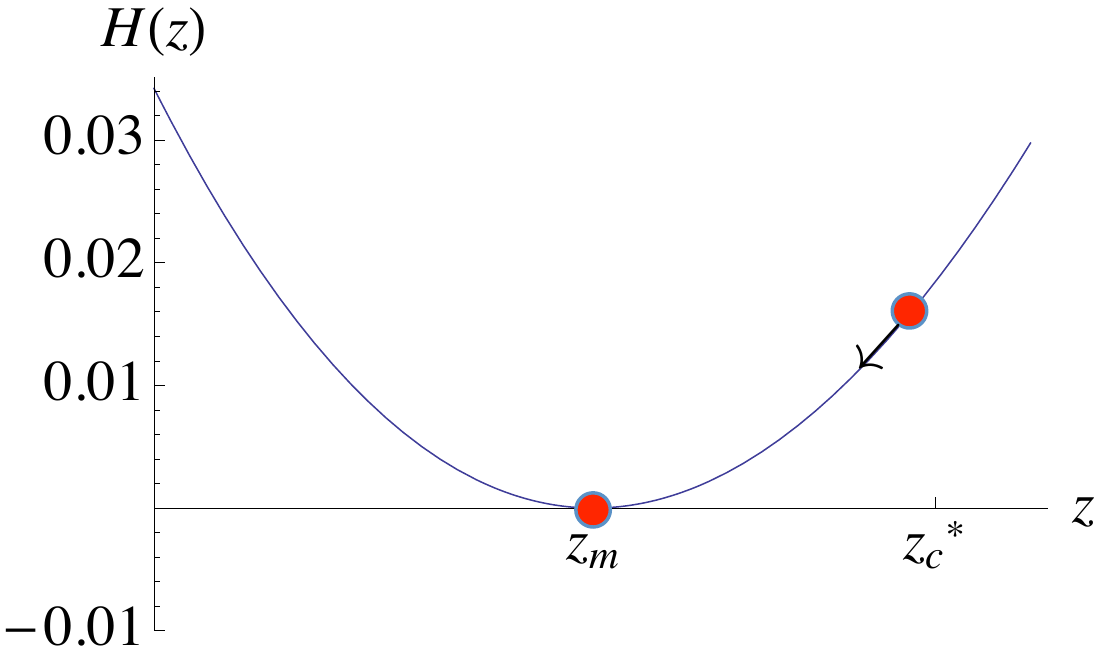}} 
\subfigure[]{\includegraphics[scale=0.4]{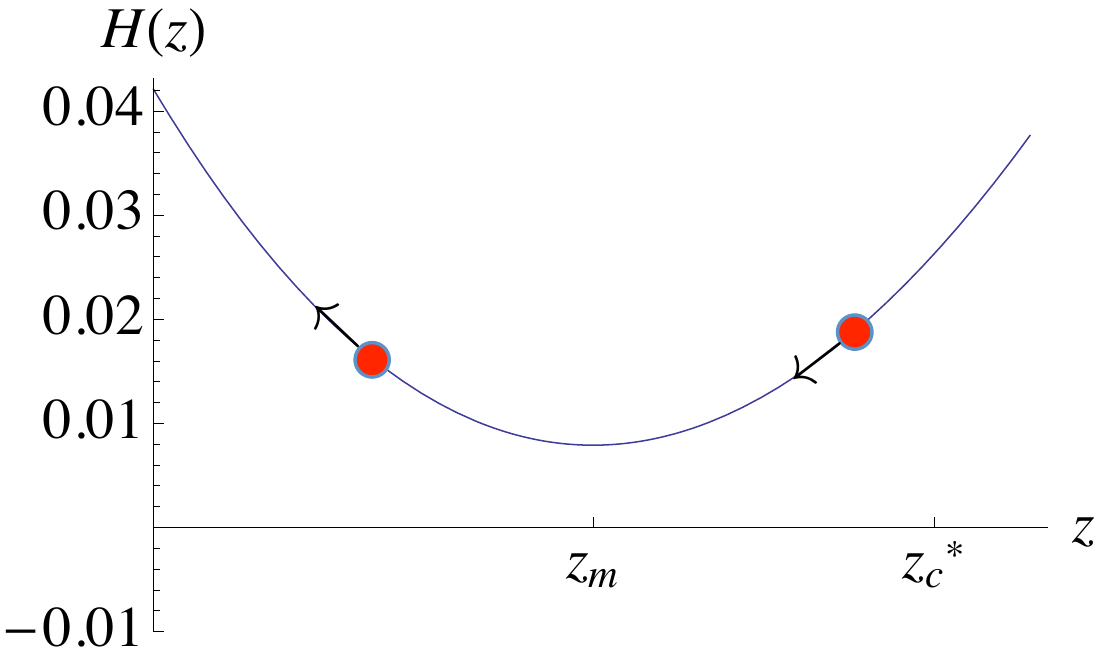}} 
\subfigure[]{\includegraphics[scale=0.4]{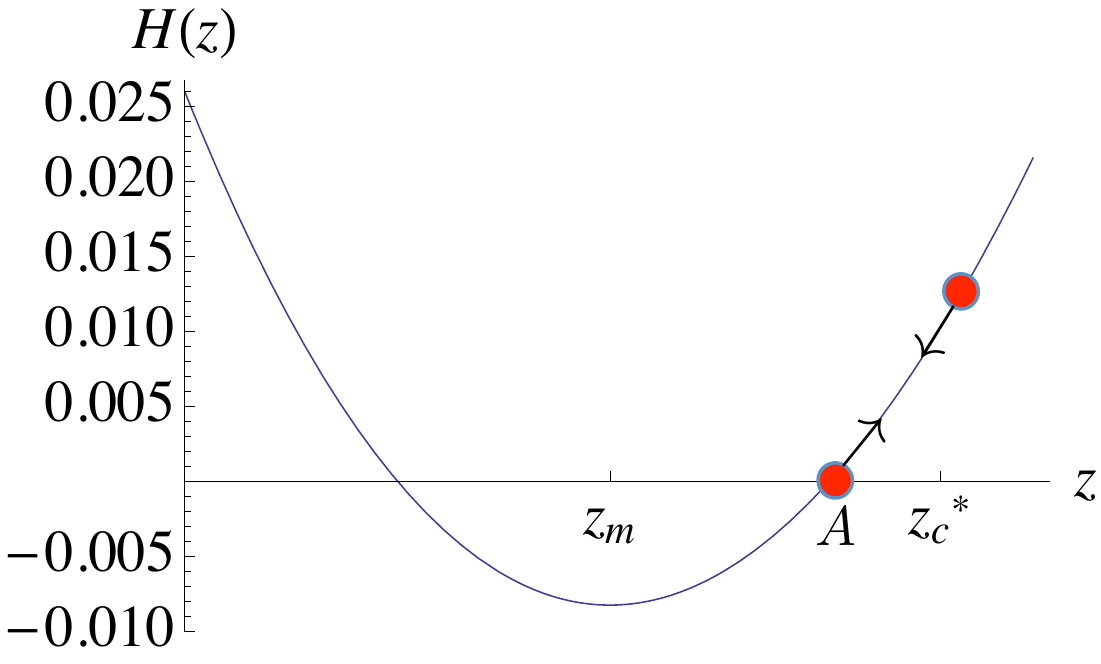}} 
\caption{ $H(z)$ for $z_t < z_t^{(s)}$. In this case $z_c^* > z_m$ and $z_+' < 0$ for $z_c \lesssim z_c^*$. 
(a): $z_c = z_c^*$. 
$z(x)$ decreases then asymptotes to $z=z_m$. 
(b): $z_c = z_c^* - \ep$ for $\ep >0$. Since ${d E^2 \ov d z_c} \bigr|_{z_c^*} < 0$,  
$H(z_m) > 0$.  $z' $ remains negative throughout and $\Ga_{\Sig}$ can reach the boundary. If $\ep$ is small, then $H(z_m)$ is small (positive) and $\Ga_{\Sig}$ hangs near the critical extremal surface $z = z_m$ for a long interval in $x$ before eventually reaching the boundary. 
(c): $z_c = z_c^* + \ep$. Since 
${d E^2 \ov d z_c} \bigr|_{z_c^*} > 0$, now  
$H(z_m) < 0$. $z (x)$ first decreases to point A, then turns around and never reaches the boundary.  
} 
\label{fig:stripevo1}
\end{figure}

\begin{figure}[h!]
\centering
\subfigure[]{\includegraphics[scale=0.4]{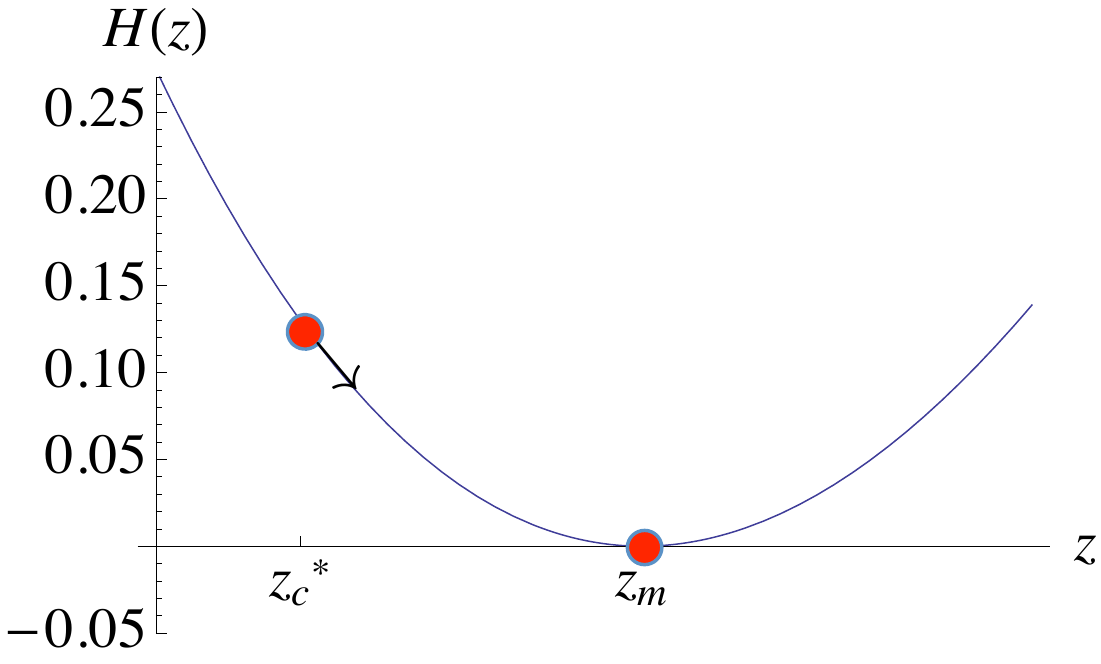}} 
\subfigure[]{\includegraphics[scale=0.4]{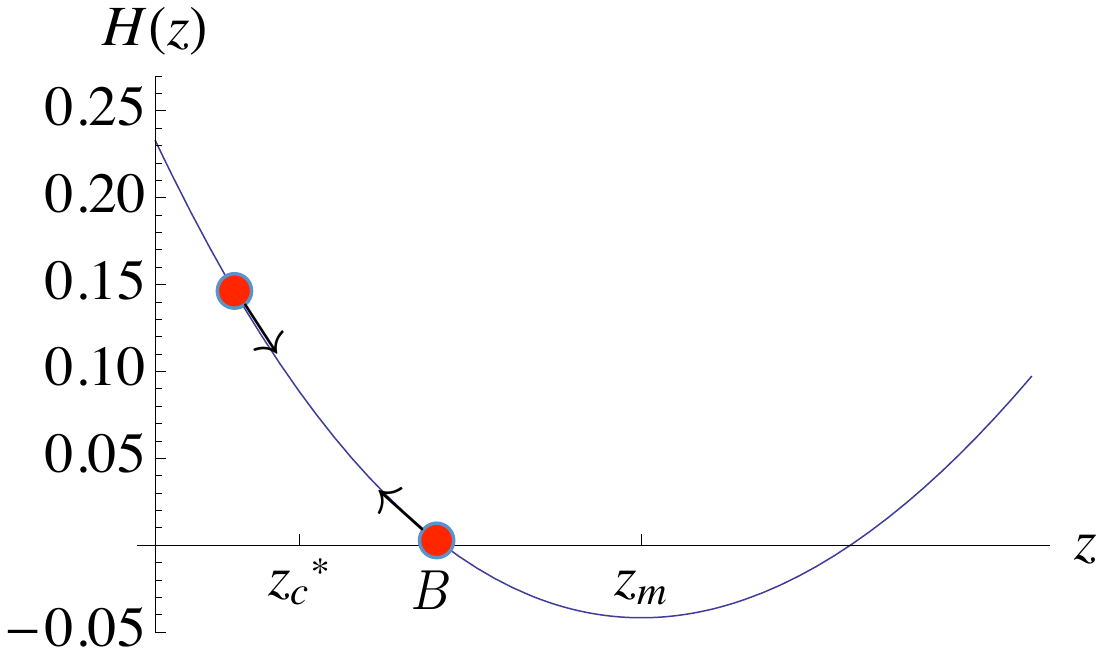}} 
\subfigure[]{\includegraphics[scale=0.4]{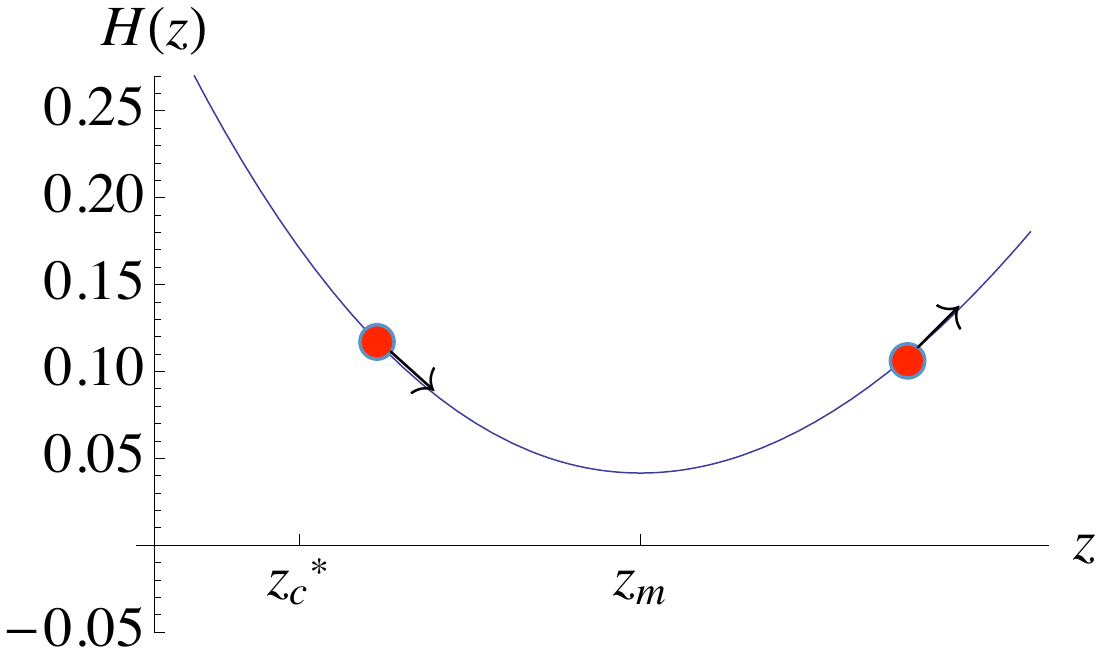}} \caption{ $H(z)$ for $z_t > z_t^{(s)}$. In this case {$z_c^*< z_m$. If $z_c < z_s$, $z'_c < 0$ and $z (\rho)$ monotonically} decreases to zero. 
{These} plots show what happens {when $z_c > z_s$ so that $z_+' > 0$}. 
(a): $z_c = z_c^*$.  $z(\rho)$ {increases} and asymptotes to $z=z_m$. 
(b): $z_c = z_c^* - \ep$ for a positive $\ep >0$. With 
${d E^2 \ov d z_c} \bigr|_{z_c^*} > 0$,  
 $H(z_m) < 0$. $z(x)$ first increases, then turns around at point $B$ and monotonically decreases to zero. 
 If $\ep$ is small, then $H(z_m)$ is small (negative), and $\Ga_{\Sig}$ hangs near the critical surface $z = z_m$ (i.e. near $B$) for a long interval in $x$ before eventually reaching the boundary. 
(c): $z_c = z_c^* + \ep$. With $H(z_m) > 0$, $z (x)$ only increases
and never reaches the boundary.
} 
\label{fig:stripevo2}
\end{figure}

With the above properties established, the behavior of $z(\rho)$ can be read off from 
Fig.~\ref{fig:stripevo1}--Fig.~\ref{fig:stripevo2}. In particular, for a given $z_t$, 
$\Ga_{\Sig}$ only reaches the boundary for $z_c < z_c^* (z_t)$,
and at $z_c=z_c^*(z_t)$, it asymptotes to a critical extremal surface $z = z_m$. 
Note that this conclusion holds in the presence of other roots to~\eqref{findzc} as long as the following are satisfied:

\ben 

\item In region I there is no other root lying between $z_m$ and $z_c^*$. 

\item In region II there is no other root lying between $z_s$ and $z_c^*$. 

\begin{figure}[t]
\centering
\subfigure{\includegraphics[scale=0.52]{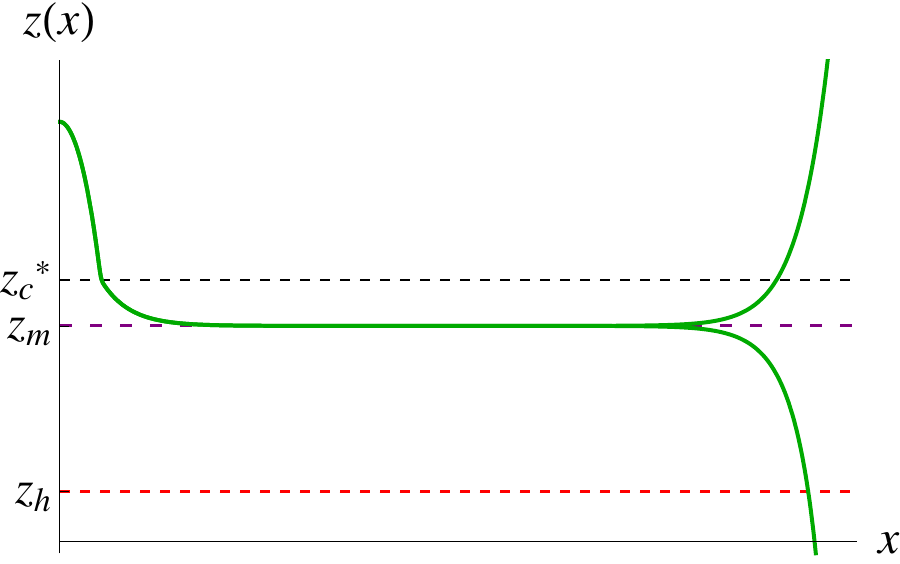}}
\subfigure{\includegraphics[scale=0.52]{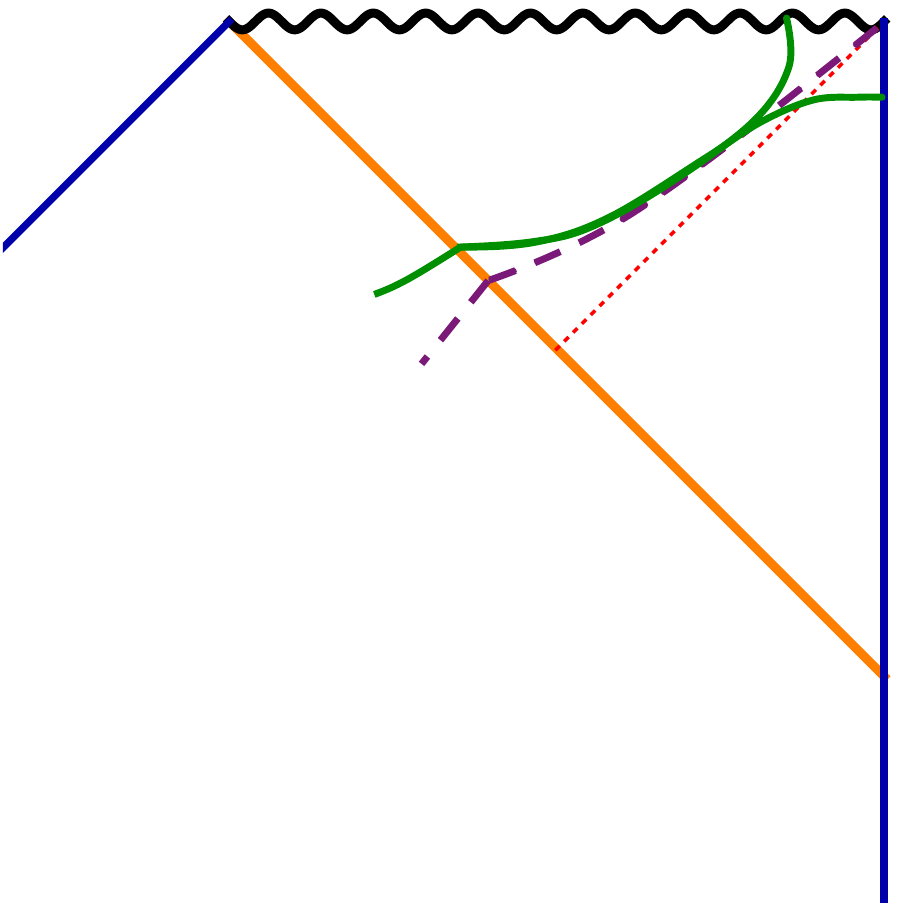}}
 \caption{Left: Behavior of near-critical surfaces with $\ep=\pm 10^{-12}$ for 
$z_t=1.3 z_h < z_t^{(s)}$, for Schwarzschild with $d=3$, $n=2$, and $\Sig$ a strip. The critical surface runs to infinite $x$ along $z=z_m$. For small $\ep$, the solution runs along the critical surface for a while before reaching the boundary or black hole singularity, depending on the sign of $\ep$. Right: Cartoon of the near-critical surfaces on the Penrose diagram. Dashed curve is constant $z = z_m$ slice. 
} 
\label{fig:stripevo11}
\end{figure}

\begin{figure}[h!]
\centering
\subfigure{\includegraphics[scale=0.52]{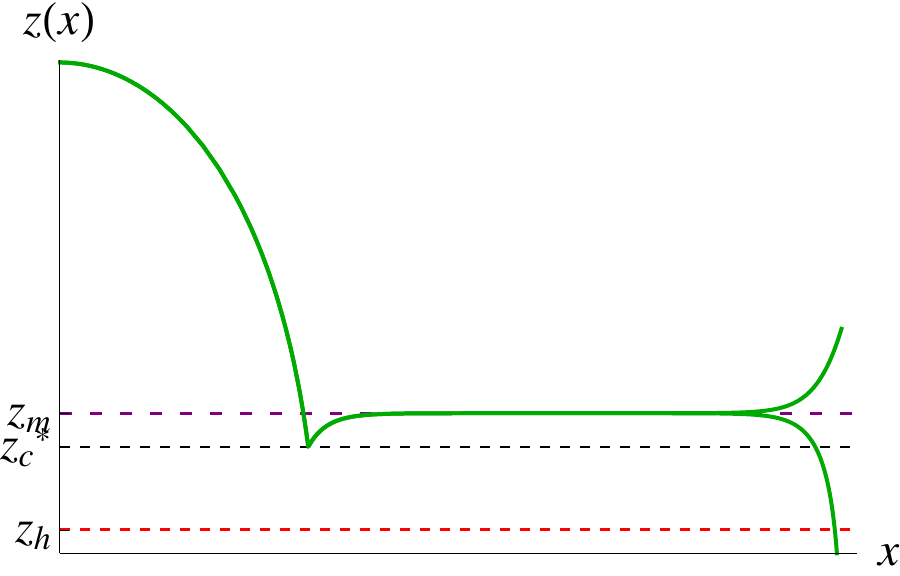}}
\subfigure{\includegraphics[scale=0.52]{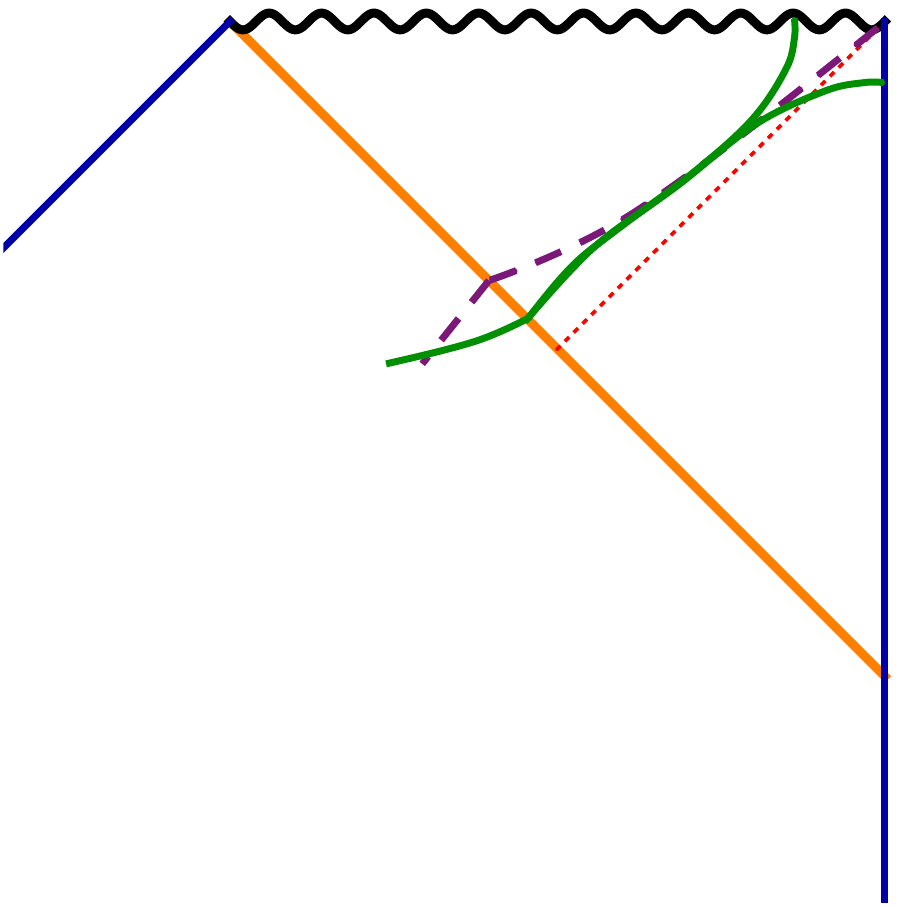}}
\caption{Left: Behavior of near-critical surfaces with $\ep=\pm 10^{-12}$ for $z_t=3 z_h >z_t^{(s)}$, for Schwarzschild $g(z)$ with $d=3$, $n=2$, and $\Sig$ a strip. Right: Cartoon of the near-critical surfaces on the Penrose diagram.
} 
\label{fig:stripevo21}
\end{figure}

\een
It can be readily checked that these conditions are satisfied by Schwarzschild and Reissner-Nordstrom $g(z)$ for general $d$. 
In Fig.~\ref{fig:stripevo11}--\ref{fig:stripevo21} we plot some examples of near-critical surfaces with $z_c \approx z_c^*$.

Now let us mention some explicit results. For Schwarzschild $g(z)$~\eqref{schw} and $d=2$, the case discussed in Sec.~\ref{sec:2dsto}, one finds 
\be 
z_s = \sqrt{2} z_h, \qquad z_t^{(s)} =2 z_h, \qquad  z_m = \sqrt{z_t z_h} \ ,
\ee
and\footnote{Note that in this case there are two positive 
roots to equation~\eqref{findzc}. {The root below is the branch chosen by~\eqref{rigro}.}}
\be \label{2dcrl1}
z_c^*  = \ha \le(\sqrt{z_t^2 + 4 z_tz_h - 4 z_h^2} - z_t \ri)  + z_h ,
\ee
{where}
\be \label{zcrl}
\quad z_c^* \to 2 z_h \ , \quad {\rm as}\quad z_t \to \infty \ .
\ee
Using~\eqref{relw} and~\eqref{relw1}, one finds that the critical line~\eqref{2dcrl1} {is precisely equivalent} to~\eqref{2dcrl}. 
Similarly,~\eqref{zcrl} maps to~\eqref{lzerm}.  

For Schwarzschild $g(z)$~\eqref{schw} in general $d$, one has 
\be 
z_s = 2^{1 \ov d} z_h \ , \qquad z_t^{(s)} = \le({d \ov d-n}\ri)^{1 \ov 2n} 2^{1 \ov d} z_h \ .
\ee
but the expressions for $z_m$ and $z_c^*$ get complicated. In the following discussion we will mostly be interested in the $z_t \to \infty$ limit, for which introducing
\be \label{defeta}
\eta \equiv  {2 n \ov d} \ ,
\ee
we find:
\ben 

\item For $\eta > 1$, 
\be \label{mner}
z_m =  \le(\eta \ov \eta -1 \ri)^{1 \ov d} z_h\ , \quad z_c^* = 
\le({4 (\eta-1)^{\eta -1} \ov \eta^\eta}\ri)^{1 \ov 2 (d-n)} z_h  
\ .
\ee
Note that both $z_m$ and $z_c^*$ remain finite as $z_t \to \infty$ 
and ${z_c^* \ov z_m} = \le({4 (\eta-1) \ov \eta^2}\ri)^{1 \ov 2 (d-n)} < 1$. 

\item {For $\eta <1$,
\be \label{mner1} 
z_m = (1-\eta)^{1 \ov 2n} z_t\ , \qquad z_c^* \sim z_t^{ {d-2n \ov 2 (d-n)}}  \ll z_m\ .
\ee
Note that both $z_m$ and $z_c^*$ approach infinity as $z_t \to \infty$.}

\item For $\eta=1$, i.e. $n = {d \ov 2}$, 
\be\label{mner2} 
z_m = \sqrt{z_t z_h}\ , \qquad z_s < z_c^* = 2^{1 \ov d-n} z_h  \ll z_m \ .
\ee
In this case $z_m$ approaches infinity, but $z_c^*$ remains finite as $z_t \to \infty$. 

\een
For Reissner-Nordstrom $g(z)$~\eqref{eq:RN}, we find that for $n=d-1$ and in the limit $z_t \to \infty$,
\be  \label{rnex}
z_m = \le({2 (d-1) \ov d-2} {1 \ov 1 + Q^2 z_h^{2d-2}}\ri)^{1 \ov d} z_h
\ee
and $z_c^*$ is also finite but is given by a complicated expression which is not particularly illuminating. 
Also note that in the extremal limit,
\be
z_m \to z_h\ , \qquad z_c^* \to \le(2-{2 \ov d}\ri)^{1 \ov d-2}z_h\ .
\ee
and that for sufficiently large $Q$, $z_m$ never reaches $z_s$ for all $z_t$.

\subsection{Sphere} \label{sec:critB}

We now examine the case of $\Sig$ being a sphere with $n\geq 2$ (thus $d \geq 3$). The analysis is more complicated as the equation of motion for $z (\rho)$ giving the black hole portion of $\Ga_{\Sig}$ is now a second order nonlinear differential equation,~\eqref{modeom}. We copy it here for convenience,
\bea \label{modeom1}
&&\le( h +E^2 B^2\ri)z''  +  \le(h + z'^2  \ri)\le({n-1 \ov \rho} z' + {n h \ov z} \ri)   \nn
&&+\le(E^2B^2-z'^2\ri) {\p_z h  \ov 2}=0
\eea
with 
\be \label{varp2}
E=-{1 \ov 2}\le({\rho_c \ov z_c}\ri)^{n}{g(z_c) \ov z_t}\ , \qquad B \equiv {z^{n} \ov \rho^{n-1}}  \ .
\ee

\begin{figure}[h!]
\begin{center}
\includegraphics[scale=0.6]{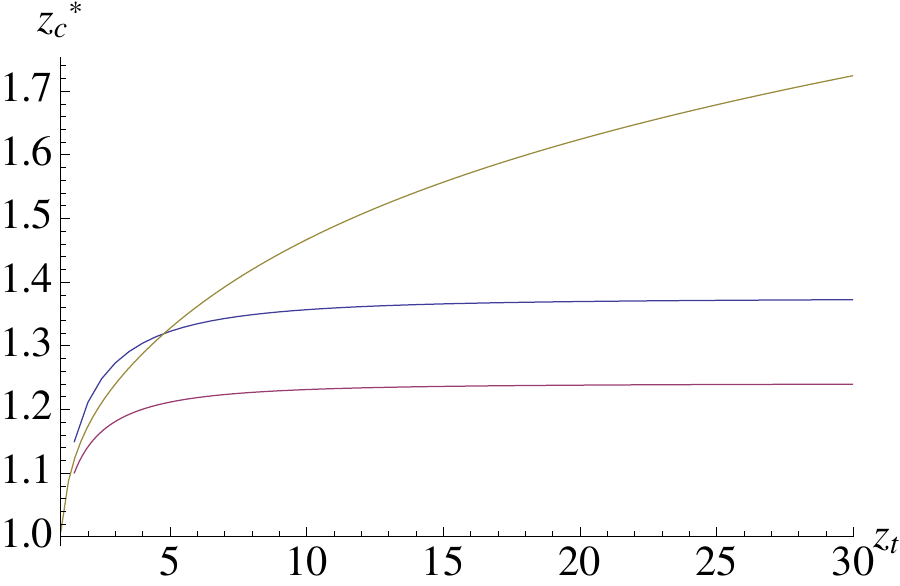} 
\end{center}
\caption{Plot of $z_c^* $ as a function of $z_t$ for $d=3$ Schwarzschild with $n=2$ (blue), $d=4$ with $n=3$ (red), and $d=6$ with $n=2$ (yellow). We plot in the unit $z_h =1$. 
For the last case $z_c^*$ appears to grow with $z_t$ as $z_t^{1 \ov 7}$. This should also be compared with the strip case~\eqref{mner1} where $z_c^*$ grows with $z_t$ as $z_t^{1\ov 4}$. 
}
\label{fig:zcplot}
\end{figure}

 We again expect that for a given $z_t$, there is a critical $z_c^*$ beyond which $\Ga_{\Sig}$ never reaches the boundary. For a given $h(z)$, $z_c^* (z_t)$ can be readily found by numerically solving~\eqref{modeom1}. From the strip analysis~\eqref{mner}--\eqref{mner2}, a natural guess for Schwarzschild $h(z)$ is that for $\eta = {2n \ov d} \geq 1$, $z_c^*$ remains finite as $z_t \to \infty$. This appears to be supported by numerical results. In Fig.~\ref{fig:zcplot} we show some examples. 

\begin{figure}[h!]
\centering
\subfigure{\includegraphics[scale=0.52]{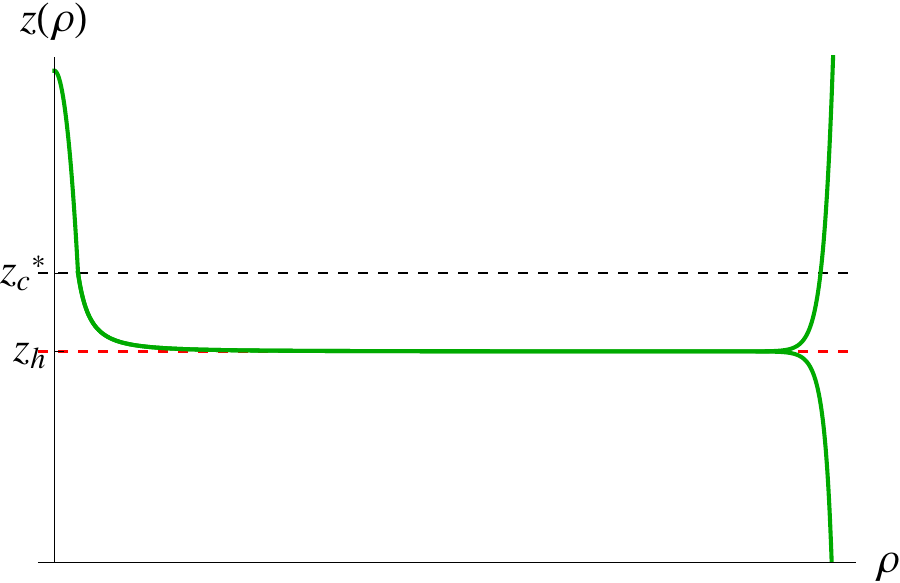}}
\subfigure{\includegraphics[scale=0.52]{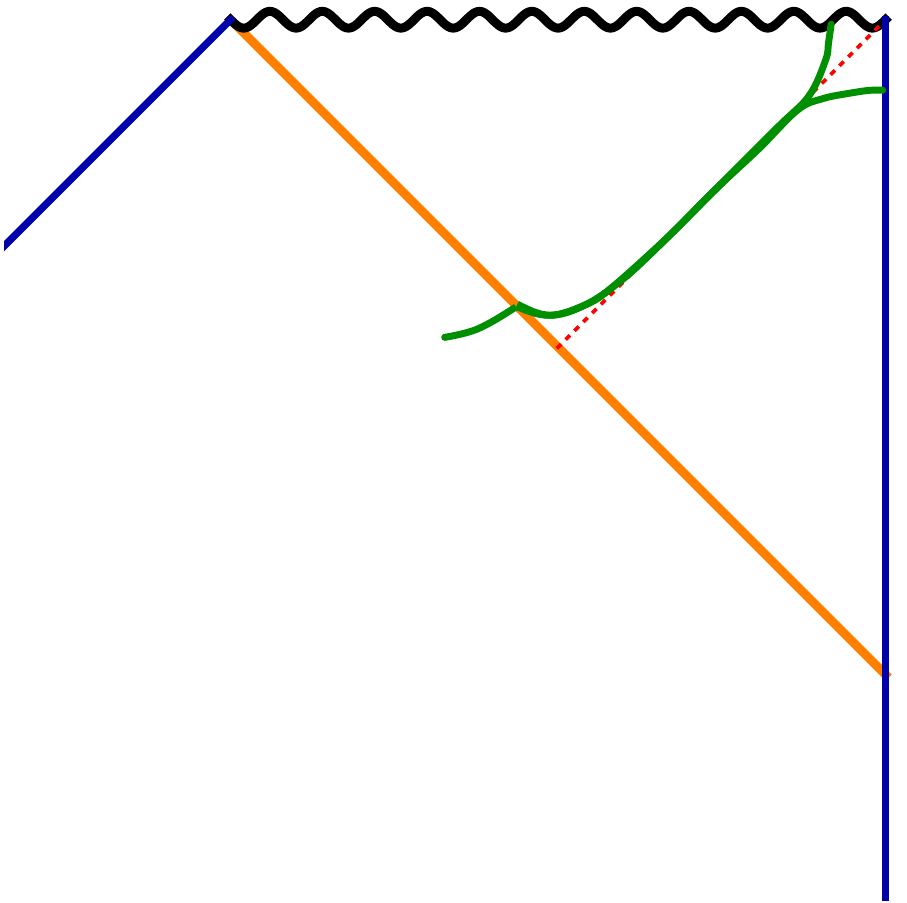}}
\caption{Left: Behavior of near-critical surfaces with $\ep=\pm 10^{-58}$ for $z_t=1.5 z_h \sim z_c^{*}$, for Schwarzschild  with $d=4$, $n=3$, and $\Sig$ a sphere. Note the surfaces now run along the horizon (c.f. Figs.~\ref{fig:stripevo11}, \ref{fig:stripevo21}). Right: Cartoon of the near-critical surfaces on the Penrose diagram.} 
\label{fig:critical}
\end{figure}

At $z_c = z_c^*$  
the critical solution $z^* (\rho)$ should reach $ \rho = \infty$ along some constant $z$ surface.
Now, solving~\eqref{modeom1} for a constant $z$ in the limit $\rho \to \infty$, one finds the unique solution
\be \label{sosle}
z = z_h \ .
\ee
In other words, independent of the choice of $z_t$ and {the} function $g(z)$, the critical extremal surface approaches and runs along at the horizon to $\rho=\infty$. Expanding about the solution~\eqref{sosle} in the equation~\eqref{modeom1}, one finds a perturbation which grows exponentially in $\rho$ (in Sec.~\ref{gscale} we work this out explicitly).  
By tuning $z_c$ to $z_c^*$, one ensures that this exponentially growing perturbation is absent and $ z \to z_h$ 
as $\rho \to \infty$. For $z_c = z_c^* (1- \ep)$, $\ep \ll 1$, the perturbation acquires a small coefficient, and $z(\rho)$ runs along the horizon for a while before eventually breaking away. Depending on the sign of $\ep$, it either approaches the boundary ($\ep >0$) or turns away from it ($\ep < 0$).  See Fig.~\ref{fig:critical}. 

\begin{figure}[h]
\begin{center}
\includegraphics[scale=0.8]{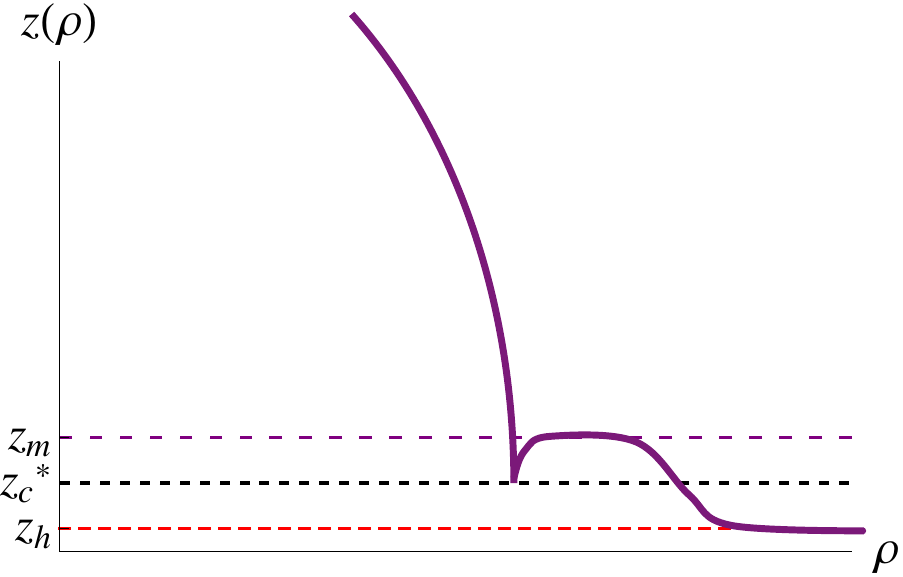} 
\end{center}
\caption{ Cartoon of $z^* (\rho)$ for $z_t \gg z_c^*$, with $z_c^* \sim O(1)$ as $z_t \to \infty$ : There is an intermediate plateau 
 at $z = z_m$ for $\rho \sim \rho_c$. The critical surface eventually approaches the horizon for $\rho \gg \rho_c$.
}
\label{fig:critical1}
\end{figure}

When $z_t$ is large and $z_c^*$ remains finite 
in the large $z_t$ limit, the critical extremal solution $z^* (\rho)$ has another interesting feature 
which will play an important role in our discussion of the linear growth of entanglement entropy in Sec.~\ref{sec:polin}. 
From~\eqref{varp00}, for a finite $z_c \approx z_c^*$,
\be 
\rho_c = z_t + O(1/z_t), \qquad z_t \to \infty \ .
\ee
Then for the range of 
$\rho$ satisfying $\rho \geq \rho_c$ and ${\rho \ov \rho_c} \approx 1$, equation~\eqref{modeom1} can be 
solved approximately by 
$z^* (\rho) \approx z_m$ with $z_m$ given by
\be  \label{nnor}
{n h^2 (z_m) \ov z_m} + \le({z_m \ov z_c^*}\ri)^{2n} {g^2 (z_c^*) h' (z_m) \ov 8 }=0 \ .
\ee
The above equation is obtained from~\eqref{modeom1} by setting $z (\rho) = z_m$, $z_c = z_c^*$
and ${\rho_c^{2n} \ov z_t^2 \rho^{2(n-1)}} = 1$. This results in a plateau at $z= z_m$ for a range of $\rho \sim \rho_c$ 
as indicated in Fig.~\ref{fig:critical1}. Note equation~\eqref{nnor} agrees prescisely with equations~\eqref{ohh1}--\eqref{ohh2} for a strip. That is, provided the $z_c^*$ in~\eqref{nnor} agrees with that of the strip, the $z_m$ determined 
from~\eqref{nnor} agrees precisely with the location of the critical surface for a strip. We will show in Sec.~\ref{sec:linsph} this is indeed the case.


\subsection{Summary} 

\begin{figure}[!h]
\begin{center}
\includegraphics[scale=0.6]{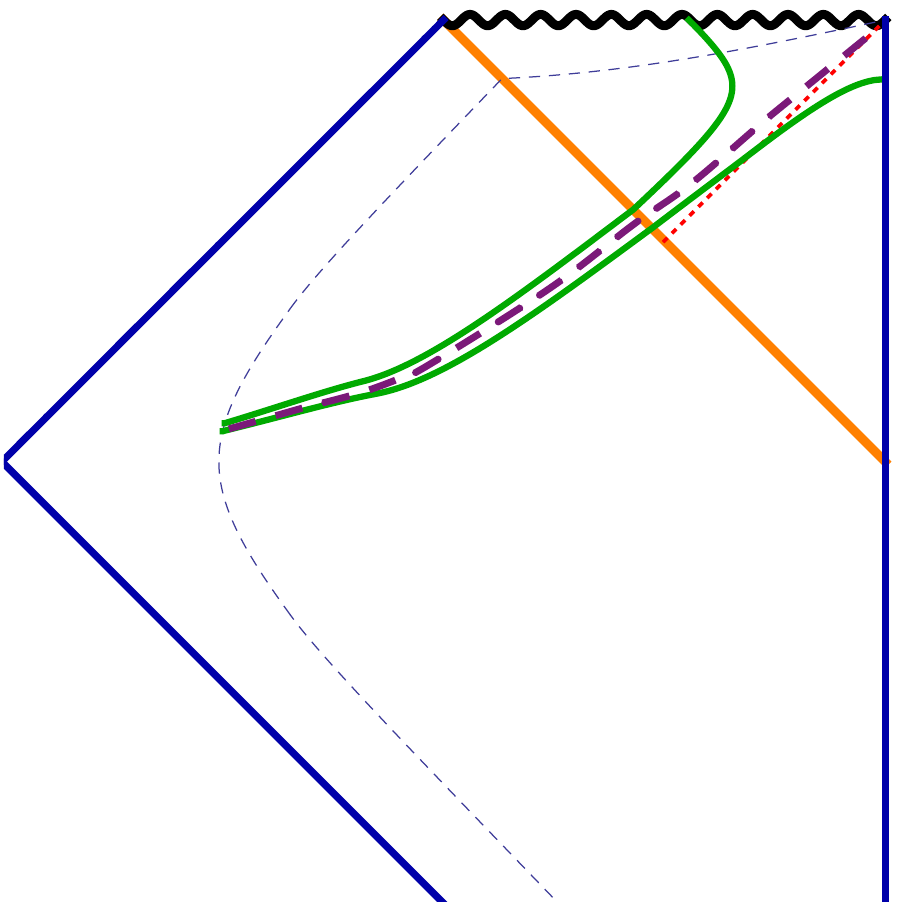} \;
\end{center}
\caption{
The dotted line denotes a curve at constant $z$, along which $v$ increases from $-\infty$ to 
$+\infty$ from bottom (not shown) to top. The dashed purple line corresponds to $\Ga_{\Sig}^*$, the critical extremal surface, while the green lines 
correspond to $\Ga_{\Sig}$ with $v_t$ just above and below $v_t^*$. 
}
\label{fig:criticalP}
\end{figure}

In this section, we showed explicitly for cases of $\Sig$ being a strip or sphere that in the Penrose diagram there exists a critical line
$v_t^* (z_t)$:\footnote{Recall $v_t = z_c - z_t$. Thus statements regarding $z_c^*$ can immediately be translated to those about $v_t^*$.}  $\Ga_{\Sig}$ reaches the boundary only for $v_t < v_t^*$, with the {\it critical extremal surface} $\Ga_{\Sig}^*$ corresponding to $v_t^* (z_t)$ stretching to $R, \ft = \infty$. 
See Fig.~\ref{fig:criticalP}. 
The same phenomenon should apply to general shapes.  

In the numerical plots presented in Sec.~\ref{sec:dcal} (see Fig.~\ref{fig:numevo}), we saw that for $\ft \gtrsim O(z_h)$, constant $R$ trajectories of $\Ga_{\Sig}$ in the $(z_t, z_c)$ plane collapse onto a single curve. From the above discussion, we now understand that this is a consequence of: (i) a critical  $z_c^* (z_t)$ exists on which $\Ga_{\Sig}$ asymptotes to a critical extremal surface that extends to infinite $R$ and $\ft$, and (ii) $z_c^*$ remains finite (of order $O(z_h)$) as $z_t \to \infty$.
Thus at large fixed $R$, when $\ft$ becomes sufficiently large, i.e. of order $O(z_h)$, $(z_t, z_c)$ quickly approaches the critical line $z_c^* (z_t)$. This is clearly exemplified in the $(1+1)$-dimensional story in Sec.~\ref{sec:2dscal}. There, $\ep$, parameterizing the distance to the critical line~\eqref{2dcrl} (or~\eqref{2dcrl1}), gave the leading large $\ell$ and $\tau$ behavior, while $\phi$ in~\eqref{2dcrl} (or $z_t$ in~\eqref{2dcrl1}), parametrizing the location on the critical line, mapped to $\ell - \tau$ or $\tau_s-\tau$.

In short, for large $R, \ft \gg z_h$, with corresponding $(z_t, v_t)$ lying very close to the 
critical line $v_t^* (z_t)$,  $\Ga_\Sig$ closely follows $\Ga_{\Sig}^*$ before deviating away to reach the boundary. The evolution of $\sA_\Sig$ can then be largely determined from that of $\Ga_{\Sig}^*$. 
Again, this is seen in the discussion in $(1+1)$-dimensions of Sec.~\ref{sec:2dsto}.
In higher dimensions, with much less analytic control, this feature provides a powerful tool for extracting the evolution of $\sA_\Sig (\ft)$. 

For $\Sig$ a strip, the critical extremal surfaces asymptotes to a constant-$z$ hypersurface $z=z_m$ lying inside the horizon, i.e. $z_m > z_h$ with $z_m$ depending on the function $h(z)$ in~\eqref{BHre1}. 
It is important to keep in mind that $z_t$ changes during the time evolution, and so does $z_m$. 

For $\Sig$ a sphere, the critical extremal surface for large enough $z_t$ forms an intermediate plateau at some $z=z_m$ before running along the horizon $z=z_h$ all the way to $\rho, v = \infty$, see Fig.~\ref{fig:critical1}. For moderate $z_t > z_h$, the critical extremal surface runs along the horizon $z=z_h$ to $\rho, v = \infty$ with no plateau at $z=z_m$, see Fig.~\ref{fig:critical}. 

We will see below that for a sphere, the plateau at $z= z_m$ governs a linear growth in $\sA$ at early times, while the plateau at the horizon governs a memory loss effect at late times.

\section{Linear growth: strip}

In this section, we show that with $\Sig$ given by a strip $\sA (R, \ft)$ grows linearly with $\ft$ 
for $R \gg \ft \gg z_h$.  
The evolution can be straightforwardly worked out from the discussion of Sec.~\ref{sec:critA} and as we will see is largely controlled by the critical extremal surface discussed in the last section. 
The same growth also applies to a sphere and other shapes as will be discussed in the next section.


\subsection{Linear growth} \label{sec:linstr}


To obtain the behavior for $R \gg \ft \gg z_h$,
we consider $z_c$ close to $z_c^*$ for some $z_t$, 
\be \label{eoo}
z_c = z_c^* (1- \ep )\ , \qquad  \ep  \ll 1  \ 
\ee
%
and assume that 
\be \label{jejen}
{z_c^* \ov z_t}, {z_m \ov z_t} \ll 1\ , \qquad {z_c^* \ov |\log \ep|}  \ll 1 \ .
\ee
In this regime we can expand  $\ft$, $R$, and $\sA$ in a double expansion of $1/z_t$ and $\ep$.

We now proceed to evaluate the boundary quantities $\ft$, $R$, and $\sA$ using~\eqref{raidu1}--\eqref{aver3}. 
Note that these equations 
should be modified when $z(x)$ is not monotonic, which happens, for example, 
for $z_t > z_t^{(s)}$. Then from~\eqref{oen2}, $z_c \approx z_c^* < z_m$, i.e. after intersecting the shell, $z (x)$ first moves to larger values of $z$ before turning around as illustrated in Fig.~\ref{fig:stripevo2}(b) and Fig.~\ref{fig:stripevo21}. In this case equation~\eqref{raidu1} should be modified to
\be \label{eeo}
R=\int_{z_c}^{z_t}dz\, {1 \ov \sqrt{{z_t^{2n} \ov z^{2n}}-1}}+\le(\int_{z_c}^{z_r}dz+\int_{0}^{z_r}dz\ri){1 \ov \sqrt{H(z)}}
\ee
and similarly for others. In the above equation $z_r$ is the root of $H(z)$ which is slightly smaller than $z_m$ (i.e. point $B$ of the second plot of Fig.~\ref{fig:stripevo2}),
and $z_r =z_m$ for $\ep =0$.

It is useful to separate  $z(x)$ into four regions (see Fig.~\ref{fig:stripevo21}): (i) AdS region from $z_t$ to $z_c$, (ii) from $z_c$ to near $z_m$, (iii) running along $z_m$, and (iv) from near $z_m$ to boundary $z=0$. 
One can then check that contributions to $\ft$, $R$, and $\sA-\sA_{\rm vac}$ from regions (ii) and (iv) are at most  $O(z_c^*)$.
\footnote{When $z_m \to \infty$ as $z_t \to \infty$, as for example in the case~\eqref{mner2}, one has to be careful because the integration range from $z_c^*$ to $z_m$ is large. One can check that divergent contributions from (ii) and (iv) cancel.
} 

Now let is look at region (iii).  Near $z =z_m$, with $z_c = z_c^* (1- \ep)$, we have 
\be 
H (z) =  H_2 (z - z_m)^2 + b \ep\ ,
\ee 
where
\be
H_2 =\ha H''(z_m)\ , \qquad  b = - z_c^* {d E^2 \ov d z_c} \biggr|_{z_c^*} \ .
\ee
Note $H_2 > 0$ and that $b < 0$ ($b >0$) for $z_t > z_t^{(s)}$ ($z_t < z_t^{(s)}$). In~\eqref{biute} (or its non-monotonic version), there is no contribution from region (i), while region (iii) contributes at order $\log \ep$, leading to 
\be \label{rooet}
\ft = - {E (z_c^*) \ov h(z_m) \sqrt{H_2}} \log \ep + \cdots\ .
\ee
In~\eqref{raidu1} (or~\eqref{eeo}) there is an $O(z_t)$ contribution from (i) in addition to a $\log\ep$ term from (iii), 
\be \label{rexpep}
R = a_n 
 z_t  - {1 \ov \sqrt{H_2}} \log \ep + \cdots \ ,
\ee
where $a_n$ was introduced~\eqref{stripexm} (c.f.~\eqref{shep}). 
Using~\eqref{rooet}, we can then rewrite~\eqref{rexpep} as 
\be \label{ueeo}
z_t = {1 \ov a_n} \le(R - {h(z_m) \ov E(z_c^*)} \ft \ri) + \cdots \ . 
\ee

Now consider the evaluation of $\sA$ using~\eqref{aver1}--\eqref{aver3}.
After subtracting the vacuum value $\sA_{\rm vac}$, the diverging contribution near $z=0$ in region (iv)
cancels and the dominant contribution is again from region (iii),  
\be \label{appex} 
{1 \ov \tilde K} \De \sA = {1 \ov  \tilde K} (\sA - \sA_{\rm vac}) = - { z_t^n  \ov z_m^{2n} \sqrt{H_2}} \log \ep + O(1)\ .
\ee

{Collecting \eqref{rooet} and \eqref{appex}, we find
\be \label{finsr1}
\De \sA = \tilde K \lam \ft + \cdots
\ee
with 
\be 
\lam=  { z_t^n  \ov z_m^{2n}} {h (z_m) \ov E (z_c^*)}  =  {\sqrt{- h(z_m)} \ov z_m^n} + \cdots
\ee
where in the second equality we have used~\eqref{eio2} to express $E (z_c^*)$ as 
\be \label{ebbn}
E (z_c^*) = - \sqrt{- h (z_m) \le({z_t^{2n} \ov z_m^{2n}} -1 \ri)} = - \sqrt{- h (z_m) }
{z_t^{n} \ov z_m^{n}}  + \cdots \ .
\ee
Upon substituting the explicit form of $\tilde K$ \eqref{coesr}, we have
\be \label{finsr11}
\De \sA =  \sqrt{-\ga (z_m)} A_{\rm strip} \ft + \cdots
\ee
where $\ga (z_m)$ is the determinant of the induced metric on the critical extremal surface at $z_m$, which is spanned 
by $v$ and $x_{2}, \cdots , x_{n}$, i.e. directions along $\Sig$}. Using the equilibrium ``density'' $\fa_{\rm eq}$ introduced in~\eqref{eqarea}, we can also write \eqref{finsr1} as 
\be \label{striplin}
\De \sA = \fa_{\rm eq} A_{\rm strip}  v_{n}  \ft + O(1)
\ee
where the velocity $v_{n} $ is given by
\be \label{allnsv}
v_{n} =  \le({z_h \ov z_m}\ri)^{n} \sqrt{- h(z_m)} \ .
\ee

In particular, for $n=d-1$, we have the entanglement entropy
\be \label{linsrip}
\De S ={\De \sA \ov 4 G_N} = s_{\rm eq} A_{\rm strip} v_E  \ft + O(1)
\ee
where $s_{\rm eq}$ is the equilibrium entropy density in~\eqref{eqent}, and 
\be \label{stropv}
v_E  \equiv v_{d-1} = \le({z_h \ov z_m}\ri)^{d-1} \sqrt{- h(z_m)} \ .
\ee

In the regime of~\eqref{jejen} we can approximate the value of $z_m$ in various equations above 
by that at $z_t = \infty$. So to leading order in large $R$ limit, the evolution is linear. Note in order for~\eqref{jejen} to be satisfied we need $\ft$ to be large enough so that $z_c$ is sufficiently close to $z_c^*$, but not too large such that $z_t$ becomes comparable to $z_c^*$ (see~\eqref{ueeo}) to invalidate~\eqref{jejen}.

\subsection{Example: Schwarzschild} \label{sec:nee}

Let us now consider the Schwarzschild case for explicit illustration. 
From~\eqref{mner}--\eqref{mner2}, depending on the value of $\eta = {2n \ov d} $, $z_c^*$ and 
$z_m$ behave differently in the limit of a large $z_t$. Below we consider these situations separately. 
While we are considering Schwarzschild, the discussion only depends whether 
$z_c^*$ and $z_m$ have a finite limit as $z_t \to \infty$. So we will still keep $h(z)$ general in our discussion.

\subsubsection{$\eta > 1$} \label{sec:etae}

For $\eta > 1$, which covers the case of entanglement entropy $n = d-1$ in $d>2$, 
both $z_c^*$ and $z_m$ remain finite of order $O(z_h)$ in the limit of large $z_t$. 
The assumptions~\eqref{jejen} then apply when $R\gg \ft \gg O(z_h)$. 

In this case we can  show that the linear growth~\eqref{striplin} in fact persists all the way to saturation, which happens via a discontinuous transition. We do this by assuming the conclusion, strongly suggested by Fig.~\ref{fig:numevo}, and checking self-consistency. 

With the linear growth~\eqref{striplin}, $\sA$ will reach its equilibrium value~\eqref{eqarea} at time
\be \label{ansq1}
\ft_s={R \ov v_{n}}={R \ov \le({z_h  \ov z_m}\ri)^n \sqrt{-h(z_m)}}\ ,
\ee
when, from \eqref{ueeo} and~\eqref{ebbn}, 
\be \label{ansq2}
z_t = {R \ov a_n} \le(1 - \le({z_m^2 \ov z_t z_h }\ri)^n \ri)  + \cdots \ .
\ee
From~\eqref{mner}, for $\eta>1$ the second term in parentheses is small for large $z_t$, so 
we find that when the system reaches the equilibrium value, $z_t$ is still very large. 

When $\ft$ is greater than~\eqref{ansq1}, equation~\eqref{striplin} exceeds its equilibrium value, and the extremal surface with smallest area is no longer a near-critical extremal surface to which~\eqref{striplin} applies, but one that lies solely in the black hole region. Thus the extremal surface jumps at $\ft_s$, and the saturation is discontinuous. Note that for entanglement entropy, the saturation time is 
\be \label{eske}
\ft_s = {R \ov v_{E}}
\ee
where $v_{E}$ was given in~\eqref{stropv}.

\subsubsection{$\eta =1$} \label{sec:eta1}

For $\eta =1$, which covers the case of entanglement entropy in $d=2$ examined earlier in Sec.~\ref{sec:2dsto} and that of a spacelike Wilson loop in $d=4$, $z_c^*$ remains finite but $z_m$ increases with $z_t$ in the large $z_t$ limit. In this case, there is still a linear regime, with 
\be 
v_n = 1 \ .
\ee
Furthermore, due to~\eqref{mner2}, the expression inside parentheses in~\eqref{ansq2} becomes zero 
at the time~\eqref{ansq1}, i.e. $z_t$ becomes comparable to $z_c$ 
before~\eqref{ansq1} is reached. Thus the system exits the linear growth regime before saturation.
This is consistent with what we saw in  Sec.~\ref{sec:2dsto} for the $d=2$ case. 
In Sec.~\ref{sec:satur} and Sec.~\ref{gscale} we discuss the behavior of the system
after exiting the linear regime in higher dimensions.

\subsubsection{$\eta < 1$}

For $\eta < 1$, from~\eqref{mner1} both $z_c^* \sim z_t^{\al} $ (with $\al < 1$) and $z_m \propto z_t$ grow with $z_t$ in the limit $z_t \to \infty$. 
Then since $z_c^*$ is also very large for large $z_t$, it may take a long time for $z_c$ to reach $z_c^*$.  If $z_t $ is still $O(R)$ as $z_c$ first approaches $z_c^*$, the linear regime could still exist. Supposing such a regime exists, equation~\eqref{allnsv} gives for Schwarzschild $h(z)$ 
\be 
v_n^{\rm (S)} = \le({z_m \ov z_h}\ri)^{{d \ov 2} - n} \to \quad \infty\ ,
\ee
which is physically unreasonable and suggests that a linear regime does not exist. 
Explicit numerical calculation appears to be consistent with this expectation~\cite{eliz}.

\section{Linear growth: general shapes} \label{sec:polin}

In this section we generalize the linear growth found in the last section for a strip to general shapes. 
We show that for $\ft$ in the range $R \gg  \ft \gg z_h$, $\sA_\Sig (t)$ generically exhibits 
linear growth in $\ft$ with a slope independent of the shape of $\Sig$. 
Again the technical requirement is that $z_c^*$ should remain finite as $z_t \to \infty$, which for Schwarzschild $g(z)$ amounts to $2 n \geq d$.   


We first revisit the strip story and rederive the linear growth from a scaling limit, which we can extend straightforwardly to general shapes.  We will also extend results to 
the wider class of metrics~\eqref{vaidya1}. 

\subsection{Revisiting strip: a scaling limit} \label{sec:sjer}

The linear growth of the last section
occurs when $z_t$ is large but $z_c^*$ remains finite in the limit $z_t \to \infty$. 
In this regime, with $z_c \approx z_c^*$ we have~(from~\eqref{exe0})
\be \label{oo1}
 x_c = x (z_c) = a_n z_t - {z^{n+1}_c \ov n z_t^n} + \cdots \ .
\ee
Also from~\eqref{ueeo} and~\eqref{ebbn}
\be \label{oo2}
a_n z_t = R - O(z_t^{-n})    \ . 
\ee
The above equations suggest that in the black hole region we should consider a scaling coordinate 
\be  \label{bbe}
y = (R - x)  {z_t^n } \ . 
\ee
Indeed, in terms of $y$ equation~\eqref{yuem} (or~\eqref{eio2}) has a scaling form independent of $z_t$ to leading order as $z_t \to \infty$,
\be \label{91}
\le({dz \ov dy}\ri)^2 = { h (z) \ov z^{2n}} +a^2  , \quad a^2 =  {g_c^2 \ov 4 z_c^{2n}} \ .
\ee 
Similarly, to leading order in $1/z_t$, equation~\eqref{vekp1} becomes  
\be  \label{vfje}
{d v \ov dz} =  {1 \ov h} \le({a \ov \sqrt{{h (z) \ov z^{2n}} + a^2 } }- 1\ri) \ .
\ee

\begin{figure}[!h]
\begin{center}
\includegraphics[scale=0.4]{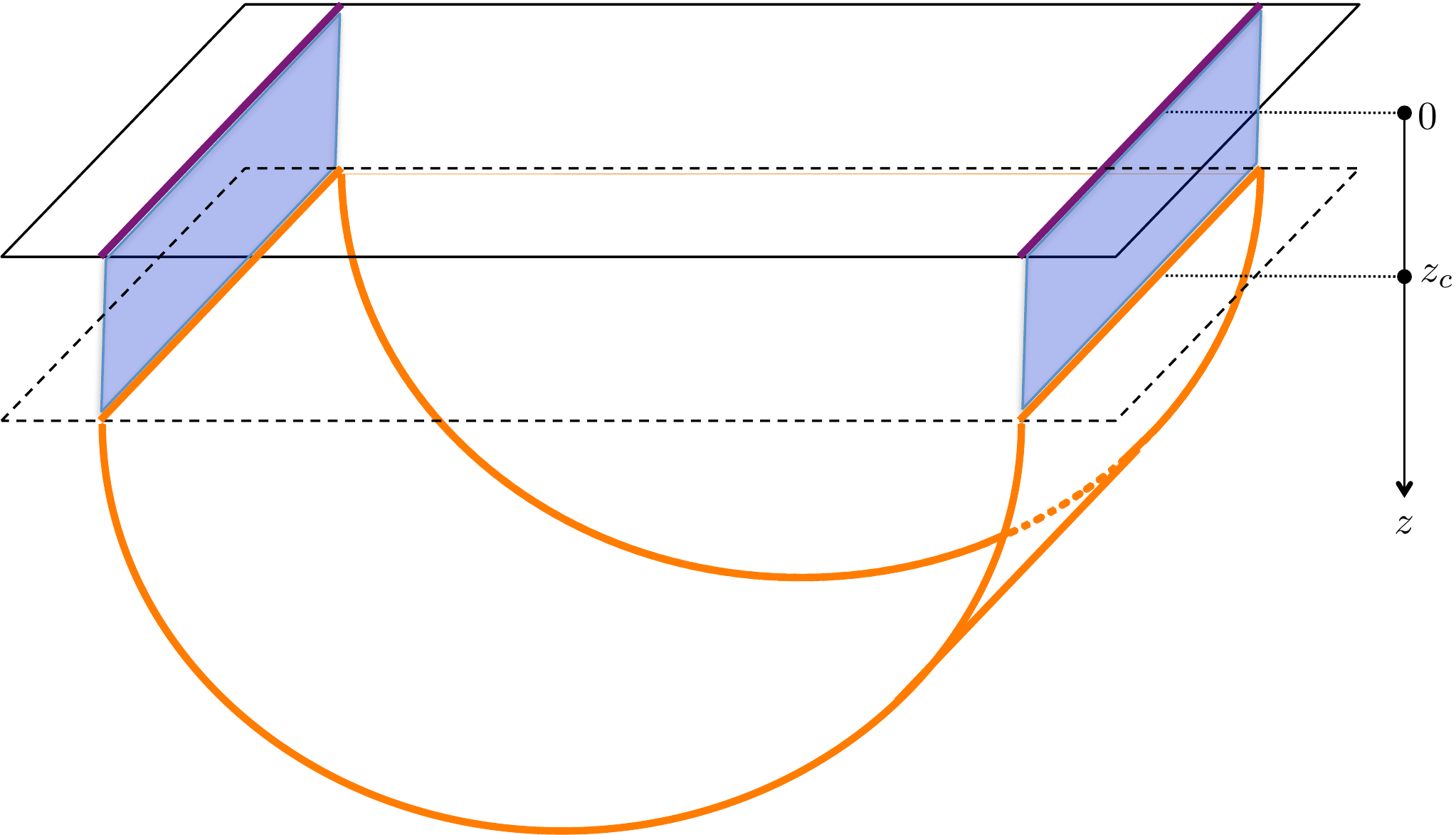} \;
\end{center}
\caption{
In the limit of a large $z_t$ and a finite $z_c \approx z_c^*$, the evolution in the black hole region is essentially 
solely in the time direction, with two sides of the strip evolving independently.
}
\label{fig:linear0}
\end{figure}

From~\eqref{91} and~\eqref{vfje}, we conclude 
\be 
{dx \ov dz} \sim {1 \ov z_t^n} \ , \qquad {dv \ov dz} \sim O(1) \ .
\ee
Then using $z$ as the independent variable, the action~\eqref{striparea} in the black hole region is\bea 
\sA_{\rm BH} &=& L^n A_{\rm strip} \int_0^{z_c} dz \, {1 \ov z^n} \sqrt{\le({dx \ov dz}\ri)^2 - 2 {dv \ov dz} - h \le({dv \ov dz}\ri)^2} 
\cr
\label{jjer}
&=& L^n A_{\rm strip}  \int_0^{z_c} dz \, {1 \ov z^n} \sqrt{- 2 {dv \ov dz} - h \le({dv \ov dz}\ri)^2}  
\eea
where in the second equality we have dropped the term $\le({dx \ov dz}\ri)^2 \sim O(z_t^{-2n})$.  
It may look odd that in~\eqref{jjer} $x(z)$ completely drops out. This in fact has a simple geometric interpretation: from~\eqref{oo1}--\eqref{oo2}, by the time the extremal surface reaches $z_c$, $x (z_c) = 
R - O(z_t^{-n})$ has essentially reached its boundary value $R$, while $v (z_c)$ is zero and still far away from its boundary value $v(z=0) = \ft$. Thus  the evolution of the extremal surface in the black hole region (for $z < z_c$) is almost completely in the time direction. See Fig.~\ref{fig:linear0} for an illustration. 
For purposes of calculating the area $\sA$ to leading order in $1/z_t$, we can simply ignore the evolution in $x$-direction. 
As a consistency check, we indeed recover~\eqref{vfje} by variation of~\eqref{jjer}. 

Integrating~\eqref{vfje} we find that 
\be \label{ii1}
\ft = \int_0^{z_c}\, {dz \ov h} \le({a \ov \sqrt{{h (z) \ov z^{2n}}  + a^2 } }- 1\ri) 
\ee
and further substituting~\eqref{vfje} into~\eqref{jjer} we have 
\be\label{ii2}
\sA_{\rm BH} = L^n A_{\rm strip} \int_0^{z_c} dz \, {1 \ov z^{2n} \sqrt{{h (z) \ov z^{2n}}  + a^2}}  
\ee
The linear growth of $\sA (\ft)$ can now be immediately understood from~\eqref{ii1} and~\eqref{ii2}. 
As before, for $z_c = z_c^*$ with $z_c^*$ given by~\eqref{ohh2}, ${h (z) \ov z^{2n}}  + a^2$ has a double zero 
at  its minimum $z_m$ which precisely coincides with~\eqref{ohh1}\footnote{${h (z) \ov z^{2n}}+ a^2$ differs from $H(z)$ of~\eqref{yuem} only by an overall scaling and thus has the same minimum and zero.}. 
For $z_c = z_c^* (1-\ep)$ with $\ep \to 0$, both the integrals for $\ft$ and $\sA_{\rm BH}$ are then dominated by region around $z_m$, and we precisely recover~\eqref{striplin}. 

Note that the action~\eqref{jjer} as well as the linear growth 
of $\sA$ is in fact identical to that of~\cite{Hartman:2013qma}, where 
entanglement entropy between half spaces lying on two asymptotic boundaries
of an eternal AdS black hole was considered. The agreement can be easily understood from Fig.~\ref{fig:linear0}; in the large $z_t$ limit, each half of the strip evolves independently in the black hole region solely in the time direction, which coincides with the set-up of~\cite{Hartman:2013qma}. 






\subsection{General shapes} \label{sec:linsph}

\begin{figure}[!h]
\begin{center}
\includegraphics[scale=0.4]{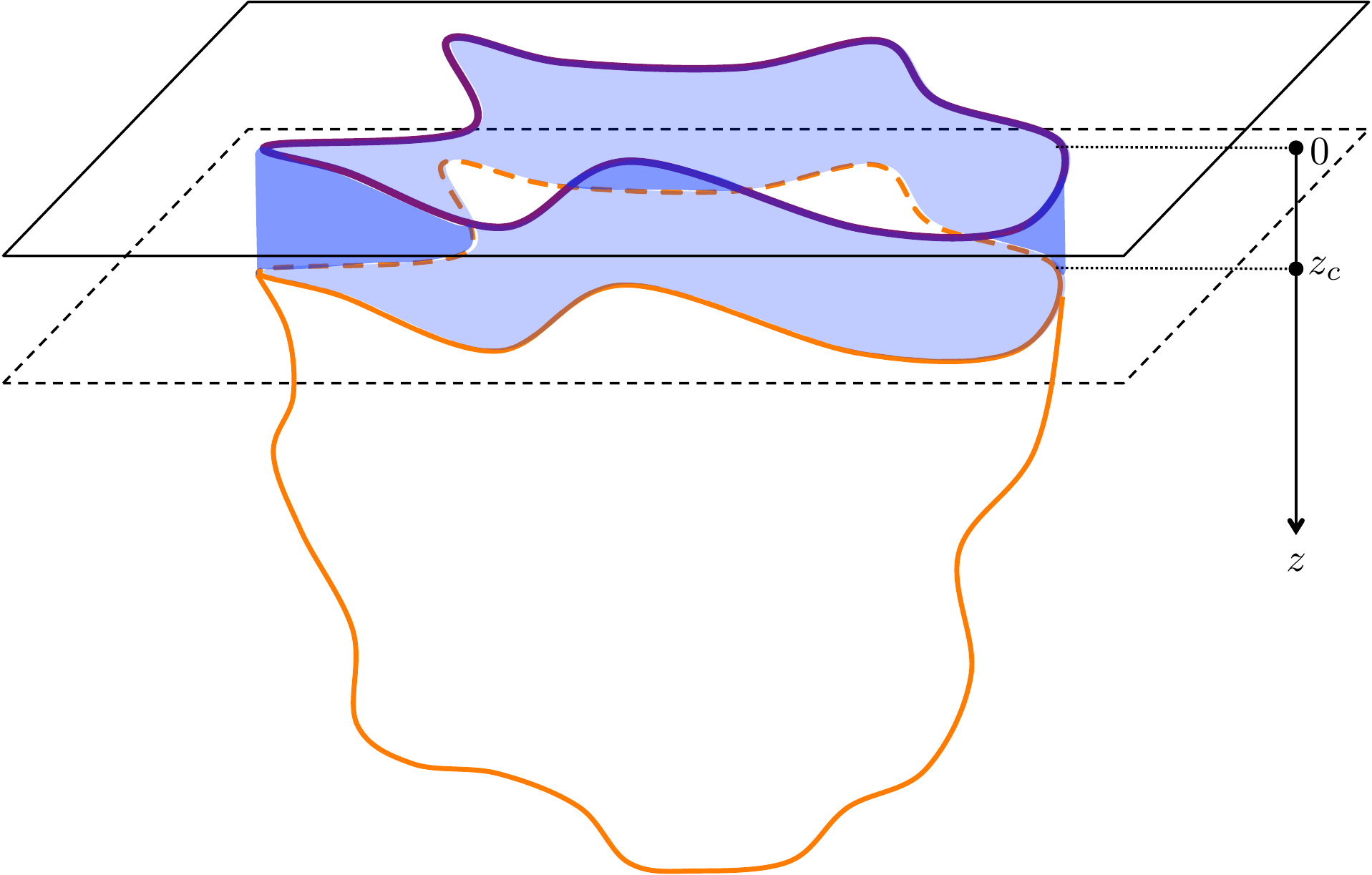} \;
\end{center}
\caption{
A cartoon of an extremal surface for $\Sig$ with some arbitrary shape, in the large size limit and $\ft$ in the linear regime. Upon entering the black hole region, the extremal surface has essentially attained its boundary shape $\Sig$. The evolution in the black hole 
region is essentially solely in the time direction and is the same as that for a strip. }
\label{fig:linear1}
\end{figure}

The intuition obtained from the above discussion for a strip and Fig.~\ref{fig:linear0} can now be generalized 
to arbitrary shapes. For arbitrary $\Sig$, we again expect that in the limit $R \gg \ft \gg z_h$, the evolution of the extremal surface after entering the shell will be essentially solely in the time direction, as indicated in Fig.~\ref{fig:linear1}. In other words, in the large size limit, when $z_c$ is much smaller than the size of $\Sig$, the curvature of $\Sig$ should not matter in the black hole and each point of the extremal surface essentially evolves like one on a strip. Below we present arguments that this is indeed the case.

Consider a smooth entangling surface $\Sig$ which can be parameterized in terms of
polar coordinates~\eqref{polat} as 
\be \label{sue1}
\rho = R r (\Om), \qquad x_a =0  \ 
\ee
where $\Om$ denotes collectively  the angular coordinates parameterizing $\Sig$, $R$ is the size of $\Sig$, and the function $r (\Om)$ specifies the shape of $\Sig$ . 
The bulk extremal surface can then 
be parameterized in terms of $\rho (z, \Om), v (z, \Om)$ with boundary conditions 
\be 
\rho(z=0,\Omega)=R \, r (\Omega), \qquad v (z=0, \Om) = \ft
\ee
and regularity at the tip of the surface. 

Writing (see~\eqref{polat})
\be 
d \Om^2_{n-1} = \sum_i g_i (\Om) d \th_i^2 \ , \quad d^{n-1}\Om = \prod_{i} \sqrt{g_i} d \th_i
\ee
the area of $\Sig$ can be written as 
\be 
A_\Sig = R^{n-1} \int d^{n-1} \Om \, r^{n-1} (\Om) \sqrt{1 + {1 \ov r^2} \sum_i {r_i^2 \ov g_i}}
\ee
where 
\be 
r_i \equiv  \p_{\th_i} r (\Om)  \ .
\ee
Meanwhile, in the Vaidya geometry, the action for an $n$-dimensional extremal surface ending on the above $\Sig$ can be written as 
\be \label{hnnw}
\sA_\Sig = L^n \int_\de^{z_t} d z \int d^{n-1} \Omega \ {\rho^{n-1} \ov z^{n}} \sqrt{Q}
\ee
with
\be \label{hnnw1}
Q = \rho'^2 - 2 v' - f (v,z) v'^2 + {1 \ov \rho^2} \sum_{i} {1 \ov g_i } G_i - {1 \ov \rho^4} 
\sum_{i,j} {(\rho_i v_j - \rho_j v_i)^2 \ov g_i g_j } 
\ee
where we have used the notation
\be
\rho' \equiv \p_z \rho, \quad \rho_i \equiv \p_i \rho , \quad v' \equiv \p_z v, \quad v_i \equiv \p_i v  
\ee
and 
\be 
G_i = - f (v,z) (\rho' v_i - \rho_i v')^2  + 2 \rho_i (\rho'  v_i - \rho_i v') - v_i^2 \ .
\ee
In~\eqref{hnnw} $\delta$ is a short-distance cutoff. It is readily found that in the black hole region $\rho$ and $v$ have the following small $z$ expansion (for $z \ll z_h$)
\bea \label{vdne}
\rho (z, \Om) &= & R r (\Om) - {z^2 \ov R} \tilde r (\Om) + \cdots  \\
v (z, \Om) & = & \ft - z + O(z^{n+1})  \ 
\label{vdne1}
\eea
where $\tilde r (\Om)$ is a function which can be determined from $r (\Om)$.



For $R \gg \ft$, to leading order in $1/R$, the part of the extremal surface in the AdS region can be approximated by that in pure AdS, which we denote $\rho^{(0)} (z, \Om)$ (and for which $t$ constant). 
For $z/R \ll 1$, $\rho^{(0)}$ has the the expansion
\be  \label{mhhe}
 \rho^{(0)} (  z,\Omega) = R r(\Omega)  + O(R^{-1})
\ee
Note that in contrast to~\eqref{vdne} which applies only to $z \ll z_h$, due to the scaling symmetry of pure AdS and that $\Sig$ as defined in~\eqref{sue1} has a scalable form,
equation~\eqref{mhhe} in fact applies to any $z/R \ll 1$ and in particular $z \sim z_c \approx z_c^*$. 
Thus we conclude that when the extremal surface enters the shell at $z_c$,  
\be \label{epee}
\rho (z_c, \Om) = Rr (\Om) - O(R^{-1})   \ .
\ee

From~\eqref{vdne}--\eqref{vdne1} and~\eqref{epee}, the extremal surface in the black hole region should then have the following scaling 
\be 
\rho' \sim O(R^{-1}), \quad \rho_i \sim O(R), \quad v_i \sim O(R^{-1}), \quad 
v' \sim O(1)  \ .
\ee 
Plugging in the above scaling into the action~\eqref{hnnw} we find that to leading order in $1/R$, 
\bea 
\sA_{\Sig, \rm BH} &=& L^n R^{n-1} \int_\de^{z_c} d z \int d^{n-1} \Omega \ r^{n-1} (\Om) \nn
&&{1 \ov z^{n}} 
\sqrt{- 2 v' - h v'^2} \sqrt{1 + {1 \ov r^2} \sum_i {r_i^2 \ov g_i}} \cr
& = & L^n A_\Sig \int_\de^{z_c}  {dz \ov z^{n}} 
\sqrt{- 2 v' - h v'^2} \label{enrl}
\eea
which reduces to~\eqref{jjer}. In particular, all evolution in $\rho$ and $\Om$ directions
have dropped out. Thus we conclude that~\eqref{striplin} in fact applies to all shapes 
with $A_{\rm strip}$ replaced by $A_{\Sig}$. 

\begin{figure}[!h]
\begin{center}
\includegraphics[scale=0.8]{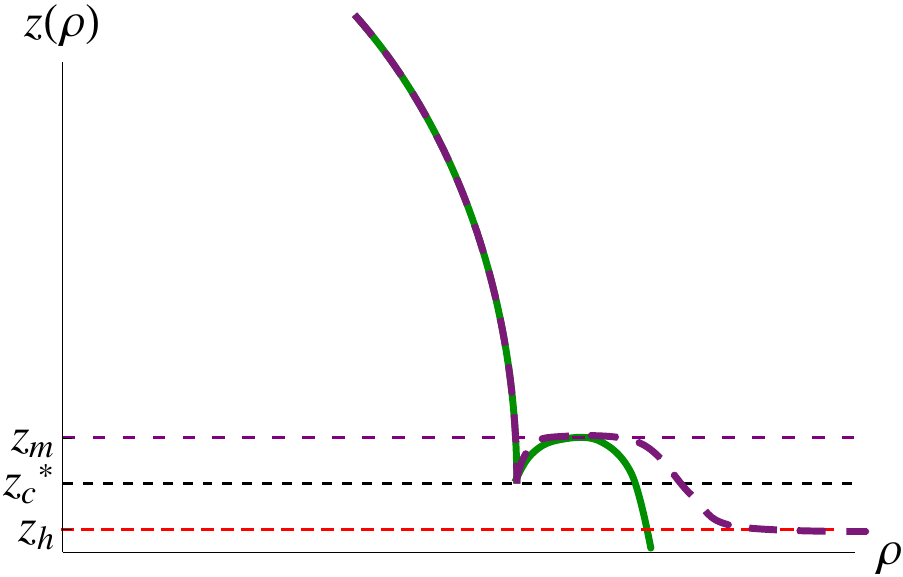}
\end{center}
\caption{Cartoon: For a sphere, in the linear regime the extremal surface follows the critical extremal surface for a while but exits near the first plateau. The dashed curve is the critical extremal surface.}
\label{fig:plat}
\end{figure}

The above discussion encompasses the case of $\Sig$ being a sphere for which $r (\Om) =1$. 
In that case one can derive the above scaling limit explicitly from equations~\eqref{1}--\eqref{2}. In particular, the linear growth regime is controlled by the first plateau of the critical extremal surface 
as indicated in Fig.~\ref{fig:plat}.  

\subsection{More general metrics}

The above discussion can be readily extended to more general metrics of the form~\eqref{vaidya1}--\eqref{bhnem}. The action~\eqref{enrl} is replaced by
\be \label{sd1}
\sA_{ \rm BH}= L^n A_\Sig \int_0^{z_c} \, {1 \ov z^{n}} \sqrt{- h (z) v'^2 - 2 k (z) v'} \ ,
\ee
from which $v (z)$ satisfies the equation
\be 
{1 \ov z^n} {h v' + k \ov \sqrt{- h v'^2 - 2 k v'}} ={\rm const} \ 
\ee
which can be solved as ($b$ is a positive constant)
\be \label{mm1}
v' = {k(z) \ov h (z)} \le({b \ov \sqrt{{h(z) \ov z^{2n}} + b^2}} -1 \ri)  
\ee
with 
\be \label{mm2}
{d \sA \ov dz} = L^n A_\Sig {k (z) \ov z^{2n}} {1 \ov \sqrt{{h(z) \ov z^{2n}} + b^2}} \ .
\ee
Other than a prefactor $k(z)$ appearing in both equations, equations~\eqref{mm1}--\eqref{mm2} are 
identical to~\eqref{ii1}--\eqref{ii2}. The constant $b$ should be determined by matching conditions at the null shell, i.e. be expressible in terms of $z_c$ alone in the limit $z_t \to \infty$. Its precise form is not important. As far as a $z_c^*$ exists such that ${h(z) \ov z^{2n}} + b^2$ is zero at its minimum $z_m$, $\sA$ will have a linear growth regime for $z_c$ close to $z_c^*$.  

Since in the linear regime the leading behavior is given by 
the behavior of the RHS of~\eqref{mm1}--\eqref{mm2} near $z_m$, the factor $k(z_m)$ cancels 
when we relate $\sA$ to $\ft$ and we conclude $\sA$ is still given by~\eqref{striplin} with the same $v_n$, i.e. the additional function $k(z)$ in~\eqref{sd1} cannot be seen in the linear regime.


\section{Linear growth:  an upper bound?} \label{sec:ling}

In previous sections we found that for any metric of the form~\eqref{vaidya1} and for $\Sig$ of any shape, provided that $z_c^*$ remains finite in the limit $z_t \to \infty$, there 
is a linear growth regime
\be \label{striplin2}
\De \sA (\ft) = \fa_{\rm eq} A_{\rm \Sig}  v_{n}  \ft + O(1)
\ee
for $R \gg \ft \gg z_h$. In the above equation $\fa_{\rm eq}$ is the equilibrium density introduced 
in~\eqref{eqarea}, $A_\Sig$ is the area of $\Sig$, and 
the velocity $v_{n} $ is given by
\be \label{allnsv2}
v_{n} =  \le({z_h \ov z_m}\ri)^{n} \sqrt{- h(z_m)} \ .
\ee
Here $z_m$ is the minimum of ${h(z) \ov z^{2n}}$ and lies inside the black hole 
event horizon. In particular, for entanglement entropy we have $n = d-1$ and 
\be \label{line}
\De S_\Sig (\ft) =s_{\rm eq} A_{\rm \Sig}  v_{E}  \ft + O(1)\ , \quad v_E = v_{d-1} 
\ee
where $s_{\rm eq}$ is the equlibrium entropy density. 

Now let us specialize to the evolution of entanglement entropy which has the cleanest physical interpretation. 
The linear growth regime~\eqref{line}  sets in for $\ft \gtrsim z_h \sim O(\ell_{\rm eq})$, i.e.  after local equilibration has been achieved. This explains the appearance of the equilibrium entropy density $s_{\rm eq}$ in the prefactor.  In contrast, the pre-local-equilibration quadratic growth~\eqref{quda} is proportional to the energy density $\sE$. Indeed, at very early times before the system has equilibrated 
locally, the only macroscopic data characterizing the state is the energy density. 

It is natural that in both regimes $\De S_\Sig$ is proportional to $A_\Sig$, as the time evolution in our system is generated by a {\it local} Hamiltonian which couples directly only to the degrees of freedom near $\Sig$, and the entanglement has to build up from $\Sig$. When $R$ is large, the curvature of $\Sig$ is negligible at early times, which explains the area law and shape-independence of~\eqref{quda} and~\eqref{line}. 

Note that if we stipulate that before local equilibration $S_{\Sig} (\ft)$ should be proportional to $A_{\Sig}$ and $\sE$, the quadratic time dependence in$~\eqref{quda}$ follows from dimensional analysis. Similarly, if we require that after local equilibration, $S_{\Sig} (\ft)$ is proportional to $A_{\Sig}$ and $s_{\rm eq}$, linearity in time follows. 

\begin{figure}[!h]
\begin{center}
\includegraphics[scale=0.45]{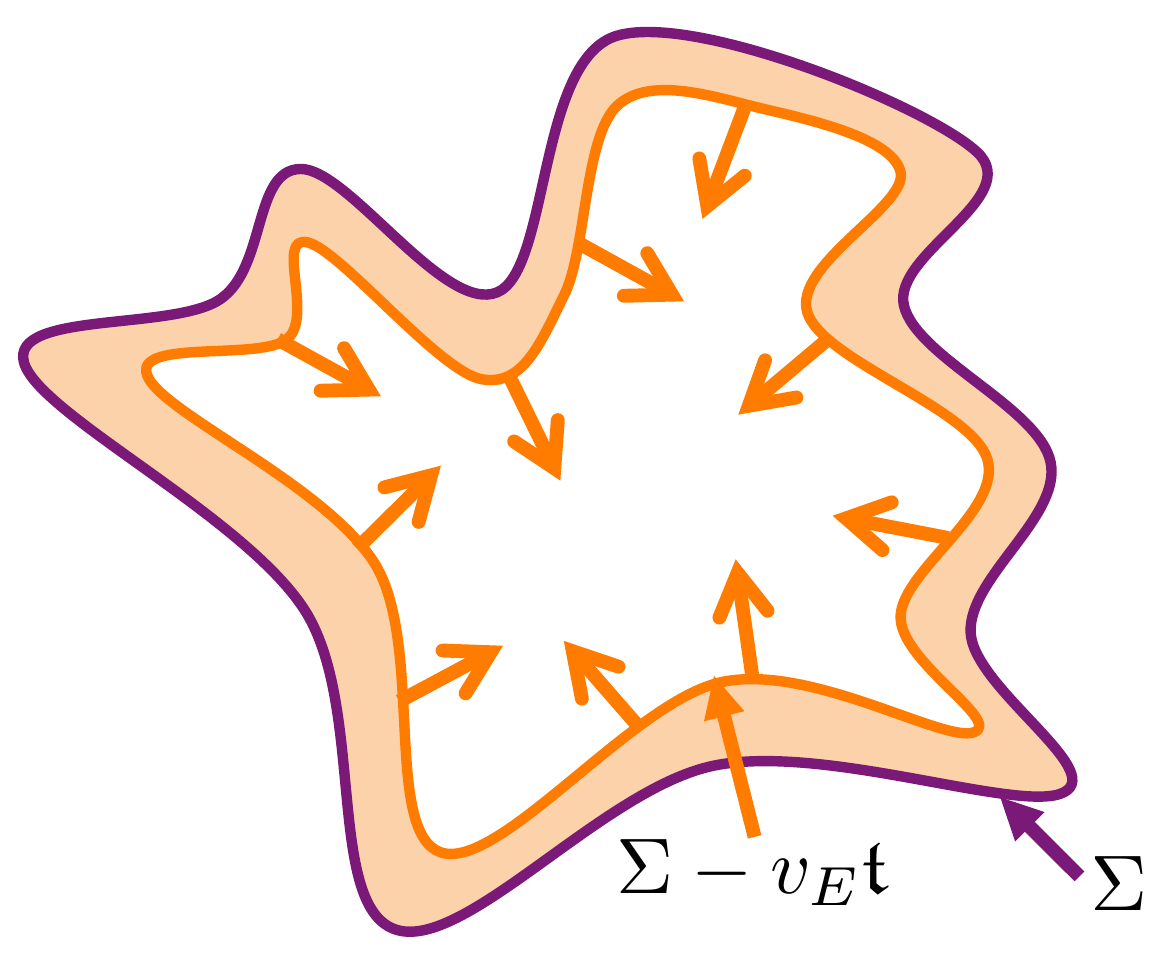} \quad
\end{center}
\caption{
The growth in entanglement entropy can be visualized as occuring via an ``entanglement tsunami" with a sharp wave-front carrying entanglement inward from $\Sig$.  The region that has been covered by the wave (i.e. yellow region in the plot) is entangled with the region outside $\Sig$, while the white region is not yet entangled. 
 }\label{fig:eevelo}
\end{figure}

As discussed in~\cite{Liu:2013iza}, equations~\eqref{quda} and~\eqref{line} suggest a 
simple geometric picture: entanglement entropy increases as if there was a wave with a sharp wave-front propagating inward from $\Sigma$, with the region that has been covered by the wave entangled with the region outside $\Sig$, and the region yet to be covered not yet entangled. See Fig.~\ref{fig:eevelo}. 
This was dubbed an ``entanglement tsuanmi" in~\cite{Liu:2013iza}. In the linear regime, the tsunami has a constant velocity given by $v_E$, while in the quadratic regime the front velocity increases linearly with time. 
The tsunami picture  highlights the {\it local} nature of the evolution of entanglement.
For quadratic and linear growth regimes, when the curvature of $\Sig$ can be neglected, different parts of the tsunami do not interact with one another. But as the tsunami advances inward, curvature effects will become important, and the propagation will become more complicated. 



 In a relativistic system, $\Vee$ should be constrained by causality, although in a general interacting quantum system relating it directly to the speed of light appears difficult. In the rest of this section we examine $\Vee$  for known black hole solutions and also various $h(z)$ satisfying null energy conditions. We find support that 
\be \label{emrp}
\Vee  \leq \Vee^{(\rm S)} = {(\eta -1)^{\ha (\eta -1)} \ov \eta^{\ha \eta}} 
= \bca 1 &  d=2 \cr
          {\sqrt{3} \ov 2^{4 \ov 3}} = 0.687 & d=3 \cr
          {\sqrt{2} \ov 3^{3 \ov 4}} = 0.620 & d=4 \cr
          \ha & d = \infty 
          \eca \ 
\ee
where $\Vee^{(\rm S)}$ is the value for a Schwarzschild black hole with $\eta = {2 (d-1) \ov d}$. 

There are reasons to suspect that the Schwarzschild value in~\eqref{emrp} may be special. The gravity limit corresponds to the infinite coupling limit of the gapless boundary Hamiltonian, in which generation of entanglement should be most efficient.  From the bulk perspective,
it is natural to expect that turning on additional matter fields (satisfying the null energy condition) will slow down thermalization. From the boundary perspective, the corresponding expectation is that when there are conserved quantities such as charge density, the equilibration process becomes less efficient. 


 
 
With M\'ark Mezei, we generalized the free-streaming model of~\cite{Calabrese:2005in} to higher dimensions and find that at early times there is linear growth as in~\eqref{line} with $s_{\rm eq}$ interpreted as giving a measure for quasiparticle density.  For $d\geq 3$,  quasiparticles can travel in different directions, and as a result although they travel at the speed of light the speed of the entanglement tsunami turns out to be smaller than $1$~\cite{speed},
\be \label{vstrem}
v^{\rm (streaming)}_E = { \Ga({d-1 \ov 2}) \ov \sqrt{\pi}  \Ga ({d \ov 2}) } <  \Vee^{(S)} < 1 \ .
\ee
Comparing with the Schwarzschild value~\eqref{emrp}, we conclude that in strongly coupled systems, the propagation of entanglement entropy is faster than that from free-streaming particles moving at the speed of light!

It is important to examine whether~\eqref{emrp} could be violated from higher derivative corrections
to Einstein gravity. As a preliminary investigation, at the end of this section we consider the example of a Schwarzschild black hole in Gauss-Bonnet gravity in $d=4$, but as we explain there one cannot draw an immediate conclusion from it.


\subsection{Schwarzschild, RN and other black holes}

Let us now consider some examples. For Schwarzschild $h(z)$~\eqref{schw}, plugging~\eqref{mner} into~\eqref{allnsv2} we find 
\be \label{vsrie}
v_{n}^{\rm (S)}  = {(\eta -1)^{\ha (\eta -1)} \ov \eta^{\ha \eta}}\ , \qquad \eta = {2n \ov d} \ .
\ee
Recall that our current discussion only applies to $\eta \geq 1$ 
and it can be readily checked from~\eqref{vsrie} that
\be
v_{n}^{\rm (S)} < 1\qquad {\rm for} \qquad \eta > 1\ , \qquad  v_{{d \ov 2}}^{\rm (S)} = 1\ .
\ee
$v_{n}^{\rm (S)}$ is a monotonically decreasing function of $\eta$. The maximal value of $\eta$ is for entanglement entropy, for which $\eta = {2 (d-1) \ov d}$ and
\be \label{schset}
v_{E}^{\rm (S)} = {d^\ha (d-2)^{ \ha-{1\ov d} } \ov (2(d-1))^{1-{1 \ov d}}} \ .
\ee
The above expression and~\eqref{stropv} were also obtained earlier in~\cite{Hartman:2013qma} in a different 
set up. 

For Reissner-Nordstrom $h(z)$, from~\eqref{rnex} the velocity for entanglement entropy is given by
\be \label{stvrn}
v_{E}^{\rm (RN)} =  \sqrt{d \ov d-2} \le(\le(1- {d\, u \ov 2 (d-1)}  \ri)^{2 (d-1) \ov d} - (1-u) \ri)^\ha 
\ee
where $u$ was defined in~\eqref{udef}--recall that $1 \geq u \geq 0$ with $u=1, 0$ being the Schwarzschild and extremal limits, respectively. We note $v_E$ decreases with increasing chemical potential.
For the extremal black hole, one finds $v_E =0$ which implies that the linear growth regime no longer exists. 

\begin{figure}[!h]
\begin{center}
\includegraphics[scale=0.5]{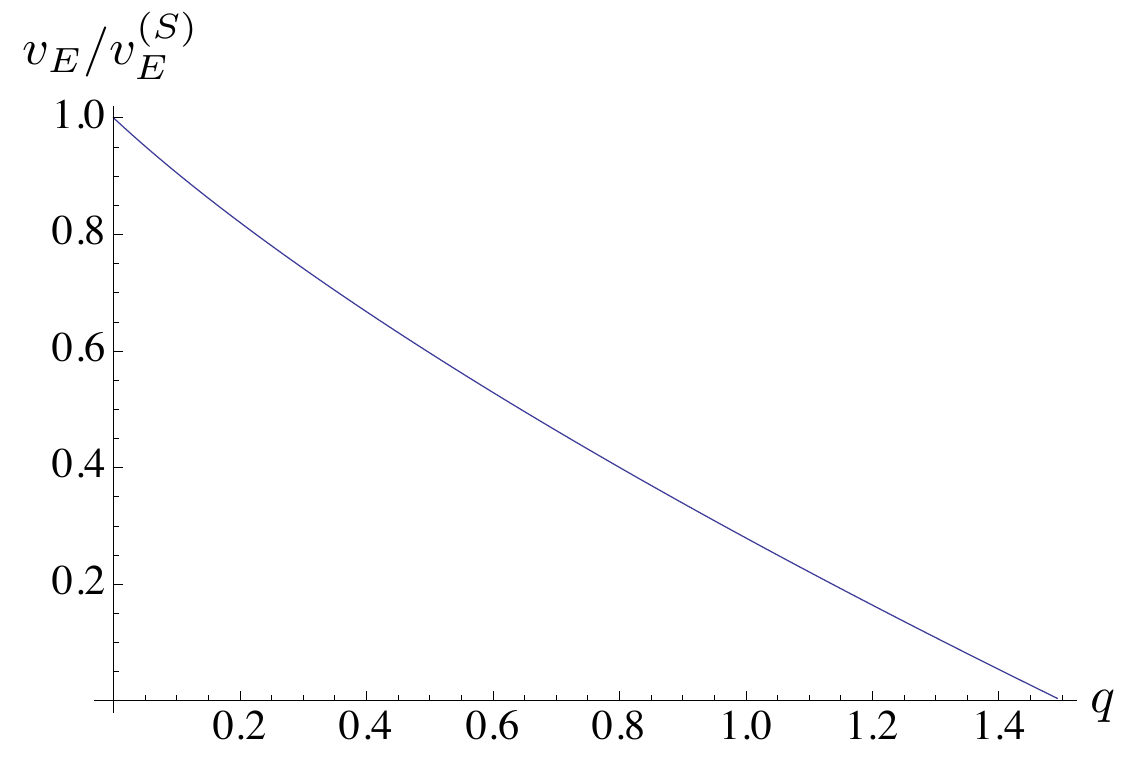} 
\includegraphics[scale=0.7]{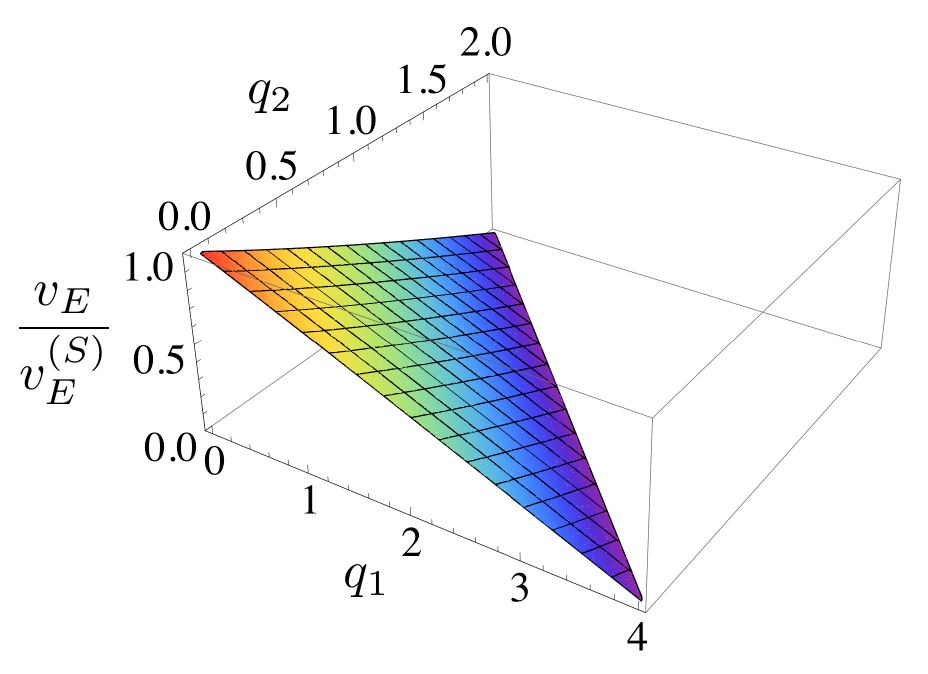} 
\end{center}
\caption{Plots of $v_E/v_E^{(S)}$ in examples of $h(z)$ with parameter space restricted by the NEC and the existence of a horizon. {\it Upper}: For \eqref{REg} with $d=3$ and $p=2$. {\it Lower}: For \eqref{tpe} with $d=4$.
 }\label{fig:ranex1}
\end{figure}

We now consider the behavior of $\Vee$ for more general black holes. Other than Schwarzschild and RN black holes there are no known examples of explicit supergravity solutions of the form~\eqref{BHre1}. 
Given that~\eqref{allnsv2} depends on some location $z=z_m$ behind the horizon, which could be shifted around by modifying $h(z)$, one may naively expect that $\Vee$ could easily be increased by changing $h(z)$ arbitrarily. However, in the examples we studied, the null energy condition 
\be \label{nec}
z^2 h'' - (d-1) z h' \geq 0 \
\ee
appears to constrain $v_E \leq v_E^{\rm (S)}$. Here are some examples: 
\bi
\item Consider 
\be \label{REg}
h(z) = 1 - M z^d + q z^{d+p} \ , \qquad p>0\ .
\ee 
The null energy condition~\eqref{nec} requires $q \geq 0$ and in order for the metric to have a horizon (and not a naked singularity), $q  \leq {d \ov p}$. (Here and below we set $z_h=1$). This constrains $v_{E} \leq v_E^{(S)}$, an example of which we show in Fig.~\ref{fig:ranex1}. Note that for $q < 0$, $\Vee$ does exceed $\Vee^{(S)}$. 

\item A three-parameter example with
\be \label{tpe}
h(z) = 1 - M z^d + q_1 z^{d+1} + q_2 z^{d+2} \ .
\ee
The null energy condition~\eqref{nec} requires both $q_1$ and $q_2$ to be non-negative, and the existence of a horizon requires $q_1 + 2 q_2 \leq d$. Then again $v_E \leq v_E^{(S)}$, an example of which is shown in Fig.~\ref{fig:ranex1}.
\ei
We have also looked at some non-polynomial examples and found $v_E \leq v_E^{\rm (S)}$. The phase space we have explored is not big, nor do we expect that the null energy condition is the only consistency condition. Nevertheless, the examples seem suggestive. 

\subsection{Other supergravity geometries}

{\it 1. Charged black holes in $\sN =2$ gauged supergravity in AdS$_5$~\cite{Behrndt:1998jd}:}
\be \label{c5}
ds^2={L^2 H^{1 \ov 3}(y) \ov y^2}\le(- h(y)dt^2+ d\vec{x}^2+{dy^2 \ov f(y)}\ri)
\ee
where
\be
h(y) = {f(y) \ov H(y)}, \; f(y)=H(y)  -{\mu y^4}, \;  H(y)=\prod_{i=1}^{3}\le(1+{q_i y^2 }\ri). 
\ee
We normalize $y$ so that the horizon is at $y_h =1$, then $\mu = \prod_{i=1}^{3}\le(1+q_i \ri)$. 
From~\eqref{allnsv2} we find 
\be \label{D5ve}
v_E^2 = {2 +  \ka_1 y_m^2 - \ka_3 y_m^6 \ov 1 + \ka_1 + \ka_2 + \ka_3} y_m^{-6}
\ee
with 
\be \label{deek}
\ka_1 = q_1 + q_2 + q_3, \; \ka_2 = q_1 q_2 + q_1 q_3 + q_2 q_3 , 
\; \ka_3 = q_1 q_2 q_3,
\ee
and 
\be 
y_m^2 = {\ka_1 + \sqrt{\ka_1^2 + 3 (1 + \ka_1 + \ka_3)} \ov 1 + \ka_1 + \ka_3} \ .
\ee
Note for the temperature to be non-negative requires 
\be 
\ka_3 \leq \ka_1 + 2 \ .
\ee
It can be readily checked analytically that for one- and two-charge cases with $q_3=\ka_3 =0$, the bound is satisfied for any $(q_1,q_2)$, including regions which are thermodynamically unstable. After numerical scanning we find that \eqref{D5ve} satisfies $v_E \leq v_E^{(S)}$ in the full three-charge parameter space.

{\it 2. Charged black holes in $\sN =8$ gauged supergravity in AdS$_4$~\cite{Duff:1999gh}:}
\be\label{c4}
ds^2={L^2 H^\ha (y) \ov y^2}\le(-h(y)dt^2+ d\vec{x}^2+{dy^2 \ov f(y)}\ri)\ ,
\ee
where
\be
h(y) = {f(y) \ov H(y)} , \; f(y)=H(y)  -{\mu y^3 }, \; H(y)=\prod_{i=1}^{4}\le(1+{q_i y }\ri) .
\ee
We again set $y_h=1$. Then $\mu = \prod_{i=1}^{4}\le(1+q_i \ri)$ and
requiring non-negative temperature gives
\be
\ka_4 \leq 2 \ka_1+\ka_2+3 \ 
\ee
where $\ka_i$ are defined analogously to~\eqref{deek}, with e.g. $\ka_4 = q_1 q_2 q_3 q_4$.
 We then find that
\be 
v_E^2 = {3 + 2 \ka_1 y_m +\ka_2 y^2_m - \ka_4 y_m^4 \ov 1 + \ka_1 + \ka_2 + \ka_3 + \ka_4} y_m^{-4}
\ee
where $y_m$ is the smallest positive root of the equation
\be
(1+\ka_1+\ka_2 + \ka_4) y^3 - 2 \ka_2 y^2 - 3 \ka_1 y - 4 =0 \ .
\ee
It can again be readily checked that for a single charge $q_1 \neq 0$ $v_E \leq v_E^{(S)}$ is satisfied for any $q_1$. One finds after numerical scanning that the bound is in fact satisfied in the full four-parameter space.

{\it 3. Metrics with hyperscaling violation:} Now let us consider metrics with hyperscaling 
violation~\cite{Gouteraux:2011ce,Dong:2012se}. Since we are interested in theories which have a  
Lorentz invariant vacuum, we restrict to examples with dynamical exponent unity,
\be\label{hysv}
ds^2={L^2 \ov y^2}\le({y \ov y_F}\ri)^{2 \th \ov d-1} \le(- f(y)dt^2+{dy^2 \ov f(y)}+d\vec{x}^2\ri)
\ee
where $f(y)=1-\le({y \ov y_h}\ri)^{\tilde d }$ and $\tilde d \equiv d - \th$. $y_F$ is some scale and $\th$ is a constant. Example of~\eqref{hysv} include dimensionally reduced near-horizon Dp-brane spacetimes for which $d=p+1$ and $\th = - {(d-4)^2 \ov 6-d}$. With boundary at $y=0$, such metrics are no longer asymptotically AdS, but our discussion can still be applied. 
We find in this case 
\be 
v_E^2 = {\le(\tilde \eta-1\ri)^{\tilde \eta-1} \ov \tilde \eta^{\tilde \eta}}\ , \qquad 
\tilde \eta ={2 (\tilde d -1) \ov \tilde{d}}\ .
\ee
The null energy condition now reads~\cite{Dong:2012se}  
\be
\tilde d \th \leq 0
\ee 
which implies either $\th \leq 0$ or $\tilde d \leq 0$. The former leads to $\tilde d \geq d$ and thus $v_E \leq v_E^{(\rm S)}$, while the latter is inconsistent with small $y$ describing UV physics. For examples coming from Dp-branes, $\th$ is clearly negative with $d \leq 6$, while for higher $d$ the metric no longer describes a non-gravitational field theory.

\subsection{$v_E$ from a Schwarzschild BH in Gauss-Bonnet gravity}

In this subsection as a preliminary investigation of the effect of higher derivative 
gravity terms, we compute the $v_E$ from a Schwarzschild black hole in Gauss-Bonnet gravity~\cite{Zwiebach:1985uq}, 
 \begin{eqnarray}
\label{action}
 I  &&= \frac{1}{16\pi G_N} \mathop\int d^{5}x \,
\sqrt{-g} \, [R + {12 \ov L^2}  \nonumber \\
 && +
{\lam \ov 2} L^2
(R^2-4R_{\mu\nu}R^{\mu\nu}+R_{\mu\nu\rho\sigma}R^{\mu\nu\rho\sigma})
] \ .
\end{eqnarray}
We consider the following the Vaidya metric
\be 
ds^2 = {\tilde L^2 \ov z^2} \le(- f (v,z) dv^2 - 2 dv dz + d \vec x^2 \ri)
\ee
with $f (v< 0, z) = 1$, $f(v > 0, z) = h(z)$, and~\cite{PhysRevLett.55.2656,Cai:2001dz} 
\bea \label{oep}
\tilde L^2 &=& a^2 L^2, \qquad a^2 \equiv \ha \le(1 + \sqrt{1 - 4 \lam} \ri)\ , \nn
h(z) &=& {a^2 \ov 2 \lam} \le(1 - \sqrt{1 - 4 \lam \le(1 - {z^4 \ov z_h^4} \ri)} \ri)\ .
\eea
Various thermodynamical quantities are given by 
\be 
T = {a^2 \ov \pi z_h}\ ,  \qquad s = {1 \ov 4 G_N} { \tilde L ^3 \ov z_h^3} \ , \qquad 
\sE = {3 \ov 4} Ts \ .
\ee
The entanglement entropy is obtained by extremizing the action~\cite{deBoer:2011wk,Hung:2011xb}
\be
\sA = \int d^3 \sig \, \sqrt{\ga} \le(1 + \lam L^2 \sR \ri)
\ee
where $\ga$ is the induced metric on the extremal surface and $\sR$ is the intrinsic scalar curvature
of the extremal surface. We have also suppressed a boundary term which will not be relevant for our discussion
below. 

As $v_E$ is shape-independent, it is enough to examine the extremal surface for a strip, whose 
induced metric can be written as 
\be 
ds^2 = {\tilde L^2 \ov z^2} \le(Q dx^2 + d \vec y^2 \ri)
\ee
with 
\bea
Q &=& 1- f v'^2 - 2 v' z' \ , \qquad \sqrt{\ga} = {\tilde L^3 \ov z^3} \sqrt{Q} \ , \nn
\sR &=& -{2 \ov Q^2 \tilde L^2} \le(3 Q z'^2 + z Q' z' - 2 Q z z''\ri)\ ,
\eea
where primes denote differentiation with respect to $x$. 
We need to extremize the action 
\be 
\sA =  K \int_0^R dx \, {\sqrt{Q} \ov z^3} \le(1 + \lam L^2 \sR \ri)
\ee
with 
\be 
K = \tilde L^3 A_{\rm strip}\ .
\ee
It is convenient to split the Lagrangian as 
\be 
\sL = \sL_0 + \sL_1\ , \qquad \sL_0 = {\sqrt{Q} \ov z^3} \ , \qquad \sL_1 = \lam L^2 {\sqrt{Q} \ov z^3} \sR\ .
\ee
Note that $\sL_0$ depends on $\lam$ through $h(z)$. 
We focus on the black hole region where equations of motion can be written as 
\bea \label{strip100}
&& {z' + h v' \ov z^{3} \sqrt{Q}} + \sO_v = {\rm const} \ , \\
 &&  \p_x \le( {v' \ov z^{3} \sqrt{Q}} \ri) = {1 \ov z^{3} \sqrt{Q}} \le(
3 {Q \ov z} + \ha h'(z) v'^2 \ri) + \sO_z  , \nn
\label{strip200}
\eea
with 
\bea 
\sO_v &= &- {\p \sL_1 \ov \p v'} + \p_x \le({\p \sL_1 \ov \p v''} \ri)  \\
\sO_z &= & - {\p \sL_1 \ov \p z} +  \p_x \le({\p \sL_1 \ov \p z'} \ri) 
-   \p_x^2 \le({\p \sL_1 \ov \p z''} \ri) 
\eea

To identify the linear regime, we look for a solution with 
\be 
z = z_m = {\rm const}, \qquad v' = {\rm const}, \qquad Q = {\rm const}
\ee
One can check explicitly that 
\ben 

\item Every term in $\sO_v$ contains at least a factor of $z'$ or $z''$. It will thus contribute zero. 

\item  Every term in $\sO_z$ contains at least a factor of $z'$ or $z''$ or $Q'$. It will thus contribute zero. 

\een 

So to find the value of $z_m$ and $v'$ we can simply ignore $\sL_1$, and the story is exactly the same as before except that $h(z)$ is now given by~\eqref{oep}. That is, $z_m$ is determined by 
\be 
z_m h' (z_m) - 6 h(z_m) = 0
\ee
and 
 \be 
 Q = - h(z_m) v'^2  \ .
 \ee
We find
\be 
{d \sA \ov dv } = K {\sqrt{Q} \ov z_m^3 v'}  =K {\sqrt{- h(z_m)} \ov z_m^3}
\ee
and 
\be 
v_E = {z_h^3 \sqrt{- h(z_m)} \ov z_m^3}\ .
\ee
Expanding in small $\lam$, we thus have
\be \label{gbn}
v_E ={\sqrt{2} \ov 3^{3 \ov 4}} - {3^{1 \ov 4} \ov \sqrt{2}} \lam + O(\lam^2) \ .
\ee
Entanglement entropy in Gauss-Bonnet gravity was studied numerically in~\cite{Li:2013cja} and their results are consistent with the above.

While in principle $\lam$ can take both signs, in all known 
examples $\lam$ appears to be positive~\cite{Buchel:2008vz}. We should also note that in all known examples where the Gauss-Bonnet term arises, there are probe branes and orientifolds which back-react on the metric and give rise to additional contributions at the same (or a more dominant) order.\footnote{See~\cite{Chang:2013mca,Jensen:2013lxa,Kontoudi:2013rla} for recent progress in computing contributions 
to entanglement entropy from probe branes.} Thus 
it seems one cannot draw a conclusion based on~\eqref{gbn} alone.

\section{Saturation} \label{sec:satur}

In this section we consider the saturation time and {critical behavior in the case of} continuous saturation. 
{The basic strategy was outlined in Sec.~\ref{sec:dcal} near~\eqref{satcp} -- for continuous saturation, $z_t - z_c \to 0$ as one approaches the equilibrium, and one can} expand $R, \ft$ and $\sA$ in terms of small $z_t - z_c$. 
Such an expansion also provides a simple diagnostic of whether saturation is discontinuous. For continuous saturation, $\ft - \ft_s$ must be negative in the limit $z_t-z_c$ goes to zero. If it is positive, then saturation is discontinuous, and equation~\eqref{satcp} 
does not give the saturation time. 



\subsection{Strip}  \label{sec:stripsat}

We {already saw in Sec.~\ref{sec:etae} that for Schwarzschild $g(z)$ and $\eta=2n/d > 1$ (which includes the case of entanglement entropy for $d \geq 3$)} saturation is discontinuous {-- at saturation time given by~\eqref{ansq1}, $\Ga_{\Sig}$ jumps directly from a near-critical extremal surface whose area grows linearly in time, to one residing entirely in the black hole and corresponding to equilibrium.} 
Here we consider general $g(z)$ and $n$.

Let us start by supposing that saturation is continuous with saturation time given by~\eqref{satcp}. In the large $R$ limit, $z_b$ is close to the horizon $z_h$, and~\eqref{satcp} has the leading behavior 
\be 
\ft_s = {1 \ov  h' (z_b)}  \log (z_h - z_b) + \cdots  \ .
\ee
In this limit $z_b$ can be found as in Appendix~\ref{app:equ} (see~\eqref{teipn} and~\eqref{bher}), from which
\be \label{nnec}
 \ft_s = {1 \ov c_n} R + O(R^0)\ ,  \quad c_n = \sqrt{z_h |h'(z_h)| \ov 2n} ={ \sqrt{2 \pi z_h T \ov n}}  \ .
\ee


Next, introducing the expansion parameter $\ep$
\be \label{satepdef}
z_c =  z_t \le(1 - {\ep^2 \ov 2n}\ri)\ ,
\ee 
we find that $\ft$ given by \eqref{biute} has the expansion (see Appendix~\ref{app:act1} for details)
\be \label{enrm}
\ft - \ft_s = u_1 \ep + O(\ep^2) + \cdots  
\ee
where
\be  \label{u1}
u_1 = \ha  g(z_b)  \le({z_b \ov n h^2 (z_b) F'(z_b)} - H (z_b) \ri)   
\ee
with
\bea
F(z_b) &\equiv& \int_0^1 {dy \ov \sqrt{y^{-2n}-1}} {z_b \ov \sqrt{h(z_b y)}}\ , \nn
H(z_b) &\equiv&  \int_0^{1} {dy \ov \sqrt{h (z_b y) (y^{-2n}-1)}}{z_b \ov h(z_b y)}   \ .
\eea
Note that $u_1 < 0$ implies $\ft < \ft_s$ as $z_c \to z_t$, as one expects for continuous saturation, while $u_1 > 0$ implies $\ft > \ft_s$ as $z_c \to z_t$, indicates that the saturation is discontinuous. 

The sign of $u_1$ as given in~\eqref{u1} is not universal and depends on $d$, $n$, and $g(z)$. 
In the case of Schwarzschild $g(z)$, 
for $d=2$ and $n=1$, $u_1 =0$, which agrees with the result of Sec.~\ref{sec:d2sat}. 
For $d=3,4$, we find that $u_1 < 0$ for $n=1$, but $u_1 > 0$ for $n > 1$. 
Thus for Schwarzschild $g(z)$, correlation functions in $d=3,4$ have continuous saturation, but  
a rectangular spacelike Wilson line and the entanglement entropy for a strip region 
have discontinuous saturation. 
For Reissner-Nordstrom $g(z)$ and $d=3,4$, $u_1$ can have either sign for $n=1$ but again $u_1>0$ for $n>1$, implying discontinuous saturation for Wilson lines and entanglement entropy.


Meanwhile, for $\sA$ given by \eqref{aver1}--\eqref{aver3}, one finds the small $\ep$ expansion (see Appendix~\ref{app:act1}) 
\be \label{aene}
{\De\sA -\De \sA_{\rm eq} \propto \ep^2}
\ee
which for a generic continuous transition (i.e. one with $u_1 < 0$) gives 
\be 
{\De\sA_{\rm eq} -\De \sA \propto (\ft_s - \ft)^2  \ .}
\ee
{In the language of phase transitions, such a quadratic approach corresponds to mean-field behavior.}

\begin{figure}[!h]
\begin{center}
\includegraphics[scale=0.8]{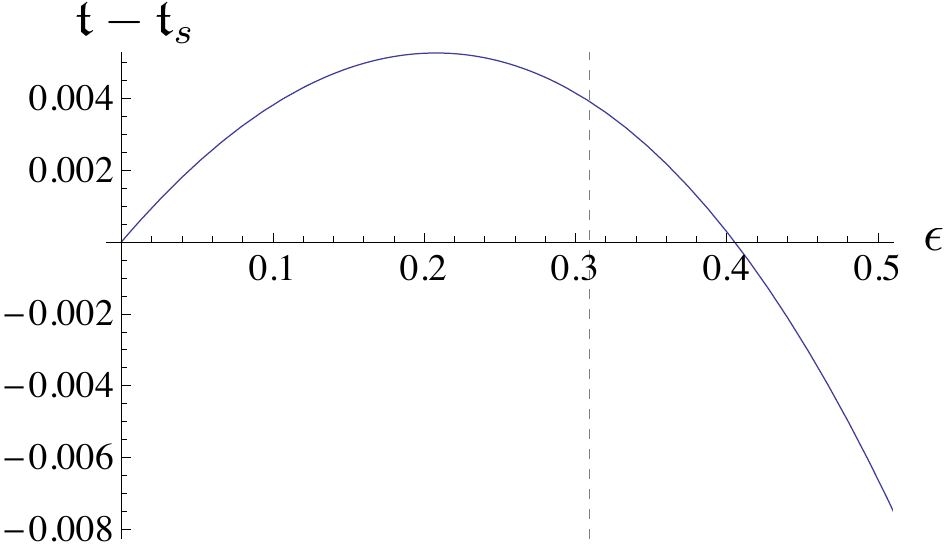} 
\includegraphics[scale=0.8]{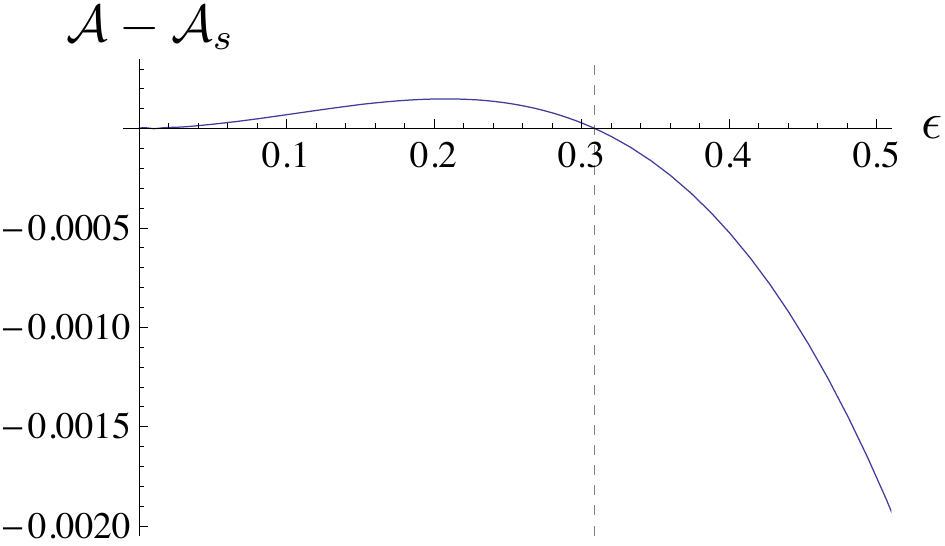}
\end{center}
\caption{Plots of $\ft-\ft_s$ and $\sA-\sA_s$ as functions of $\ep$ in \eqref{satepdef}, with $d=4$ Schwarzschild, $n=3$, and $z_b=0.8$. $\ft_s$ is the time when continuous saturation would have occurred, but true saturation $\ft_s^{(\rm true)}$ occurs at the dashed line, for which $\ft_s^{(\rm true)}>\ft_s$. 
 }\label{fig:nonmo}
\end{figure}

Note that for a given $R$, a solution which lies fully in the back hole region exists only for $\ft > \ft_s (R)$, so for a discontinuous saturation the ``genuine'' saturation time $\ft_s^{(\rm true)}$ is always larger than that  given by~\eqref{ansq1}.
 See Fig.~\ref{fig:nonmo} for an explicit example. 

To summarize, for $\Sig$ a strip the saturation leading to equilibrium is non-universal, with possibilities of both discontinuous and continuous saturation. When the saturation is continuous one finds that $\De \sA$ approaches its equilibrium value quadratically in $\ft_s - \ft$ irrespective of $n$. In contrast, we will see below that for $\Sig$ a sphere, saturation is almost always continuous (except when $n=2$) and there is a nontrivial $n$-dependent critical exponent.

\subsection{Sphere}

{Again let us first assume that saturation is continuous}. Then from~\eqref{satcp} and~\eqref{sphrrp}, we find that in the large $R$ limit 
\be \label{satsph}
 \ft_s =  {1 \ov c_n} R - {n-1 \ov 4 \pi T} \log R + {O(R^0)}\ 
\ee
where $c_n$ was given earlier in~\eqref{nnec}. For entanglement entropy we then have 
\be \label{fjske}
\ft_s (R)= {1 \ov c_E } R - {d-2 \ov 4 \pi T} \log R + O(R^0)
\ee
where $c_E$ is the dimensionless number 
\be \label{uber}
c_{E} = \sqrt{z_h |h'(z_h)| \ov 2(d-1)} = \sqrt{2 \pi z_h T \ov d-1} \ .
\ee

To find the critical behavior during saturation we need to solve for $z (\rho)$, which we accomplish by expanding about the solution at equilibrium, $z_0(\rho)$.
After a somewhat long calculation (outlined in Appendix~\ref{app:sea}), we find that using the expansion parameter $\ep$ defined by 
\be 
\rho_c = z_c \ep \ ,
\ee
$\ft$ given by \eqref{texpl} has the expansion
\be  \label{sphtmts}
\ft-\ft_s=\begin{dcases} - \ha \le(z_b +  {g(z_b) z_b \ov h(z_b)} \le({b_1 \ov b_2}  + I_0 \ri) \ri) \ep^2+ \cdots  & n=2 \\
- \ha z_b \ep^2 + \cdots & n>2
\end{dcases}\ 
\ee
where $b_1, b_2$ and $I_0$ are some constants which are defined in Appendix~\ref{app:sea}. 
Thus for $n >2$, saturation is always continuous, while for $n=2$ it {is} model dependent. 
Computing $b_1, b_2, I_0$ in \eqref{sphtmts} explicitly, one finds that the coefficient before $\ep^2$ is positive for Schwarzschild $g(z)$ (saturation is continuous), but becomes negative for Reissner-Norstrom $g(z)$ at sufficiently large chemical potential and for sufficiently large $R$ (saturation is discontinuous). Meanwhile, $\sA$ given by \eqref{speac} has the expansion
\bea
&&\De\sA-\De\sA_{\rm eq}\nn
&=&\begin{dcases}
K {g^2(z_b) \ov 8 h(z_b)}\ep^4 \log \ep + O(\ep^4)  & n=2 \nn
-K {g(z_b) \ov 2 (n-2)}\le({n-2 \ov n+2}+{g(z_b) \ov 4 h(z_b)}\ri)\ep^{n+2}+\cdots &
 n>2
\end{dcases}\ .
\eea
We thus find 
\be \label{exexp}
\De\sA_{\rm eq} -\De \sA \propto \bca  - (\ft_s - \ft)^{2} \log (\ft_s - \ft) + \cdots & n=2 \cr
(\ft_s - \ft)^{{n\ov 2} +1} + \cdots & n > 2
\eca  \ .
\ee

Characterizing continuous saturation with a nontrivial scaling exponent 
\be \label{svew2}
 S (R,\ft)- S^{\rm (eq)} (R) \propto - (\ft_s - \ft)^\ga\ , \qquad \ft_s  - \ft \ll \eql\ ,
 \ee
we thus find that for an $n$-dimensional extremal surface
\be 
\ga_n = {n+2 \ov 2} \ .
\ee
Note that the above exponent depends only on $n$ and is independent of the boundary spacetime dimension $d$.
Also note that in~\eqref{exexp}, the $n=2$ expression applies to cases of continuous saturation. There is a logarithmic prefactor by which the scaling barely avoids the ``mean-field" exponent $\gamma=2$. 
For $d=2$, only $n=1$ is possible and $\ga = {3 \ov 2}$ which was previously found 
in~\cite{Hubeny:2013hz}.
 For entanglement entropy, $n = d-1$, giving 
\be 
\ga_{\rm E} = {d+1 \ov 2} \ . 
\ee

\subsection{More on the saturation time}

Let us now collect the results we have obtained so far on saturation time. 
For a strip we showed in Sec.~\ref{sec:nee} (see~\eqref{ansq2}) that for $z_t \gg z_m^2$, 
the linear regime persists all the way to discontinuous saturation, with saturation time in the large $R$ limit given by 
\be 
\ft_s = {R \ov v_n} + \cdots , \qquad v_n = \le({z_h \ov z_m}\ri)^{n} \sqrt{- h(z_m)} \ .
\ee
This happens, for example, for Schwarzschild with $\eta = {2n \ov d} > 1$. 

For continuous saturation we found earlier in this section that up to logarithmic corrections, for both a strip and sphere 
\be \label{jen}
\ft_s = {R \ov c_n} + \cdots, \quad c_n =  \sqrt{z_h |h'(z_h)| \ov 2n}  \ .
\ee
It is tempting to speculate that the above result applies to continuous saturation for all shapes. 

For Schwarzschild and RN black holes $c_n$ is given by
\be 
c_n^{(\rm S)} = 1/\sqrt{\eta}\ ,   \qquad c_n^{(\rm RN)} = \sqrt{u/ \eta} \leq c_n^{(S)}  \ .
\ee
In particular for entanglement entropy we have 
\be 
c_E^{(S)} = \sqrt{d \ov 2 (d-1)} \ .
\ee 

It can be readily checked that for $\eta > 1$
\be \label{ooem}
v_n^{(\rm S)} < c_n^{(\rm S)}  < 1
\ee
For a sphere, equation~\eqref{ooem} may be understood heuristically from the tsunami picture of 
Fig.~\ref{fig:eevelo}--the volume of an annulus region of unit width becomes smaller as the tsunami advances inward.

For $\eta = 1$
\be 
v_{d \ov 2}^{(\rm S)} =  c_{d \ov 2}^{(\rm S)}  = 1 \ .
\ee
As discussed earlier for $n=1$ in $d=2$, the saturation is continuous, but is discontinuous for 
$n=2$ in $d=4$. In the latter case the ``true'' saturation time should be greater than~\eqref{ansq1}
which at leading order in the large $R$ expansion gives~\eqref{jen}. Numerical results suggest that 
the difference is $O(1)$ in the large $R$ limit and thus at leading order 
the ``true'' saturation time is still given by $\ft_s^{(\rm true)} = R$. 

For $\eta < 1$, as in the case of equal-time correlation functions in $d=3,4$,
the saturation is continuous and 
\be 
c_n^{(\rm S)}  > 1 \ .
\ee
That $\ft_s < R$ has been observed before numerically in 
e.g.~\cite{Balasubramanian:2011ur,Caceres:2012em}.
Recall that in this case $\sA$ appears in an exponential with a minus sign. Since $c_n$ does not correspond directly to any physical propagation, there is no obvious constraint on it from causality.

\section{Memory loss regime}\label{gscale}

In this section, we examine implications of the critical extremal surface for the evolution of $\sA (R, \ft)$ for a strip and sphere in the regime $\ft_s \gg \ft_s - \ft \gg z_h$. In $(1+1)$-dimensions, we saw in Sec.~\ref{sec:2dscal} 
that in this regime the difference between $\sA (R, \ft)$ and the equilibrium value $\sA_{\rm eq} (R)$ is a function of $\ft_s (R) - \ft = R - \ft$ only and not of $R$ and $\ft$ separately. In other words, at late times in the evolution, the size $R$ has been ``forgotten''.  We emphasize that since $\ft_s \propto R \to \infty$ in the large $R$ limit,  such memory loss can happen long before saturation. 

We will generalize this result to higher dimensions. At a heuristic level the existence of such a scaling regime 
is expected, as for large $R$ and $\ft$ $(z_t, z_c)$ very closely follows the critical line $z_c^* (z_t)$ 
as time evolves. Thus in the limit $R, \ft \to \infty$ the system is controlled by a single parameter along the line $z_c^* (z_t)$ rather than two separate variables $R$ and $\ft$.  Recall that in the $(1+1)$-dimensional story in Sec.~\ref{sec:2dscal}, $\ep$, parameterizing the distance to the critical line~\eqref{2dcrl} (or~\eqref{2dcrl1}), gave the leading large $\ell$ and $\tau$ behavior, while $\phi$ in~\eqref{2dcrl} (or $z_t$ in~\eqref{2dcrl1}), parametrizing the location on the critical line, mapped to $\ell - \tau$ or $\tau_s-\tau$. In the limit $\ell, \tau \to \infty$ with their difference finite, $\ep$ drops out to leading order and $\sA - \sA_{\rm eq}$ is determined by 
a single parameter $\phi$ only.

In general dimensions, the story becomes technically much more involved. For example, for $\Sig$ a sphere, even determining the scaling variable (the analogue of $\ell -\tau$ in~\eqref{scalf}) is a nontrivial challenge. We will leave the explicit scaling functions (the analogue of $\lambda$ in \eqref{scaol}), which requires working out the $O(1)$ counterparts of~\eqref{eie1}--\eqref{sace}, for future investigation.

\subsection{Strip}

For definiteness we will restrict our discussion to Schwarzschild. 
With a given $R$, as $\ft$ increases, $z_t$ decreases. For $\eta > 1$, as discussed in Sec.~\ref{sec:nee} 
$z_t$ remains large compared to $\log \ep$ term  in~\eqref{rexpep} all the way to saturation, in which case 
the linear regime persists to the saturation. 
But this is no longer so for $\eta \leq 1$. For $\eta =1$, in Sec.~\ref{sec:eta1} we showed that 
before saturation $z_t$ will become comparable to $z_h$ and the system will eventually exit the linear growth regime.
For $\eta < 1$, for which the linear regime appears not to exist, from discussion of Sec.~\ref{sec:stripsat}, we saw at least for $d=3,4$, the saturation is continuous which implies that $z_t$ again has to become comparable to $z_h$ before saturation. 

We will now focus on $\eta \leq 1$.  We show below that for $\eta =1$ there is another scaling regime prior to the saturation when $z_t$ is $O(1)$ (i.e. no longer scales with $R$).

We again consider $z_c = z_c^* (1-\ep), \, \ep \to 0$. Following a discussion 
similar to that of Sec.~\ref{sec:linstr} we find that 
\bea\label{teb1}
\ft &=&  - {E (z_c^*) \ov h(z_m) \sqrt{H_2}} \log \ep  + O(1)\  , \\ 
\label{tebb}
R &=& - {1 \ov \sqrt{H_2}} \log \ep + O(1)\ , \\
{1 \ov \tilde K} \De \sA &=&  -{z_t^{n} \ov z_m^{2n} \sqrt{H_2}} \log \ep + O(1) \ .
\label{teb2}
\eea
Note that $z_t$ is now considered to be $O(1)$, which varies with $R, \ft$, and both 
$z_c^*, z_m$ are functions of $z_t$. 

For Schwarzschild with $h(z) = 1- {z^d \ov z_h^d}$, we find from~\eqref{ejek}
\be \label{ec1}
z_t^{2n} = {d z_m^{d+2n} \ov 2n z_h^d + (d-2n) z_m^d} 
\ee 
and from~\eqref{findzc} 
\be \label{ec2}
E (z_c^*) =  - \sqrt{- h (z_m) \le({z_t^{2n} \ov z_m^{2n}} -1 \ri)} 
= {h (z_m) \ov \sqrt{1 + (\eta^{-1} -1) {z_m^d \ov z_h^d}}} .
\ee

For $\eta =1$, we then have 
\be 
z_t^{n} = {z_m^{2n} \ov z_h^n} , \qquad E (z_c^*) = h(z_m)  \ .
\ee
Using these equations in~\eqref{teb1}--\eqref{teb2} we find that 
\bea\label{teb3}
\ft &=&  - {1\ov  \sqrt{H_2}} \log \ep  + O(1)\  , \\ 
R &=& - {1 \ov \sqrt{H_2}} \log \ep + O(1)\ , \\
{1 \ov \tilde K} \De \sA &=&  {R  \ov z_h^{n}}  + O(1) \ .
\label{teb4}
\eea
Note that $O(1)$ terms are evaluated in the $\ep \to 0$ limit $z_c \to z_c^* (z_t)$ and therefore are functions only of $z_t$. In other words,
\be 
R - \ft = \chi (z_t)\ , \quad  \De\sA = \Delta\sA_{\rm eq} + \al (z_t) \quad {\rm as }\quad \ep \to 0 \quad
\ee
where {$\chi$ and $\al$} are some functions whose {explicit form} we have not determined for general $n$, and in the second equation we have used~\eqref{eqarea}.  We thus conclude that for $ \ft, R \gg R- \ft \gg z_h$, $\sA (R, \ft)$ has the scaling behavior 
\be \label{nnme1}
 \sA (R, \ft) - \sA_{\rm eq} (R) =\lam (R -\ft)  + \cdots
\ee
where $\lam (x) = \al (\chi^{-1} (x))$ and $\cdots$ are terms suppressed in the large $R, \ft$ limit. Here we will not attempt to find these function explicitly for general $d$. For $d=2$, functions $\chi$, $\al$, and $\lam$ are given in~\eqref{eorp}--\eqref{scaol}. 
The above discussion does not apply near saturation when $R - \ft \lesssim O(z_h)$. Recall from Sec.~\ref{sec:d2sat} that in $d=2$ ($n=1$) saturation is continuous. But in $d=4$ with $n=2$, the results in Sec.~\ref{sec:stripsat} show that saturation is discontinuous. In both cases the saturation time is given by $\ft_s = R$ for large $R$ and thus~\eqref{nnme1} can also be written as 
\be \label{nnme2}
 \sA (R, \ft) - \sA_{\rm eq} (R) =\lam (\ft_s -\ft)  + \cdots \ .
\ee

For $\eta < 1$, from~\eqref{teb1}--\eqref{teb2} we find that 
\be 
R - {h (z_m) \ov  E (z_c^*)} \ft = O(1), \quad {\De \sA \ov \tilde K} = {z_t^n \ov z_m^{2n}}  R + O(1)
\ee
but in this case from~\eqref{ec1}--\eqref{ec2} the prefactor ${h (z_m) \ov  E (z_c^*)} $ before $\ft$ as well as the prefactor before $R$ on the right side of the second equations depends on $z_t$. Thus a scaling 
regime does not appear to exist.

\subsection{Sphere} \label{sec:mosp}

We now consider $\Sig$ being a sphere. 
Since the discussion is rather involved, here we only outline the basic steps and final results, leaving details to Appendix~\ref{app:gensca}.

The basic strategy is the same as in previous sections; we consider $z_c$ close to the critical line, 
\be  \label{gscep} 
z_c = z_c^* (1- \ep )\ , \qquad  \ep  \ll 1  \ ,
\ee
and expand the quantities $\ft$, $R$ and $\sA$ in $\ep$. In contrast to the linear 
regime, where $R \gg \ft \sim z_h |\log \ep| \gg z_h $ and we expressed all quantities in a double expansion 
of $1/R$ and $\ep$, here we have instead
\be \label{newg}
R \to \infty\ , \quad - \log \ep \sim O(R) \to \infty\ , \quad z_t , \rho_c, z_c^* \sim O(1) \ .
\ee
That is, evolution of the extremal surface happens largely after the surface has entered the black hole 
region. 

We denote the critical extremal surface for $z_c = z_c^*$ as $z^* (\rho)$. As discussed earlier in Sec.~\ref{sec:critB}, $z^*$ asymptotes to the horizon $z_h$ for sufficiently large $\rho$. In the regime of $z_t \sim z_c^*$, an example of $z^*$ was given in Fig.~\ref{fig:critical}. 
 More explicitly, for large $\rho \gg z_t$ we can write $z^*$ as (see Appendix~\ref{app:gensca} for more details)
\be \label{eor1}
z^* (\rho)= z_h + \chi_* (\rho)
\ee
where $\chi_*$ has the asymptotic behavior
\be
\label{assol1}
\chi_* (\rho) =  {\al \ov \rho^{n-1}}+ O(\rho^{-n}), \quad \rho \gg \rho_c 
\ee
with $\al$ some constant. 

With~\eqref{gscep}, we can expand solution $z$ about $z^*$,
\be \label{poec}
z (\rho) = z^* (\rho) - \ep z_1 (\rho)  + O(\ep^2)  \ .
\ee
At the shell $z_1$ satisfies the boundary conditions 
\be \label{bdma}
z_1 (\rho_c) = z_c^*\ , \quad 
z_1' (\rho_c) =  {\rho_c \ov z_c^*} \le(1 - \ha g(z^*_c) 
+  \ha z_c^* g' (z_c^*) \ri) 
\ee
which can be obtained from the matching conditions discussed in Sec.~\ref{app:eomsp}. 
Focusing on large $\rho$ for which $z^*$ asymptotes to the horizon, we have
\be \label{horex}
z (\rho) = z_h + \chi^* (\rho) - \ep z_1 (\rho) + O(\ep^2) \ .
\ee

The equation for $z_1$ can be obtained by inserting~\eqref{horex} into~\eqref{modeom} and expanding in $\ep$.  Due to $h(z_h)=0$, this expansion differs depending on the relative magnitudes of $\chi^*$ and $\ep z_1$, and as a result, the near-horizon region for $z$ can be further subdivided into three regions in which $z_1$ can have distinct behavior (see Appendix~\ref{app:gensca} for details):

\ben 

\item Region I: $\chi^* \gg  \ep z_1$. {In this region, $z$ is well 
approximated by $z^*$ and approaches the horizon from the inside. Solving for $z_1$, we find it has the leading large $\rho$ behavior}
\be \label{solI}
z_1(\rho) = A_1 e^{\ga_n \rho} \rho^{- \beta_n} \le(1 + O\le(\rho^{-1} \ri) \ri) 
\ee
where 
\be \label{b1}
\beta_n = n-1 + {b_1 \ov 2 \ga_n}\ , \qquad b_1 = \de_{n,2}{|E| (h_2 -h_1) \ov \sqrt{h_1}}\ .
\ee
Here $A_1 (\rho_c)$ is a positive $O(1)$ constant determined by the boundary conditions~\eqref{bdma}, and $\ga_n$, $h_{1,2}$ are some constants given in~\eqref{cndef} and~\eqref{hnot}. 
Equation~\eqref{solI} applies in the region
\be
{\al \ov \rho^{n-1}} \gg \ep A_1 e^{\ga_n \rho}\rho^{-\beta_n} 
\ee
which translates into
\be 
\rho_c \ll \rho \ll - {1 \ov \ga_n} \log \ep + {b_1 \ov 2 \ga^2_n} \log \log {1 \ov \ep} + O(1) + \cdots 
\ee
which, when written using $R$ (see \eqref{epr} below), is
\be \label{regime1}
\rho_c \ll \rho \ll  R - {1 \ov \ga_n} \le(n-1 -{b_1 \ov 2 \ga_n} \ri) \log R + O(1) \ .
\ee

\item Region II: $\chi^* \sim \ep z_1$. Since $z_1$ {grows exponentially} with $\rho$, at a certain point $\ep z_1$ surpasses $\chi^*$ and $z$ crosses the horizon. Close to this crossing  $\chi^*$ and $\ep z_1$ are comparable and need to be treated on equal ground, making the equation for $z_1$ complicated.

\item Region III: $\chi^* \ll \ep z_1 \ll 1$. In this region, $z_1$ has grown sufficiently large that it dominates over $\chi^*$, and has leading large $\rho$ behavior 
\be \label{sol3}
z_1(\rho)=   A_2 \rho^{-(n-1)} e^{\ga_n \rho} \le(1 + O(\rho^{-1}) \ri) 
\ee
with $A_2(\rho_c)$ a positive $O(1)$ constant. The domain of the region is
\be
\label{regime2pre}
{1 \ov \rho^{n-1}}  \ll \ep z_1 \ll 1 \
\ee
or more explicitly  
\be
  - {1 \ov \ga_n} \log \ep \ll  \rho \ll - {1 \ov \ga_n} \log \ep + 
{n-1 \ov \ga_n} \log \log {1 \ov \ep}  \ .
\ee
Note that $\ep z_1$ should become $O(1)$ when $\rho \approx R$, and $z (\rho)$ then quickly deviates from the horizon to reach the boundary, i.e.
\be 
\ep z_1 (R) \sim O(1) 
\ee
which leads to 
\be \label{epr} 
- \log \ep = \ga_n R - (n-1) \log R + O(1) \ .
\ee
This relation can be established rigorously by carefully matching~\eqref{sol3} with an expansion of $z$ near the boundary following techniques developed in~\cite{Liu:2013una}. Using~\eqref{epr}, we can rewrite~\eqref{regime2pre} as
\be \label{regime2}
R - {n-1 \ov \ga_n} \log R \ll \rho \ll R \ .
\ee
\een
Note that for $n > 2$, {$b_1 =0$ in \eqref{b1}, and the leading behavior~\eqref{solI} and~\eqref{sol3} in regions I and III match up to an overall constant factor. Consistently, the domain of the regions~\eqref{regime1} and~\eqref{regime2} are adjacent to each other, i.e. the width of region II is $O(1)$ as $\ep \to 0$ or equivalently, $R \to \infty$.}
In contrast, for $n=2$, {$b_1 \neq 0$ so that the power of $\rho$ in~\eqref{solI} and~\eqref{sol3} do not match, and region II should be of width $O(\log R)$.}\footnote{This is {evidently the case} when $b_1 < 0$, for example for Schwarzschild $h(z)$. {However, $b_1$ can also be positive, for example for Reissner-Norstrom $h(z)$ at sufficiently large chemical potential.} When $b_1$ is positive, even though naively it appears that~\eqref{regime1} and~\eqref{regime2} overlap with each other, 
it is likely that the width of region II is still $O(\log R)$ in order for the exponent of $\rho$ to evolve from 
that of~\eqref{solI} to that of~\eqref{sol3}.}

One can {proceed to use $z_1$ obtained as above in the three regions} to calculate the boundary quantities {$\ft$~\eqref{texpl} and $\sA$~\eqref{speac}} (see Appendix~\ref{app:gensca} for details). We find that for $n> 2$ 
\be \label{teie}
\ft = \ft_s (R)  + O(1)\ ,
\ee
where $\ft_s (R)$ is the saturation time and was given before in~\eqref{satsph}, 
and 
\be \label{aaep}
\De\sA -\De\sA_{\rm eq}= O(1) \ .
\ee
Working in the $\ep \to 0$ limit, the $O(1)$ terms in \eqref{teie} and \eqref{aaep} can be functions of $z_t$ only. 
Eliminating $z_t$-dependence between~\eqref{teie} and~\eqref{aaep}, we find the scaling behavior 
\be \label{evem}
\sA (\ft, R) - \sA_{\rm eq} = - \fa_{\rm eq}  \lam \le(\ft_s (R)  - \ft \ri)
\ee
for some function $\lam$. In~\eqref{evem} we have included a prefactor $\fa_{\rm eq}$ as 
$\sA_{\rm eq} (R) \propto \fa_{\rm eq}$ and a minus sign, so that $\lam$ is positive and has the dimension 
of volume enclosed by $\Sig$.  Finding the explicit form of $\lam$ requires computing the $O(1)$ terms 
in~\eqref{teie}--\eqref{aaep}, which is a rather intricate task and will not be attempted here. 

For $n=2$ (which gives the entanglement entropy in $d=3$), we cannot rule out a possible additional $\log R$ term in~\eqref{teie}, due to complications in region II mentioned earlier. Thus we do not yet have a clean answer in that case. 

\subsection{Memory loss} 

Let us again specialize to the case of entanglement entropy with $n= d-1$. 
Given that $S_{\rm eq} (R) = V_{\Sig} s_{\rm eq}$, one can interpret $\lam$ in~\eqref{evem}
as the volume which has not yet been entangled. Equation~\eqref{evem} then implies that the ``left-over'' volume only depends on the difference $\ft_s - \ft$ and not on $R$ and $\ft$ separately. In other words, at late times of evolution, the size $R$ has been ``forgotten''.  We again emphasize that with \eqref{evem} valid for $\ft_s \gg \ft_s - \ft \gg \eql$, such memory loss can happen long before saturation.

\begin{figure}[!h]
\begin{center}
\includegraphics[scale=0.45]{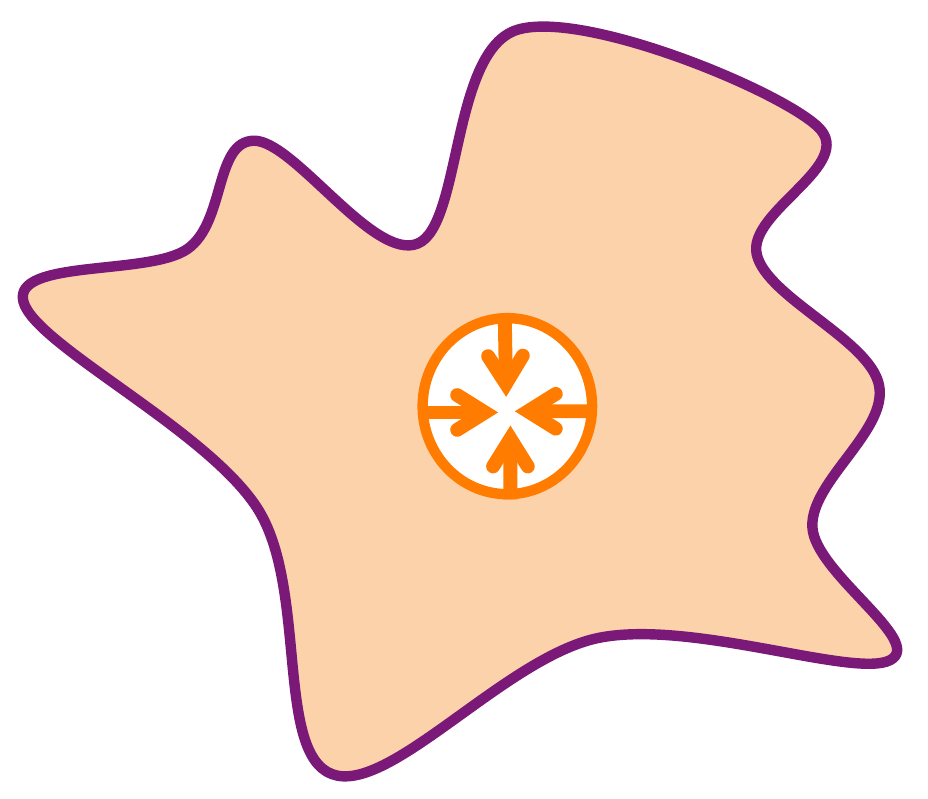}
\end{center}
\caption{
A cartoon picture for late-time memory loss. The (hypothetical) tsunami picture discussed in Sec.~\ref{sec:ling} can be used to visualize the memory loss regime--for a wide class of compact $\Sig$, in the limit of large size, at late times the wave front may approach that of a spherical $\Sig$.
 }\label{fig:loss}
\end{figure}

Note that the existence of the memory loss regime itself is not related to the tsunami picture discussed earlier. However, the tsunami picture does lend a natural geometric interpretation to the regime, as the memory loss of the wave front of the entanglement tsunami. 
It is tempting to speculate that due to interactions among different parts of the tsunami wavefront, for a generic surface $\Sig$ in the limit of large $R$, memory of {\it both} the size and shape of $\Sig$ could be lost during late times in evolution. See Fig.~\ref{fig:loss} for a cartoon. 
It would be interesting to understand whether this indeed happens. 

If such ``memory loss'' as indicated in Fig.~\ref{fig:loss} indeed occurs,  
we expect that in the infinite size limit, the space of all possible $\Sig$ separates into different 
basins of attraction, defined by various attractors (or ``fixed points'') such as the sphere and strip. 
For example, for a smooth compact $\Sig$, at late times the wave front of the tsunami may approach 
that of the sphere, while for an elongated surface $\Sig$ with topology that of a strip, it may approach that of the strip.
This would also imply that the saturation behavior for generic $\Sig$ could be classified using those of 
the ``fixed points.''



\section{Conclusions and discussions}

In this paper we considered the evolution of entanglement entropy and various other nonlocal observables 
during equilibration, in a class of quenched holographic systems. In the bulk the equilibration 
process is described by a Vaidya geometry, with different observables having a unified description as 
functions of the area of extremal surfaces of different dimension $n$. We were able to derive general scaling results for these observables without using the explicit bulk metric. Some of these lead to universal behavior in the boundary theory.

It is important to keep in mind that while the entanglement entropy is proportional to the area, for other observables the area appears in an exponential with a minus sign. So the boundary interpretation of the evolution of $\sA$ could be very different. We also see interesting differences in the evolution of $\sA$ 
for different $n$. For example, there appears to be no linear evolution for $n < {d \ov 2}$, which includes correlation functions in $d > 2$. 
 See tables~\ref{tab:d3tab}--\ref{tab:d4tab} for a list of the time-dependence of various observables 
in $d=3$ and $d=4$. 

In the rest of this section we discuss some future directions, 
using language for entanglement entropy.

\subsection{More general equilibration processes}

In this paper we restricted our discussion to the equilibration following a global quench. 
It is interesting to consider more general equilibration processes, in particular those 
with inhomogeneous or anisotropic initial states (see~\cite{Balasubramanian:2013rva,Balasubramanian:2013oga,Nozaki:2013wia} for recent related work).

There are reasons to believe some of our results may apply to these more general situations. 
 In particular, an important feature of the linear growth~\eqref{line} is that the speed $\Vee$ {\it characterizes properties of the equilibrium state}, as it is solely determined by the metric of the black hole. This highlights the local nature of entanglement propagation. At corresponding times, locally, the system has already achieved equilibrium, although for large regions non-local observables such as entanglement entropy remain far from their equilibrium values. Thus  $\Vee$ should be independent of the nature of the initial state, including whether it was isotropic or homogeneous.
 Similarly, the memory loss regime occurs long after a system has achieved local equilibration, and we again expect that it should survive more general initial states. 
 
 The pre-local-equilibration stage is likely sensitive to the nature of initial states, including the  
value of the sourcing interval $\de \ft$. Nevertheless, that the early growth~\eqref{quda} is proportional to the energy density is consistent with other recent studies of the entanglement entropy of excited states~\cite{Bhattacharya:2012mi,Blanco:2013joa,Allahbakhshi:2013rda,Wong:2013gua}. 

Finally with a nonzero sourcing interval $\de \ft$, we expect the  
wave front of ``entanglement tsunami'' to develop a finite spread, but the picture of an entanglement wave that propagates may still apply as long as $\de \ft$ is much smaller than the 
size of the region one is exploring. If $\de \ft$ is comparable to or larger than the local equilibration scale $\eql$, the pre-local-equilibration and saturation regimes likely can no longer be sharply defined.

\bwt

\begin{table}[h]
	\begin{tabular}{|c| >{\centering}p{2.5 cm}<{\centering} | >{\centering}p{2.5 cm}<{\centering} | >{\centering}p{3.1 cm}<{\centering } | >{\centering}p{4.7 cm} <{\centering} |}\hline
	 & $\ft \ll z_h$ & $z_h \ll \ft \ll R$ & $z_h \ll \ft_s-\ft \ll \ft_s$ & saturation \tabularnewline \hline
	 Equal-time two-point function & $G_{vac}\exp\le(-\# \ft^3\ri)$ & no linear regime & no scaling
	& $G_{eq}\exp(\#(\ft_s-\ft)^2)$\tabularnewline \hline 
	 Wilson loop (rectangular) & $\mathcal{W}_{vac}\exp(-\# \ft^2)$ & $\mathcal{W}_{vac} \exp(-\# \ft)$ & linear regime persists & discontinuous \tabularnewline \hline
	Wilson loop (circular) & $\mathcal{W}_{vac}\exp(-\# \ft^2)$ & $\mathcal{W}_{vac}\exp(-\# \ft)$ &undetermined & $\mathcal{W}_{eq}\exp\le(-\#\le(\ft_s-\ft \ri)^2\log (\ft_s-\ft)\ri)$ \tabularnewline \hline
	EE (strip) & $\sS_{vac}+\# \ft^2$ & $\sS_{vac}+\# \ft$ & linear regime persists & discontinuous \tabularnewline \hline
	EE (sphere) & $\sS_{vac}+\#\ft^2$ & $\sS_{vac}+\# \ft$ & undetermined & $\sS_{eq}+\# \le(\ft_s-\ft\ri)^2\log(\ft_s-\ft)$\tabularnewline\hline
	\end{tabular}
	\caption{Time-dependence of non-local variables in $d=3$ for Schwarzschild. $\#$ is used to denote some positive coefficient. To lowest approximation in the large $R$ limit, $\ft_s \propto R$, with coefficients as follows: for the equal-time two-point function, $\ft_s/R=\sqrt{2/3}$, for the rectangular Wilson loop and strip EE, $\ft_s/R=2^{4/3}/3^{1/2}$, and for the circular Wilson loop and sphere EE, $\ft_s/R=2/\sqrt{3}$.}\label{tab:d3tab}
\end{table}

\begin{table}[h]
	\begin{tabular}{|c| >{\centering}p{2.5 cm}<{\centering} | >{\centering}p{2.5 cm}<{\centering} | >{\centering}p{3.1 cm}<{\centering } | >{\centering}p{4.7 cm} <{\centering} |}\hline
& $\ft \ll z_h$ & $z_h \ll \ft \ll R$ & $z_h \ll \ft_s-\ft \ll \ft_s$ & saturation \tabularnewline \hline
	Equal-time two-point function & $G_{vac}\exp\le(-\# \ft^4\ri)$ & no linear regime & no scaling
	 & $G_{eq}\exp(\#(\ft_s-\ft)^2)$\tabularnewline \hline 
	 Wilson loop (rectangular) & $\mathcal{W}_{vac}\exp(-\# \ft^3)$ & $\mathcal{W}_{vac} \exp(-\# \ft)$ & $\mathcal{W}_{eq}\exp\le(\#\lambda\le(\ft_s-\ft\ri)\ri)$ & discontinuous \tabularnewline \hline
	Wilson loop (circular) & $\mathcal{W}_{vac}\exp(-\# \ft^3)$ & $\mathcal{W}_{vac}\exp(-\# \ft)$ & undetermined & $\mathcal{W}_{eq}\exp\le(-\# \le(\ft_s-\ft\ri)^2\log(\ft_s-\ft)\ri)$ \tabularnewline \hline
	EE (strip) & $\sS_{vac}+\# \ft^2$ & $\sS_{vac}+\# \ft$ & linear regime persists & discontinuous \tabularnewline \hline
	EE (sphere) & $\sS_{vac}+\#\ft^2$ & $\sS_{vac}+\# \ft$ & $\sS_{eq}-\#\tilde{\lambda} \le(\ft_s-\ft\ri)$ & $\sS_{eq}-\# \le(\ft_s-\ft\ri)^{5/2}$\tabularnewline\hline
	\end{tabular}
	\caption{Time-dependence of non-local variables in $d=4$ for Schwarzschild. $\#$ is used as above and the functions $\lambda$ and $\tilde{\lambda}$ are those from \eqref{nnme1} and \eqref{evem}. The saturation times are: for the equal-time two-point function, $\ft_s/R=1/\sqrt{2}$, for the rectangular and circular Wilson loops, $\ft_s/R=1$, for strip EE $\ft_s/R=3^{3/4}/\sqrt{2}$, and for sphere EE, $\ft_s/R=\sqrt{3/2}$. }\label{tab:d4tab}
	
\end{table}
\ewt

\subsection{Entanglement growth}

\begin{figure}[!h]
\begin{center}
\includegraphics[scale=0.6]{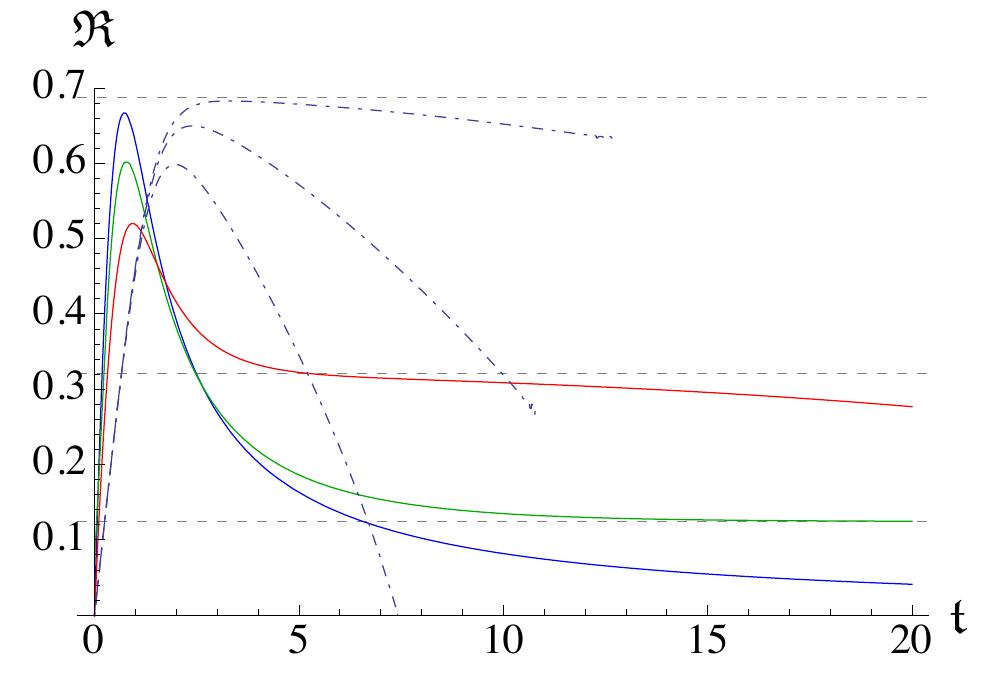} 
\includegraphics[scale=0.6]{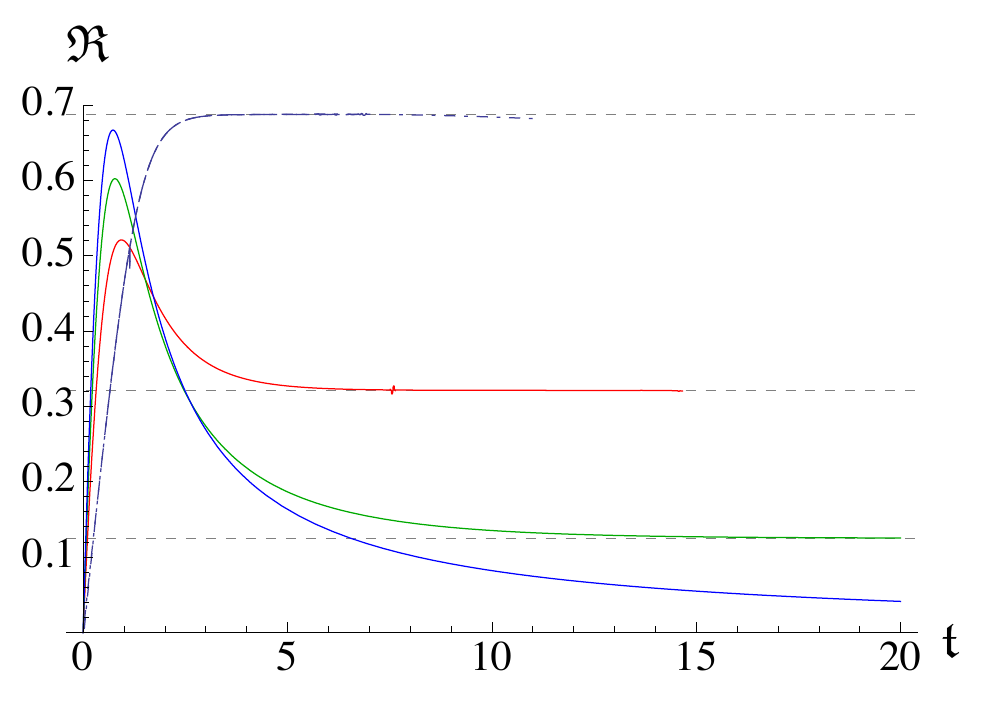}
\includegraphics[scale=0.6]{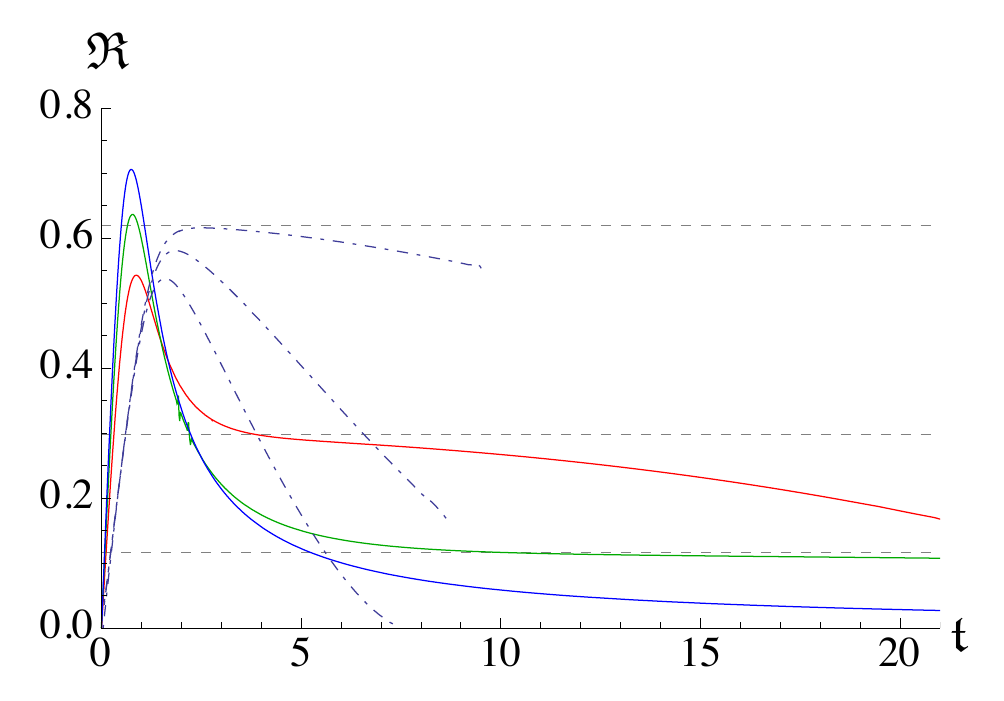}
\end{center}
\caption{$\fR_\Sig$ for $\Sig$ a sphere or strip, for Schwarzschild and RN black holes.
We use units in which the horizon is at $z_h =1$. 
{\it Upper}: $d=3$ and $\Sig$ a sphere. The dot-dashed curves are for the Schwarzschild black hole with $R = 7$, $13$, and $50$, respectively (larger values of $\ft$ for the $R=13, 50$ curves are not shown due to insufficient numerics), with the top horizontal dashed line marking $\Vee^{(\rm S)}$. 
Red, green, and blue curves are for the RN black hole with $(u =0.5, R=20)$, $(u=0.2, R=50)$, and $(u=0, R=50)$ respectively, and the two lower dashed horizontal lines mark $v_E$ for $u=0.5$ and $0.2$.  
{\it Middle}: For $d=3$ and $\Sig$ a strip. The dot-dashed curves are for the Schwarzschild black hole with $R=7, 12, 15$. It is interesting to note their evolution is essentially identical with the exception of different saturation times. The visible end of the dot-dashed curves coincides with discontinuous saturation for $R=7$. For $R=12$ and $15$ the curves have not been extended to saturation due to insufficient numerics. 
 The red, green, and blue curves are for the RN black hole with $(u=0.5, R=5)$, $(u=0.5, R=6)$, and $(u=0, R=6)$, respectively. The $u=0.5$ curve ends at saturation, but for $u=0.2$ and $0$, saturation happens at larger values of $\ft$ than shown.
{\it Lower}: For $d=4$ and $\Sig$ a sphere. The color and pattern scheme is identical to the upper plot, but the Schwarzschild curves are at $R=7$, $12$, and $50$, respectively, and $u=0.5$, $0.2$, $0$ curves are all at $R=20$.}\label{fig:frbe}
\end{figure}

It is interesting to compare the growth of entanglement entropy among different systems. 
For this purpose we need a dimensionless quantity in which the system size or total number 
of degrees of freedom has been factored out, since clearly for a subsystem with more degrees of freedom 
the entanglement entropy should increase faster. In~\cite{Liu:2013iza}, motivated by the linear growth~\eqref{line}
we introduced a dimensionless rate of growth
\be \label{groR}
\fR_{\Sig} (\ft) \equiv {1 \ov s_{\rm eq} A_{\Sig}} {d S_{\Sig} \ov d\ft} \ .
\ee
In the linear regime, $\fR_{\Sig}$ is a constant given by $\Vee$, while in the pre-local-equilibration 
regime $\ft \ll \eql$, from~\eqref{quda}, 
\be 
\fR_{\Sig} (\ft)  = {2 \pi \ov d-1} {\sE \ft \ov s_{\rm eq}} 
\ee 
grows linearly with time. In Fig.~\ref{fig:frbe} we give numerical plots of $\fR_\Sig$ for some examples.

In all explicit examples we studied, it appears that after local equilibration (i.e. after the linear growth regime has set in), $\fR_\Sig$ monotonically decreases with time. Given that we also found earlier that $v_E$ 
appears to have an upper bound at the Schwarzschild value~\eqref{emrp}, it is tempting to speculate that {\it after local equilibration}
\be \label{ineq2}
\fR_\Sig (\ft) \leq \Vee^{(\rm S)} \ .
\ee

Before local equilibration, the behavior of $\fR_\Sig$ appears to be sensitive to the initial state. In particular for a RN black hole with $\Sig$ a sphere or strip, we find $\fR_\Sig$ can exceed $\Vee^{(\rm S)}$ 
near $\eql$ (see Fig.~\ref{fig:frbe}). Also, for a highly anisotropic initial state, $\fR_\Sig$ could for a certain period of time resemble that of a $(1+1)$-dimensional system. As in $(1+1)$-dimensions $v_E^{(S)} = 1$, it then appears 
at best one can have 
\be \label{eorp1}
\fR_\Sig (\ft) \leq 1 \ .
\ee

It is clearly of great interest to explore more systems to see whether the inequalities \eqref{emrp}, \eqref{ineq2} and~\eqref{eorp} are valid, or to find a proof.

If true, the inequalities \eqref{emrp}, \eqref{ineq2} and~\eqref{eorp} may be considered as field theory generalizations of the small incremental entangling conjecture~\cite{bravyi} for ancilla-assisted entanglement rates in a {\it spin} system, which was recently proved in~\cite{franketal}. The conjecture states that ${dS \ov dt} \leq c ||H|| \log D$ where $S$ is the entanglement entropy between subsystems $aA$ and $bB$, $||H||$ is the norm of the Hamiltonian $H$ that generates entanglement between $A$ and $B$ ($a$, $b$ are ancillas), $D = {\rm min} (D_A, D_B)$ where $D_A$ is the dimension of the Hilbert space of $A$, and $c$ is a constant independent of $D$.
In our case, the Hamiltonian is local and thus couples directly only the degrees of freedom near $\Sig$--the analogue of $\log D$ is proportional to $A_\Sig$, and the entropy density $s_{\rm eq}$ in~\eqref{groR} can be seen as giving a measure of the density of excited degrees of freedom. 

\subsection{Tsunami picture: local propagation of entanglement}

In~\cite{Liu:2013iza} and Sec.~\ref{sec:ling} we discussed that the time evolution of $S_\Sig (\ft)$ 
suggests a picture of an entanglement wave front propagating inward from the boundary of the entangled
region. See Fig.~\ref{fig:eevelo}. 
We stress that at the level of our discussion so far this is merely a hypothetical picture to explain 
the time dependence of $S_\Sig (\ft)$. As mentioned earlier, from the field theory perspective, 
the existence of such an entanglement wave front may be understood heuristically as resulting from evolution 
under a local Hamiltonian. It would be very interesting to see whether is possible to ``detect'' such local propagation using other observables.  In the free streaming quasiparticle model of~\cite{speed}, the picture of an entanglement tsunami does emerge at early stages of time evolution in terms of propagating quasiparticles. But as the system evolves, in particular toward the late stage, the picture becomes more murky. 
 
On the gravity side it should be possible to make the tsunami picture more precise. It is tempting to 
interpret the black hole and pure AdS regions of the extremal surface as respectively corresponding to parts covered and not yet covered by the tsunami wave. The two bulk regions of the extremal 
surface are separated sharply at the collapsing shell and their respective sizes are controlled by 
the tip of the surface $z_t (\ft)$ and its intersection with the shell $z_c (\ft)$. It should be possible to 
describe the motion of the tsunami wave front in terms of these data.


\subsection{Application to black holes}

One striking feature of our results, which was also emphasized in~\cite{Hartman:2013qma,Maldacena:2012xp} in different contexts, is that the growth of entanglement entropy as well as the evolution of other nonlocal observables, such as correlation functions and Wilson loops,  is largely controlled by geometries inside the horizon of the collapsing black hole. In particular, the linear growth~\eqref{striplin2}--\eqref{line} is controlled by a constant-$z$ hypersurface inside the horizon while the memory loss regime discussed in Sec.~\ref{sec:mosp} 
is controlled by an extremal surface which asymptotes to the horizon from the inside.
In contrast, for a static eternal black hole an extremal surface whose boundary is at fixed time always lies outside the horizon~\cite{Hubeny:2012ry}.\footnote{While for correlation functions separated in the time direction it is possible to relate the geometry inside the horizon to certain features of boundary correlation functions via analytic continuation~\cite{Balasubramanian:1999zv,Louko:2000tp,Kraus:2002iv,Kaplan:2004qe,Fidkowski:2003nf,Festuccia:2005pi}, the relation is less direct.} 

The relation between entanglement growth and certain spatial hypersurfaces inside the horizon is 
tantalizing. In particular, possible bounds on $v_E$~\eqref{allnsv2} and the entanglement growth 
rate~\eqref{groR} impose nontrivial constraints on the geometry inside the horizon.


\vspace{0.2in}   \centerline{\bf{Acknowledgements}} \vspace{0.2in} We thank  M\'ark Mezei for many discussions, and thank E.~Berrigan, J.~Maldacena, V.~Hubeny, M.~Rangamani, B.~Swingle, T.~Takayanagi, J.~Zaanen for conversations. Work supported in part by funds provided by the U.S. Department of Energy
(D.O.E.) under cooperative research agreement DE-FG0205ER41360.

\appendix


\section{Equilibrium behavior of extremal surfaces} \label{app:equ}

{Here} we briefly review the behavior of $\Ga_{\Sig}$ in {a} black hole geometry, {corresponding to} the equilibrium behavior of various boundary observables. In {a} black hole geometry, an extremal surface
always lies outside the horizon~\cite{Hubeny:2012ry}, {i.e. denoting the location of the tip of $\Ga_{\Sig}$ by $z_b$,} $z_b < z_h$. {In our regime of interest $R \gg z_h$, $z_b$ is very close to the horizon, and we will write} 
\be \label{teipn}
z_b = z_h (1- \ep)\ , \quad \ep \ll 1 \ .
\ee 

\subsection{Strip}

{With $\Sig$ a strip, $R$ and $\sA$} in the black hole geometry can be obtained from~\eqref{raidu1} and~\eqref{aver3} by setting $E=0$ {($z_c=z_t$)} and $z_t = z_b$, 
\bea
\label{raidu2}
R & = &   \int_0^{z_b} {dz \ov \sqrt{h  \le({z_b^{2n} \ov z^{2 n}} -1\ri) }}\ ,  \\
\sA_{\rm eq} & = &  z_b^n \tilde K \int_0^{z_b} dz \,  {1 \ov z^{2n} \sqrt{h \le({z_b^{2n} \ov z^{2 n}} -1\ri) }} \ .
\label{theac}
\eea
{Thus we find that in the large $R$ limit, with $z_b$ given by~\eqref{teipn}}, 
\be \label{bher}
R =   -{1 \ov \ga_n} \log \ep + O(1)\ , \quad \ga_n \equiv  {1 \ov z_h} \sqrt{2 n h_1} \ , \quad 
h_1 \equiv - z_h h' (z_h)
\ee
and
\be 
\sA_{\rm eq} =- {\tilde K \ov z_h^n \ga_n} \log \ep + O(1)
= {L^n V_{\rm strip} \ov z_h^n} + O(R^0)
\ee
where $V_{\rm strip} = A_{\rm strip} R$ is the volume enclosed by the strip $\Sig$. 

\subsection{Sphere}

For {$\Sig$ a sphere}, the story is more complicated. One needs to solve the differential equation~\eqref{modeom} {with $E=0$} to find the relation between $z_b$ and $R$.  In the large $R$ limit, this can be done 
by matching an expansion near the horizon with an expansion near the boundary~\cite{Liu:2013una}. 
With {$z_b$ given by~\eqref{teipn}} one finds~\cite{Liu:2013una}
\be \label{sphrrp}
- \log \ep = \ga_n R - (n-1) \log R + O(R^0) 
\ee
and $z (\rho)$ can be written near the horizon as 
\be \label{eqnh}
z (\rho) = z_h - \ep z_1 (\rho) + {O\le(\ep^2\ri)}
\ee
with
\be \label{eqnhz1}
z_1 (\rho)  = {A} e^{\ga_n \rho} \rho^{- (n-1)} \le(1 + O(\rho^{-2}) \ri)
\ee
where $A$ is {some} constant. {Meanwhile, one finds that the leading contribution to the area of $\Ga_{\Sig}$, given by \eqref{bhc} with $E=\rho_c=0$, comes from near the horizon, and thus}
\bea 
{\sA_{\rm eq}} &= &K \int_{0}^R d \rho \, {\rho^{n-1} \ov z^{n}} {\sqrt{1 + {z'^2 \ov h}} } \nn
&=& {K R^n \ov n z_h^n} + \cdots  = {V_{\rm sphere} L^n \ov z_h^n} + \cdots
\eea
where $\cdots$ denotes terms {lower in the large $R$ expansion}. 

This behavior for a general shape $\Sig$ has been proved in~\cite{Liu:2013una}.


\section{Details in the saturation regime} 

\subsection{Strip} \label{app:act1}

Near saturation we expect both $z_c$ and $z_t$ {of $\Ga_{\Sig}$ to be close to $z_b$, where $z_b$ is the tip of the equilibrium $\Ga_{\Sig}$ with the same boundary $\Sig$, i.e. same $R$.} We thus write  
\be
z_c = z_t \le(1 - {\ep^2 \ov 2n}\ri) \ , \qquad z_t = z_b \le(1 + {\de \ov 2n} \ri) 
\ee
where both $\ep$ and $\de$ are small parameters. Then from~\eqref{stripnd} and~\eqref{eeval} we have 
\be 
E = - \ha g (z_t) \ep + O(\ep^3) \ .
\ee

{First, we} determine the relation between $\de$ and $\ep$ by equating~\eqref{raidu1} with~\eqref{raidu2}.   
For this purpose it is convenient to write~\eqref{raidu2} as 
{
\be \label{raidu4}
R =F(z_b)\ , \qquad F(z_b)\equiv  \int_0^1 {dy \ov \sqrt{y^{-2n}-1}} {z_b \ov \sqrt{h(z_b y)}}\ .
\ee }
To expand~\eqref{raidu1} in terms of $z_t - z_b$ and $E$, {we write it as}
\be \label{raidu3}
R = A_1 - A_2 + A_3 + F(z_t)
\ee 
where 
\be \label{sooe}
A_1 = \int_{z_c}^{z_t} {dz  \ov  \sqrt{{z_t^{2n} \ov z^{2 n}} -1}} \ , \quad A_2 =  \int_{z_c}^{z_t} {dz \ov \sqrt{h  \le({z_t^{2n} \ov z^{2 n}} -1\ri) + E^2   }} \ ,
\ee
and 
\be \label{sooe1}
A_3 = 
  \int_0^{z_t} dz \le({1 \ov \sqrt{h  \le({z_t^{2n} \ov z^{2 n}} -1\ri) + E^2   }} - {1 \ov \sqrt{h  \le({z_t^{2n} \ov z^{2 n}} -1\ri) }}\ri)\ .
\ee
For small $\ep$ we {find that $A_1$, $A_2$, and $A_3$} have the expansions 
\be \label{A12exp}
A_1 = {z_t \ep \ov n} \le(1 + O(\ep^2) \ri)\ , \qquad A_2 = {z_t \ep \ov n} \le(1 + O(\ep^2) \ri)\ ,
\ee
and 
\be 
A_3 = - {1 \ov 2n} {z_t g (z_t) \ov h (z_t)} \ep + O(\ep^2) \ ,
\ee
{where in \eqref{A12exp} we used $h(z)=1-g(z)$. Then equating~\eqref{raidu4} and~\eqref{raidu3}, we have}
\be 
\de = {g (z_b) \ov h (z_b) F'(z_b)} \ep + O(\ep^2)\ .
\ee

{Next,} let us look at~\eqref{biute} which can be written as 
{
\be 
\ft -\ft_s=  B_1 + B_2  - B_3
\ee
}
where 
\bea
B_1& =& \int_{z_b}^{z_c} {dz \ov h(z)} \ , \quad B_2 =  \int_0^{z_t} {dz \ov h(z)} {E \ov \sqrt{h \le({z_t^{2n} \ov z^{2 n}} -1\ri) + E^2}}\ , \nn
 B_3 &=&  \int_{z_c}^{z_t} {dz \ov h(z)} {E \ov \sqrt{h \le({z_t^{2n} \ov z^{2 n}} -1\ri) + E^2}}\ .
\eea
{The integrals can be expanded in small $\ep$ as}
\bea
B_1 &=& {1 \ov h(z_b)} {z_b \ov 2n} \de + O(\ep^2) \ , \quad B_2 = H(z_t) E + O(\ep^2)\ , \nn B_3 &=& {E \ov h (z_b)} {z_t \ep \ov n} + O(\ep^3)\ ,
\eea
with 
\be 
H(z_t) \equiv z_t  \int_0^{1} {dy \ov h(z_t y)} {1 \ov \sqrt{h (z_t y) (y^{-2n}-1)}} \ .
\ee
Since $B_3 \sim O(\ep^2)$, {we find} 
\be \label{enrm2}
\ft - \ft_s = u_1 \ep + {O\le(\ep^2\ri)} 
\ee
where 
\be 
u_1 = \ha  g(z_b)  \le({z_b \ov n h^2 (z_b) F'(z_b)} - H (z_b) \ri)   \ . 
\ee

Now let us look at the area {of $\Ga_{\Sig}$}. The area {of the equilibrium $\Ga_{\Sig}$~\eqref{theac}} can be written as 
\be 
{1 \ov \tilde K} \sA_{\rm eq} = G (z_b) \ , \quad G(z_b) \equiv z_b^n  \int_0^{z_b} dz \,  {1 \ov z^{2n} \sqrt{h \le({z_b^{2n} \ov z^{2 n}} -1\ri) }}  \ .
\ee
{The area of $\Ga_{\Sig}$ itself~\eqref{aver1} can be written as}
\be 
{1 \ov \tilde K} \sA = C_1 - C_2 + C_3 + G(z_t) 
\ee
where 
\bea
C_1 &=& z_t^n  \int^{z_t}_{z_c} dz \, {1 \ov z^{2n} \sqrt{{z_t^{2n} \ov z^{2n}}-1}}\ , \\
C_2 &=& z_t^n \int^{z_t}_{z_c} dz \,  {1 \ov z^{2n} \sqrt{h (z) \le({z_t^{2n} \ov z^{2n}}-1\ri) + E^2}} \\ 
C_3 &= &z_t^n \int^{z_t}_{0} {dz \ov z^{2n} }\, \le( {1 \ov \sqrt{h (z) \le({z_t^{2n} \ov z^{2n}}-1\ri) + E^2}} \ri. 
 \cr
&& \quad
\left. -{1 \ov \sqrt{h (z) \le({z_t^{2n} \ov z^{2 n}} -1\ri) }} \ri)\ .
\eea
To leading order the {expansion} of the above quantities is the same as that for~\eqref{sooe}-\eqref{sooe1},
\bea
C_1 &=& {z_t^{1-n} \ep \ov n} +O (\ep^3) \ , \qquad C_2 =  {z_t^{1-n} \ep \ov n} + O(\ep^3)\ , \nn
C_3 &=& - {1 \ov 2n} {z_t^{1-n} g (z_t) \ov h (z_t)} \ep + O(\ep^2)\ .
\eea
{Thus we} find that 
\bea \label{acry}
{1 \ov \tilde K} \le(\sA - \sA_{\rm eq} \ri) = {z_b^{1-n} g (z_b) \ov 2n h(z_b)} \le({z_b^n G'(z_b) \ov F'(z_b)} - 1\ri)\ep + O(\ep^2) \ .\nn
\eea
Note that while $G (z_b)$ is a divergent integral (i.e. depends on a cutoff at small $z$), 
{$G'(z_b)$ 
should have a well defined limit when the cutoff is taken} to zero. 
{In fact, in \eqref{acry} the coefficient of the $O(\ep)$ term is identically zero, which can be seen by writing} $G'(a)$ and $F'(a)$ as 
\bwt
\bea
&&a^n G'(a)=\lim_{\de \to 0} \le(- n \int_0^{1- \de} {y^n \ov (1 - y^{2n})^{3 \ov 2} \sqrt{h (ay)}}+ { y^{1-n} \ov \sqrt{h (a y) (1- y^{2n})}} \biggr|_{1-\de} \ri)\ ,\nn
&&F'(a) =  \lim_{\de \to 0} \le(- n \int_0^{1-\de} dy {y^n \ov  (1 - y^{2n})^{3 \ov 2} \sqrt{h (ay)}}+ {y^{1+n} \ov \sqrt{h(ay) (1-y^{2n})}} \biggr|_{1-\de} \ri)\ ,
\eea
\ewt
{from which we confirm that}
\be
a^n G'(a) = F' (a)
\ee
for any $h(z)$. However, one can check that {the $O(\ep^2)$ term in \eqref{acry} (whose coefficient is rather long and which we will not give here) is generically nonzero.}

\subsection{Sphere} \label{app:sea}

{Let $z_0(\rho)$ correspond to the equilibrium $\Ga_{\Sig}$ and denote the location of its tip as $z_b$. Then near saturation, $z(\rho)$, corresponding to the black hole portion of the actual $\Ga_{\Sig}$, can be obtained by perturbing $z_0(\rho) $,} 
\be \label{dkee}
z(\rho)=z_0(\rho)+\delta z_1(\rho)+ \delta^2 z_2 (\rho)+ \cdots
\ee
where $\de$ is a small parameter which {we will obtain precisely later on}. Note that near the boundary, $z_n$ should satisfy the boundary condition 
\be \label{uvbd}
z_n (R) = 0 \ , \quad n=1,2, \cdots
\ee
They should also satisfy the boundary condition~\eqref{ntrn1} at the shell, order by order. For small $\de$,  $z_c$ and $z_t$ are close to $z_b$, and  $\rho_c = \sqrt{z_t^2 - z_c^2}$ and $E$ are all small. It is convenient to introduce another small parameter $\ep$ by 
\be \label{cjo}
\rho_c = z_c \ep  
\ee
{after which from~\eqref{varp00}--\eqref{varp1},}
\bea \label{cjo1}
z_t &=& z_c \le(1 + {\ep^2 \ov 2} + O(\ep^4)+\cdots \ri)\ ,  \nn
E &=& - { \ep^n \ov 2} {g (z_c) \ov z_c} \le(1 + O(\ep^2) +\cdots\ri) \ .
\eea

{Note that specifying $R$ and $\ep$ fixes $\Ga_{\Sig}$ entirely. Thus} we can {expand $\ft - \ft_s$ and $\sA - \sA_{\rm eq}$ in terms of $\ep$,} then $\sA - \sA_{\rm eq}$ in terms of $\ft - \ft_s$. {In order to do so we first need to relate $z_c - z_b$ and $\de$ 
to $R$ and $\ep$. This requires solving for $z_1$ near $\rho_c$ by expanding it as a power series in small $\rho$, but only after imposing the boundary condition \eqref{uvbd} at $z=0$. We leave the detailed analysis of $z_1$ to Appendix~\ref{app:solu}, and for now merely list the results. We find that for $n=2$,} 
\bea \label{ren1}
\de&=& - {g (z_b) z_b \ov 2 r_2} \ep^2 + O( \ep^4 \log \ep)+\cdots\  , \nn 
z_c&=& z_b (1 + c_1 \ep^2 \log \ep + c_2 \ep^2 + \cdots)\ ,
\eea
with
\be 
c_1 = - {g(z_b) \ov 2}\ , \quad c_2 = -{g(z_b) \ov 2r_2} (r_1 - r_2 + r_2 \log z_b) - \ha\ ,
\ee
and for $n>2$,
\be \label{ren3}
\de= {g(z_b) z_b^{n-1} \ov 2 (n-2) r_2} \ep^n + O(\ep^{n+2})+\cdots \ , \quad 
z_c = z_b (1 + d_2 \ep^2  + \cdots)\ ,
\ee
with
\be \label{ren4}
d_2 =  {n-1 \ov 2 (n-2)} g(z_b) - \ha \ . 
\ee
In the above equations {$r_1$ and $r_2$ are numerical constants that we define in~\eqref{0bd}}. 
Note that $z_t > z_b$ while $z_c$ {does not have to be greater than $z_b$}.

{Now let us look at the boundary time~\eqref{texpl}, writing it as}
\be 
\ft = \ft_1 + \ft_2 
\ee
with 
\be 
\ft_1 = -  \int_{\rho_c}^R d \rho \,   {z' \ov h}\ , \qquad \ft_2 = \int_{\rho_c}^R d \rho \,  {EB \sqrt{1+ z'^2 \ov h(z)} \ov h \sqrt{1 + {B^2 E^2 \ov h}}} \ .
\ee
{Note that $\ft_1$ can} be written as
\be \label{ft11}
\ft_1 = \int_0^{z_c} {dz \ov h(z)} =  \ft_s + \int_{z_b}^{z_c} {dz \ov h} = \ft_s + {z_c - z_b \ov h (z_b)} + \cdots
\ee
where in the second equality we have used that $z_c - z_b$ is small. {Meanwhile, from~\eqref{cjo1} $E \sim O(\ep^n)$, and to leading order in small $\ep$} $\ft_2$ can be evaluated by replacing $z$ in its integrand by the 
equilibrium solution $z_0$. The resulting integral receives the dominant contribution from its lower end, {and we have}
\be \label{f21}
\ft_2 = {E z_b^{n} \ov h (z_b)} \begin{dcases} - \log (z_b \ep)  + I_0 +\cdots& n=2 \\
                                                         {(z_b \ep)^{2-n} \ov n-2} +\cdots& n>2
                                                         \end{dcases}
\ee
where 
\be 
I_0 = \lim_{\rho_c \to 0} \le({h(z_b) \ov z_b^{2}} \int_{\rho_c}^R d \rho \,  
{z^{2}_0 \ov \rho} {\sqrt{1 + {z_0'^2 \ov h(z_0)} } \ov h(z_0) } + \log \rho_c  \ri) 
\ee
and we have replaced $z_c$ by $z_b$ wherever it appears. Collecting~\eqref{ft11},~\eqref{f21} and using~\eqref{ren1},~\eqref{ren3}, we  find that 
{\be
\ft - \ft_s=\begin{dcases}
 - \ha \le(z_b +  {g(z_b) z_b \ov h(z_b)} \le({b_1 \ov b_2}  + I_0 \ri) \ri) \ep^2 
+ \cdots  & n=2 \\
  - \ha z_b \ep^2 + \cdots & n>2 
  \end{dcases}\ .
  \ee}

{Next, we proceed to compute the area of $\Ga_{\Sig}$ given by}~\eqref{adsc} and~\eqref{bhc}. The AdS {portion can be easily expanded as}
\be \label{adsc1}
{1 \ov K} \sA_{AdS} = {\ep^{n} \ov n} - { \ep^{n+2} \ov 2 (n+2)}  +  \cdots \ ,
\ee
{while} the black hole {portion} can be written as 
\be 
\sA_{\rm BH} = \sA_1 + \sA_2 + O(E^4)+\cdots
\ee
with 
\bea
{1 \ov K} \sA_1 &=& \int_{\rho_c}^R d \rho \, \sL_0 (z,z') \equiv  \int_{\rho_c}^R d \rho \, {\rho^{n-1} \ov z^{n}} \sqrt{1 + {z'^2 \ov h}}\ , \nn
{1 \ov K} \sA_2 &=& -\ha E^2  \int_{\rho_c}^R d \rho \, {z^{n} \ov \rho^{n-1} h} \sqrt{1 + {z'^2 \ov h}} \ .
\eea
Since $\sA_2$ is multiplied by $E^2 \sim O(\ep^{2n})$, it can be computed by replacing {$z$ with $z_0$
in its integrand, and we find the leading order results
\be  \label{bhc2}
{1 \ov K} \sA_2 =-{ E^2 \ov 2} {z_b^{n} \ov h (z_b)} \begin{dcases} - \log (z_b \ep)  + I_0 +\cdots & n=2 \\
                                                         {(z_b \ep)^{2-n} \ov n-2}+\cdots & n > 2
                                                   \end{dcases}\ .
\ee}
To compute $\sA_1$, we consider {the variation of $\sL$ under a variation about the equilibrium solution $z = z_0 + \De z$, which gives}
\be \label{nroe}
{1 \ov K} \sA_1 = \int_{\rho_c}^R d \rho \, \sL (z_0, z_0') -  \Pi (z_0)  \De z \bigr|_{\rho_c}   + \cdots, 
\quad \Pi = {\p \sL_0 \ov \p z'}  \ .
\ee
Note that in~\eqref{nroe} there is also a potential boundary term at $\rho=R$, but {that it is zero due to $z$ and $z_0$ both ending at $\rho=R$}.\footnote{This boundary term at $z=0$ should be treated with some care as $\Pi$ is divergent there.} 
{Meanwhile,}
\bea
\int_{\rho_c}^R d \rho \, \sL (z_0, z_0') &=& {1 \ov K} \sA_{\rm eq} -{{1 \ov n}\le( {\ep \ov  \sqrt{1+\ep^2}}\ri)^n} \le({z_t \ov z_b}\ri)^{n}\nn
&&- {(n+1) \, h(z_b) \ep^{n+2}\ov 2 (n+2)} + \cdots
\eea
and 
\bea
\Pi (z_0) \bigr|_{\rho_c} &=& - {\rho_c^{n} \ov z_b^{n+1}} = - {\ep^{n} \ov z_b}\ , \nn
 \De z_c &=& z_c - z_0 (\rho_c) = z_c - z_b - \ha z_0' (0) \rho_c^2 + \cdots \nn
\eea
{where} the small $\rho$ expansion of $z_0$ is given in equation~\eqref{eqiex}.  
Finally, collecting all the results above 
we have
\bea
&&\sA- \sA_{\rm eq} \nn
&&=\begin{dcases}
K {g^2(z_b) \ov 8 h(z_b)}\ep^4 \log \ep + O(\ep^4)+\cdots  & n=2 \nn
 -K {g(z_b) \ov 2 (n-2)}\le({n-2 \ov n+2}+{g(z_b) \ov 4 h(z_b)}\ri)\ep^{n+2}+\cdots  & n>2
 \end{dcases}\ .
\eea

\subsection{Discussion of $z_1$ when $\Sig$ is a sphere} \label{app:solu}

Here we give the derivation of~\eqref{ren1}--\eqref{ren4}. To first order in $\de$, $z_1$ satisfies the equation of motion
\be \label{z1eom}
z_1'' + p(\rho) z_1' + q(\rho) z_1 = s (\rho)
\ee
where
\bea
&&p(\rho)=z_0'\le({2 n \ov z_0} -{h'_0 \ov h_0}\ri)+{(n-1)\ov \rho}\le(1+{ 3 z_0'^2 \ov h_0}\ri)\ , \\
&&q(\rho)=h_0'\le({n \ov z_0}+{(n-1)z_0' \ov h_0 \rho}\ri)\nn
&&+{n \ov z_0}\le(1+{z_0'^ 2\ov h_0}\ri)\le(h_0'-{h_0 \ov z_0}\ri)+{1 \ov h_0}\le(h_0' z_0''-{1 \ov 2}h_0'' z_0'^2\ri)\ ,\nn
\eea
and
\be 
s (\rho) = - {E^2 \ov \de} {z_0^{2 n} \ov h(z_0) \rho^{2 (n-1)}} \le(z''_0  + {\p_z h (z_0) \ov 2}  \ri) \ .
\ee
While the full analytic solution $z_0$ is not known, its behavior near $\rho=R$ and $\rho=0$ can be obtained by series expansions and {the same applies to functions $p $ and $q$ - this is sufficient for our purposes. Now, near the tip $\rho=0$,} 
\bea \label{eqiex}
z_0 (\rho) &=& z_b - {h(z_b) \ov 2 z_b} \rho^2 \nn
&&+ {h(z_b) ((n+1) z_b h'(z_b)- (n+2 ) h(z_b) ) \ov 8 (n+2) z_b^3} \rho^4 + O(\rho^6) \nn
\eea
while {near the boundary $\rho=R$ with $\sig\equiv R-\rho \ll 1$,}\footnote{In terms of $\rho (z)$ we have the expansion
\be 
\rho (z) = R - {z^2 \ov 2R} + \cdots 
\ee
where {the expansion is identical to that in AdS until the $O(z^n)$ term whose coefficient is undetermined.}}
\be 
z_0 (\rho) = \sqrt{2 R \sig} + \bca O(\sig) & n=2 \cr
                      O(\sig^{3 \ov 2}) & n > 2
                      \eca\ .
\ee
Then we find that the leading terms in $p$ and $q$ are given by:

\ben
\item Near $\rho=0$,
\bea
p(\rho)&=&{n-1 \ov \rho}+{\le((n-3)h(z_b)+z_b h'(z_b)\ri) \ov z_b^2}\rho+O\le(\rho^3\ri),\nn
q(\rho)&=&{n\le(-h(z_b)+z_b h'(z_b)\ri) \ov z_b^2}+O\le(\rho^2\ri) \ .
\eea

\item Near $\rho=R$, 
\bea
p(\rho)&=&{n-3 \ov 2 \sig} + \bca O(\sig^{-\ha}) & n =2 \cr
                                                      O(1) & n > 2 
                                                      \eca \ ,\nn
q(\rho)&=&-{n \ov 4 \sig^2} + \bca O(\sig^{-{3 \ov 2}}) & n =2 \cr
                                                      O(\sig^{-1}) & n > 2 
                                                      \eca  \ .
\eea
\een

Let us first look at the homogenous part of equation~\eqref{z1eom}. Near $\rho=R$,  it is convenient to {work with a basis of solutions given by the expansions}
\bea
k (\rho) &=& (R- \rho)^{n\ov 2} \le (1 + O((R- \rho)^\ha) \ri)\ , \nn
\tilde k (\rho) &=& (R- \rho)^{-\ha} \le (1 + O((R- \rho)^\ha) \ri) \ ,
\eea
while near $\rho =0$, it is more convenient to {work with a basis of solutions given by the expansions} 
\bwt
\bea \label{0bd1}
g_1 (\rho)= 1 + \sum_{m=1}^\infty g_{1m} \rho^{2m}\ , \quad
g_2(\rho) = \bca 
 g_0 \log \rho\le( 1+\sum_{m=1}^{\infty} g_{2m}\rho^{2m}\ri)+ \rho^{-(n-2)} \sum_{m=0}^{'\infty}\tilde g_{2m}\rho^{2m}
 & n \; {\rm even}  \cr 
 \rho^{-(n-2)} \le( 1+\sum_{m=1}^{\infty} g_{2m}\rho^{2m}\ri) & n \; {\rm odd}
 \eca \ ,
\eea
\ewt
where in {$g_2$ for even $n$ the prime in $\sum'$ indicates that the sum does not include $m= (n-2) / 2$.}
Since we are dealing with a linear equation, the two {bases} are related by linear superposition, 
{\be \label{0bd}
k (\rho) = r_1 g_1 (\rho) + r_2 g_2 (\rho) \ , \quad \tilde k (\rho) = \tilde r_1 g_1 (\rho) + \tilde r_2 g_2 (\rho)\ ,
\ee
where $r_1,r_2, \tilde r_1, \tilde r_2$ are constants which can be evaluated} numerically.

{Now we consider a particular solution to the full inhomogeneous equation~\eqref{z1eom},}
\bwt
\bea
z_s(\rho) =   k (\rho)\int^{R}_{\rho} d\rho'\,{s(\rho') \tilde k (\rho') \ov W_k (\rho')} - \tilde k (\rho)
\int^{R}_{\rho}d\rho'\,{s(\rho') k (\rho') \ov W_k (\rho')} 
 =  g_2 (\rho)\int^{R}_{\rho} d\rho'\,{s(\rho') g_1 (\rho') \ov W_g (\rho')}- g_1 (\rho)
\int^{R}_{\rho}d\rho'\,{s(\rho') g_2 (\rho') \ov W_g (\rho')}\nn
\eea
\ewt
{where
\be
W_k (\rho) =  k \tilde k' - \tilde k k'   \sim (R - \rho)^{n-3 \ov 2} \quad {\rm as} \quad \rho \to R
\ee
and
\be
W_g (\rho) = g_1' g_2 - g_2' g_1   \sim \rho^{- (n-1)} \quad {\rm as} \quad \rho \to 0\ .
\ee
Noting that} 
\bea
s (\rho_c) &=& - {E^2 \ov \de} {z_n^{2n-1} (z_b h'(z_b) - 2 h (z_b) )\ov 2 h(z_b) \rho_c^{2 (n-1)}} \ , \nn
s (\rho \to R) &=& {E^2 \ov \de} R \le({2 \sig \ov R}\ri)^{n-{3 \ov 2}} + \cdots
\eea
{we find} 
\be 
z_s (\rho) \sim {E^2 \ov \de} \sig^{n + \ha}  \qquad  {\rm as} \qquad \rho \to R\ .
\ee
As {is shown after the matching in \eqref{cjo1}, \eqref{ren1}, and \eqref{ren3}, ${E^2 \ov \de} \sim \de$ and thus it is consistent to ignore $z_s$ near $\rho=R$. However, it cannot be ignored near $\rho=0$ because the source term $s$ becomes singular.} 
We find that as $\rho \to 0$, {$z_s$ has the behavior} 
\be
z_s (\rho) \sim \begin{dcases}
{E^2 \ov \de}   (\log \rho)^2  + \cdots &n=2 \\
 {E^2 \ov \de} \rho^{- 2 (n-2)} +\cdots & n>2
 \end{dcases}\ .
\ee

{The actual matching that results in \eqref{ren1}--\eqref{ren4} is performed as follows. The boundary condition~\eqref{uvbd} requires us to choose $k (\rho)$ near $\rho=R$. Then near $\rho=0$}, $z_1$ can be written as {
\be \label{0bd3}
z_1 = r_1 g_1 (\rho) + r_2 g_2 (\rho)  + z_s \ .
\ee}
Plugging in~\eqref{cjo} and~\eqref{cjo1} into~\eqref{0bd3} and \eqref{0bd1}, {we obtain $z_b$ and $\de$
in terms of  $z_t$ and $\ep$ as in~\eqref{ren1}--\eqref{ren4}. We note that at leading orders $z_s$} does not contribute.

\section{Details in the memory loss regime for $\Sig$ a sphere} \label{app:gensca}

{Here} we give the equations underlying~\eqref{solI} and~\eqref{sol3}, {and}
the derivation of~\eqref{teie} and~\eqref{aaep}. {Recall the expansion parameter and expansion given in \eqref{gscep} and \eqref{poec}.}

\subsection{Critical extremal surface}

Let us first examine in some detail the asymptotic behavior of $z^* (\rho)$ for $\rho \gg \rho_c$ where it approaches the horizon. Letting
\be \label{eor}
z^* (\rho)= z_h + \chi_* (\rho)
\ee
with $\chi_*$ small and requiring it to decrease with increasing $\rho$, we find that $z^*$ has the asymptotic behavior
\be
\label{assol}
\chi_* (\rho) = z^* (\rho) - z_h = {\al \ov \rho^{n-1}}+ {\al_1 \ov \rho^n} + {\al_2 \ov \rho^{n+1}} + \cdots, \quad \rho \gg \rho_c 
\ee
where 
\be
\al ={|E| z_h^{n+1} \ov \sqrt{2n h_1}}\ , 
\qquad \al_1 = \de_{n,2}{5 E^2 z_h^5 \ov 8 h_1} \ .
\ee
Here we have used the notation
\be \label{hnot}
h_1\equiv -z_h h'(z_1) \ , \quad h_2\equiv z_h^2 h''(z_h)\ , \quad \cdots
\ee
Note that $\al$ is positive, i.e. $z^*$ approaches the horizon from above, or inside. The leading two terms in~\eqref{assol} can be obtained by equating the two most dominant terms in~\eqref{modeom} as $\rho \to \infty$, i.e. (note $B$ is defined in~\eqref{usegp})
\be 
{n h^2 \ov z} + \ha E^2 B^2 \p_z h =0 \quad \to \quad \chi^2_* = {\al^2  \ov \rho^{2 (n-1)}} {z^{2n+1} \ov z_h^{2n+1}} \ ,
\ee
while in order to obtain terms of $O(\rho^{-(n+1)})$ and higher in \eqref{assol}, one needs to take into higher-order terms in~\eqref{modeom}. Note the leading term in~\eqref{assol} can also be written as
\be 
\chi_* = {1 \ov \ga_n} |E| B + O\le(\rho^{-n}\ri) 
\ee
or
\be
h (z^* (\rho)) = - c_n |E| B+{O\le(\rho^{-n}\ri)}
\ee
where 
\be  \label{cndef}
\ga_n = {1 \ov z_h} \sqrt{2n h_1}\ , \qquad c_n = \sqrt{h_1 \ov 2n} = {h_1 \ov z_h \ga_n}\ .
\ee

Let us now calculate $v^*(\rho;\rho_c)$ and $\sA^* (\rho;\rho_c)$ corresponding to $z^* (\rho; \rho_c)$, where we have traded $z_t$ for $\rho_c$ and made explicit in our notation that $\rho_c$ is the only parameter. Evaluating~\eqref{usegp} on $z^*$, for large $\rho$  we find that
\be 
v^{*'}=  {1 \ov c_n} - \le({n-1 \ov 4 \pi T}-\de_{n,2}{|E|(3 h_1+h_2) z_h^2 \ov 2 h_1^2}\ri){1 \ov \rho} + O(\rho^{-2})\ ,
\ee 
from which
\bea \label{eooo0}
&&v^*(\rho; \rho_c) \nn
&&=  {\rho \ov c_n} - \le({n-1 \ov 4 \pi T}-\de_{n,2}{|E|(3 h_1+h_2) z_h^2 \ov 2 h_1^2}\ri)\log \rho + O(1)\ .\nn
\eea
Note that the leading term, and for $n>2$ the next-to-leading term also, are independent of $\rho_c$. 
Similarly, evaluating the integrand of~\eqref{areadf0} on $z^*$, we find
\be \label{lagje}
\sL^* = {\rho^{n-1} \ov z_h^n} + \de_{n,2}{E^2(3h_1+h_2)z_h^2 \ov 4 h_1^2}{1 \ov \rho}  + O(\rho^{-2}) 
\ee
from which 
\be \label{eooo2}
{1 \ov K} \sA^* (\rho; \rho_c) = {\rho^n \ov n z_h^n}  + \de_{n,2}{E^2(3h_1+h_2)z_h^2 \ov 4 h_1^2} \log \rho + O(1)\ .
\ee
The leading coefficients are again independent of $\rho_c$ and there is a logarithmic term only for $n=2$. 

\subsection{Equations}

We now examine the equation for $z_1 (\rho)$ as introduced in~\eqref{poec}. 
Let us first look at the {region in which $\chi^*  \gg \ep z_1 $. Plugging~\eqref{poec} into~\eqref{modeom} we find that $z_1$ satisfies a linear differential equation 
\be \label{z1eom1}
z_1'' + p_1 (\rho) z_1' + p_2 (\rho) z_1 =0
\ee
where $p_1$ and $p_2$ are some complicated functions of $\rho$, expressed via $\chi^*(\rho)$ and $h (z_h+\chi^*(\rho))$. They have the large $\rho$ expansions
\bea
&&p_1 (\rho)={a_1 \ov \rho} + {a_2 \ov \rho^2} + \cdots\nn
&&a_1=2 (n-1)\ , \quad a_2 = \de_{n,2 }{|E| z_h^2 \ov 4 h_1^{3 \ov 2}}(13 h_1 - h_2)\nn
 \eea
and\footnote{For Schwarzschild $h(z)$, $b_1 < 0$ while for RN $h(z)$, $b_1>0$ for sufficiently large charge density.}
\be
p_2(\rho)= - \ga^2_n + {b_1 \ov \rho} + {b_2 \ov \rho^2} + \cdots \ , \quad b_1 =\de_{n,2}  {|E| (h_2 -h_1) \ov \sqrt{h_1}}\ .
\ee
Equation~\eqref{z1eom1} can then be solved in terms of an expansion 
\be \label{sol11}
z_1(\rho) = A_1 e^{\ga_n \rho} \rho^{- \beta_n} \le(1 + {c_{11} \ov \rho} + O\le(\rho^{-2} \ri) \ri) +  \cdots 
\ee
with
\bea \label{dgje}
\beta_n &=& n-1 + {b_1 \ov 2 \ga}\ , \nn
c_{11} &=& {1 \ov 8 \ga^3} \le(b_1^2 + 2 \ga b_1  + (2a_1 - a_1^2 +4 b_2) \ga^2 + 4a_2 \ga^3 \ri)\ , \nn
\eea
where $A_1 (\rho_c)$ {is a positive} $O(1)$ constant determined by boundary conditions~\eqref{bdma} at $\rho_c$, {and in~\eqref{sol11} we have} suppressed terms that are {exponentially small, i.e. those proportional to $e^{-\ga \rho}$.}

{In the region in which $\chi^*  \ll \ep z_1 \ll 1$, we can plug in~\eqref{poec} into~\eqref{modeom} while ignoring $\chi^* $ and terms in~\eqref{modeom} proportional to $E^2$. We then} find a nonlinear equation for $z_1$, 
\be \label{eope}
 {z_1'' \ov z_1} -\ha {z_1'^2 \ov z_1^2} + {n-1 \ov \rho} {z_1' \ov z_1} - {\ga^2 \ov 2}= 0 \ ,
 \ee 
which {has the solution}
\bea \label{sol13}
z_1(\rho)&=&   \rho^{-(n-2)} \le( \# I_{n-2 \ov 2} \le(\ha \ga \rho\ri)  + \# K_{n-2 \ov 2} \le(\ha \ga \rho\ri) \ri)^2\nn
&=&A_2 e^{\ga \rho} \rho^{-(n-1)}\le(1+O\le(\rho^{-1}\ri)\ri) + \cdots   %
\eea
where we have again suppressed exponentially small terms.



\subsection{Time}

{In this subsection and the next, for purposes of clarity, will use a new symbol to denote the polynomial part of the large $\rho$ limit of $z^*$,}
\be
P(\rho)\equiv \chi^*(\rho) \ , \qquad z^*(\rho)=z_h+P(\rho)\ .
\ee

{Recall the labeling of regions I, II, and III given near \eqref{regime1} and \eqref{regime2}. Delineating the regions more explicitly, the boundary time can be divided as}
\bea \label{btdiv}
\ft =v(R)&=&t_{\rm I}+t_{\rm II}+t_{\rm III}\nn
&\equiv&\le(\int_{k_0}^{R-k_1}+\int_{R-k_1}^{R-k_2}+\int_{R-k_2}^{R-k_3}\ri) d\rho  \, v'\nn
\eea
where the first equality holds up to $O(1)$ terms and
\bea
k_1&=&{1 \ov \ga_{n}}\le(n-1-{b_1 \ov 2 \ga_{n}}\ri)\log R+C_1 \ , \nn
k_2&=&{1 \ov \ga_n}\le(n-1\ri)\log R-C_2\ .
\eea
Here $C_1$, $C_2$, $k_0$, $k_3$ are all positive $O(1)$ constants and $k_0$ must be chosen sufficiently large that large $\rho$ expansions apply in region I. We now proceed to calculate \eqref{btdiv}, recalling \eqref{usegp}
\be
v'={1 \ov h}\le(-z'+EB \sqrt{Q}\ri)\ , \quad Q={1+{z'^ 2\ov h} \ov 1+{E^2 B^2 \ov h}}\ .
\ee

\subsubsection{Region I}
Here
\bwt
\be \label{exp11}
z=z_h+P+O\le(\ep z_1\ri)+\cdots\ , \quad z'=O\le({P \ov \rho}\ri)+O\le(\ep z_1\ri)+\cdots
\ee
\bea \label{exp12}
1+{z'^2 \ov h}=1+O\le({P \ov \rho^2}\ri)+O\le(\ep z_1 \ov \rho\ri)+\cdots\ , \quad
1+{E^2 B^2 \ov h}=1-{\ga_n \ov c_n}P\le(1-{2 \al_1 \ov \al}{1 \ov \rho}+\le(2n+{h_2 \ov2 h_1}\ri){P \ov z_h}\ri)+O\le(\ep z_1\ri)+\cdots\nn
\eea
\ewt
Then 
\bwt
\bea
v'&=&{1 \ov c_n}\le(1-\le({n-1 \ov \ga_n}+{\al_1 \ov \al}\ri){1 \ov \rho}+\le(n+{h_2 \ov2 h_1}+{\ga_n z_h \ov 2 c_n}\ri){P \ov z_h}\ri)+O\le({1 \ov \rho^2}\ri)+\le({\ep z_1 \ov P}\ri)+\cdots
\eea
\ewt
from which
\be \label{t1}
t_{\rm I}={R-k_1\ov c_n}-\le({n-1 \ov 4 \pi T}-\de_{n,2}{|E|(3 h_1+h_2)z_h^2 \ov 2 h_1^2}\ri)\log R+O(1)\ .
\ee
Comparing with \eqref{eooo0}, we see that the two leading terms come from the solution on the critical line, $z^*=z_h+P$.

\subsubsection{Region II} 

Here the expansions require more care than in regions I and III. {Let us assume that $n=2,3$ and that $z_1$ interpolates between the leading behavior in regions I and III given in \eqref{sol11} and \eqref{sol13}.}

First, define $D$ and $X$ by
\be \label{DXdef}
z = z_h+D\ , \qquad {\al \ov \rho^{n-2}}=P\le(1-X\ri)
\ee
and note
\bea \label{DXest}
D&=&P-\ep z_1 +O\le(\ep^2\ri)+\cdots\lesssim O(P)  \ , \nn
X&=&{\al_1 \ov \al}{1 \ov \rho}+O\le({1 \ov \rho^2}\ri) +\cdots\sim O(P) \ .
\eea
Using $D$ and $X$ we can expand
\bea \label{hEBexp}
h&=&-c_n \ga_n D + O\le(D^2\ri)+\cdots\ , \nn
 EB&=&-\ga_{n}P\le(1-X\ri)\le(1+O\le(D\ri)+\cdots\ri)\ .
\eea
Also define $Y$ by
\be \label{zpexp}
z'=-\ga_n\le(P-D\ri)\le(1+Y\ri)\ ,
\ee
noting
\bea \label{Yest}
&&Y=O\le({b_1 \ov \rho}\ri)+O\le({1 \ov \rho^2}\ri)+\cdots+O\le({D \ov \le(P-D\ri)\rho}\ri)+\cdots\ , \nn
&&Y(D=0)\sim O(P)\ .
\eea

Now we divide region II into three subregions\footnote{Note subregions II$_2$ and II$_3$ each have two connected pieces.} 
\bea
\rm{II}_1 &:& |D| \ll O\le(P^2\ri)\ ,\nn 
\rm{II}_2 &:& |D| \sim O\le(P^2\ri)\ ,\nn 
\rm{II}_3 &:& |D| \gg O\le(P^2\ri)\ ,
\eea
and focus on calculating
\be
Q_s\equiv Q-1={z'^2 -E^2 B^2 \ov f+E^2B^2}
\ee
in subregions II$_1$ and II$_3$. Then
\be
f+E^2B^2 =\begin{dcases}
\ga_n^2P^2\le(1+O\le({D \ov P^2}\ri)+O\le(X\ri)+\cdots\ri) & \le(\rm{II}_1\ri)\\
-c_n\ga_n D\le(1+O\le(D\ri)+O\le({P^2 \ov D}\ri)+\cdots\ri) & \le(\rm{II}_3\ri)
\end{dcases}\ ,
\ee
and\footnote{Here we have only made potential leading terms explicit, with the exception that in the expression for subregion $\rm{II}_1$, the leading term proportional to $D$ have been noted, although it is subleading to terms without factors of $D$. This is used in calculating $v'$ in subregion II$_1$.}
\be
z'^2-E^2B^2=\begin{dcases}
2 \ga_n^2 P^2\le(X+Y+\cdots-{D \ov P}+\cdots\ri)
&  \le(\rm{II}_1\ri)\\ 
D \ga_n^2\le(-2P+D+\cdots\ri)& \le(\rm{II}_3\ri)
\end{dcases}\ ,
\ee
from which
\be \label{Qsexp}
Q_s=\begin{dcases}
2\le(X+Y+\cdots+O\le(D \ov P\ri)+\cdots\ri) & \le(\rm{II}_1\ri)\\
{2 \ga_n \ov c_n}\le(P-{D \ov 2}+\cdots\ri) & \le(\rm{II}_3\ri)
\end{dcases}\ .
\ee
Using expansions \eqref{hEBexp}, \eqref{zpexp}, and \eqref{Qsexp}, we have
\be
-z'+EB\sqrt{Q}=\begin{dcases}
O\le(D\ri)+\cdots & \le(\rm{II}_1\ri)\\
-\ga_n D+\cdots & \le(\rm{II}_3\ri)
\end{dcases}\ ,
\ee
from which
\be
v'=\begin{dcases}
O(1)+\cdots  & \le(\rm{II}_1\ri)\\
{1 \ov c_n}+\cdots & \le(\rm{II}_3\ri)
\end{dcases}
\ee
in subregions ${\rm II}_1$ and ${\rm II_3}$. But since the differential equation \eqref{modeom} does not contain any scales other than $z_h$, $v'$ should smoothly interpolate between subregions ${\rm II}_1$ and ${\rm II_3}$, i.e. it should also be $O(1)$ in subregion ${\rm II}_2$. Thus we conclude
\be \label{t2}
t_{\rm II}=\begin{dcases}
{k_1 - k_2 \ov c_n}+O\le(\log R\ri)+O(1) &n=2 \\
{k_1-k_2 \ov c_n}+O\le(1\ri) & n=3
\end{dcases}\ .
\ee

\subsubsection{Region III}
Here
\bea \label{exp31}
z&=&z_h-\ep z_1+P+O\le(\ep^2\ri)+\cdots\ , \nn
z'&=&-\ga_n \ep z_1\le(1+O\le({1 \ov \rho}\ri)+O\le(\ep\ri)+\cdots\ri)\ ,
\eea
and
\bea \label{exp32}
&&1+{z'^2 \ov h}=1+{2n \ov z_h}\le(\ep z_1 +P\ri)+O\le({\ep z_1 \ov \rho}\ri)+O\le(\ep^2\ri)+\cdots \ , \nn
&&1+{E^2 B^2 \ov h}=1+O\le({P^2 \ov \ep z_1}\ri)+\cdots\ .
\eea
Then 
\be
v'={1 \ov c_n}\le(1+O\le(\ep z_1\ri)+O\le({P \ov \ep z_1}\ri)+\cdots\ri)
\ee
and
\be \label{t3}
t_{\rm III}={1 \ov c_n}k_2+O(1)\ .
\ee

Finally, collecting \eqref{t1}, \eqref{t2}, and \eqref{t3}, we have
\bwt
\be
\ft=\begin{dcases}
{R\ov c_n}-\le({n-1 \ov 4 \pi T}-{|E|(3 h_1+h_2)z_h^2 \ov 2 h_1^2}\ri)\log R+O(\log R)+O(1) & n=2\\
{R\ov c_n}-{n-1 \ov 4 \pi T}\log R+O(1) & n=3
\end{dcases}\ .
\ee
\ewt
Note for $n=2$ there is an $O(\log R)$ piece that we were not able to determine.





\subsection{Action}
To calculate the action, we proceed in similar fashion. The action with its equilibrium value subtracted can be divided as
\bea \label{sAdiv}
 \sA-\sA_{\rm eq}&=&\sA_{\rm I}+\sA_{\rm II}+\sA_{\rm III}\nn
&\equiv&\le(\int_{k_0}^{R-k_1}+\int_{R-k_1}^{R-k_2}+\int_{R-k_2}^{R-k_3}\ri) d\rho  \, \le(\sA' - \sA_{\rm eq}'\ri)\nn
\eea
where the first equality holds up to $O(1)$ terms including the contribution from the AdS portion of extremal surfaces, and from \eqref{bhc},
\be
\sA'={\rho^{n-1} \ov z^n}\sqrt{Q}\ , \qquad \sA_{\rm eq}'={\rho^{n-1} \ov z_{\rm eq}}\sqrt{1+{z_{\rm eq}'^2 \ov h(z_{\rm eq})}}\ .
\ee
Here $\sA'$ is evaluated on the near-horizon expansion \eqref{poec} of the near-critical solution, and $\sA_{\rm eq}$ is evaluated on the near-horizon expansion \eqref{eqnh} of the equilibrium solution, where the $\ep$'s in the two expansions can be set equal.\footnote{Although the definition of the two $\ep$'s in \eqref{gscep} and \eqref{teipn} are different, their expansions in large $R$ \eqref{epr} and \eqref{bher} show that fixing $R$, they agree up to an $O(1)$ factor. This factor than can be absorbed into $z_{1,{\rm eq}}$ in \eqref{eqnhz1}.} Note that from \eqref{eqnh}, 
\be
1+{z_{{\rm eq}}'^2 \ov h(z_{{\rm eq}})}=1+{2n \ov z_h}\ep z_{1, {\rm eq}}+\cdots
\ee
and
\be
\sA_{\rm eq}'={\rho^{n-1} \ov z_h^n}\le(1+{2n \ov z_h}\ep z_{1, {\rm eq}}+\cdots\ri)\ .
\ee

Then in region I, from \eqref{exp11} and \eqref{exp12}, 
\bwt
\bea
\sA'-\sA_{\rm eq}'={\rho^{n-1} \ov z_h^n}\le(1+O\le({P \ov \rho^2}\ri)+O\le(\ep z_1\ri)+\cdots\ri)-{\rho^{n-1} \ov z_h^n}\le(1+O\le(\ep z_{1, {\rm eq}}\ri)+\cdots\ri)
\eea
\ewt

and one can check
\bwt
\be \label{b1anom}
\int^{R-k_1}_{k_0}d\rho \, {\ep z_1 \ov P} \sim O(1) \ , \quad \int^{R-k_1}_{k_0}d\rho \, {\ep z_{1, {\rm eq}} \ov P}\sim O\le(R^{b_1/2\ga_n}\ri) \ ,
\ee
\ewt
{so assuming $b_1<0$}, we have 
\be \label{sA1} 
\sA_{\rm I}=O(1)\ .
\ee
In region II, from \eqref{DXdef} and \eqref{Qsexp},
\bwt
\bea
\sA'-\sA_{\rm eq}'&=&{\rho^{n-1} \ov z_h^{n}}\le(1-n{D \ov z_h}+\cdots\ri)\le(1+{1 \ov 2}Q_s+\cdots\ri)-{\rho^{n-1} \ov z_h^n}\le(1-2n{\ep z_{1, {\rm eq}} \ov z_h}+\cdots\ri)\nn
&=&\begin{dcases}
{\rho^{n-1} \ov z_h^n}\le(X+Y\ri)+\cdots & \le(\rm{II}_1\ri) \\
{\rho^{n-1} \ov z_h^n}\le({\ga_n \ov c_n}P-\le({n \ov z_h}-{\ga_n \ov 2 c_n}\ri)D+\cdots\ri) & \le(\rm{II}_3\ri)
\end{dcases}
\eea
\ewt
and from the order of magnitudes of $D$, $X$, and $Y$ in \eqref{DXest} and \eqref{Yest}, $\sA'-\sA_{\rm eq}'$ is $O(1)$ in subregions II$_1$ and II$_3$. But as was the case with $v'$, $\sA'-\sA_{\rm eq}'$ must interpolate smoothly between subregions II$_1$ and II$_3$, so we conclude $\sA'-\sA'_{\rm eq}$ is $O(1)$ throughout region II and that
\be \label{sA2}
\sA_{\rm II}=\begin{dcases}
O\le(\log R\ri)+O(1) & n=2\\
O(1) & n=3
\end{dcases}\ .
\ee
Lastly, in region III, from \eqref{exp31} and \eqref{exp32},
\bwt
\bea \label{Adiff3}
\sA'-\sA_{\rm eq}'&=&{\rho^{n-1} \ov z_h^n}\le(1+{n \ov z_h}\le(\ep z_1-P\ri)+O\le(\ep^2 \ri)+\cdots\ri)\le(1+{n \ov z_h}\le(\ep z_1 +P\ri)+O\le(\ep z_1 \ov \rho\ri)+O\le(\ep^2\ri)+\cdots\ri)\nn
&&\le(1+O\le({P^2 \ov \ep z_1}\ri)+\cdots\ri)-{\rho^{n-1} \ov z_h^n}\le(1+{2n \ov z_h}\ep z_{1, {\rm eq}}+O\le(\ep z_{1, {\rm eq}} \ov \rho\ri)+O\le(\ep^2\ri)+\cdots\ri)
\eea
\ewt
where from \eqref{eqnhz1} and \eqref{sol13},
\be
z_{1, {\rm eq}} \sim z_1\ .
\ee
One can check the leading terms in \eqref{Adiff3} contribute at
\be
\int_{R-k_2}^{R-k_3} d\rho \, \begin{dcases}
{\ep z_1 \ov P} &\sim O\le(e^{-\ga_{n} R} R^{2(n-1)}\ri)\nn
{1 \ov \rho} & \sim O\le({\log R \ov R}\ri) 
\end{dcases}
\ee
so we have
\be \label{sA3}
\sA_{\rm III}=O(1)\ .
\ee
Collecting \eqref{sA1}, \eqref{sA2}, and \eqref{sA3}, we arrive at 
\be
\sA-\sA_{\rm eq}=\begin{cases}
O\le(\log R\ri)+O(1) & n=2 \\
O(1) & n=3
\end{cases}\ ,
\ee
where for $n=2$ we have an undetermined $O(\log R)$ piece.\footnote{{If $n=2$ {\it and} $b_1>0$, there are also $O(R^{b_1/(2 \ga_n)})$ contributions from \eqref{b1anom}.}}

\bibliographystyle{apsrev4-1}
\bibliography{quench_final}

\end{document}